\documentclass{emulateapj}

\newcommand{\citepeg}[1]{\citep[{e.g.,}][]{#1}}
\def\Swift{\textit{Swift}}

\newcommand{\OII}{[\ion{O}{2}]}
\newcommand{\OIII}{[\ion{O}{3}]}

\newcommand{\na}{Nature}
\newcommand{\actaa}{Acta Astronomica}
\newcommand{\nar}{New Astronomy}

\shorttitle{The Hosts of Dust-Obscured Gamma-Ray Bursts}
\shortauthors{Perley et al.}

\begin{document}

\title{A Population of Massive, Luminous Galaxies Hosting Heavily Dust-Obscured Gamma-Ray Bursts:  Implications for the Use of GRBs as Tracers of Cosmic Star Formation}

\def\cit{1}
\def\hubble{2}
\def\warwick{3}
\def\leicester{4}
\def\ucb{5}
\def\goddard{6}
\def\dark{7}
\def\stsci{8}
\def\iceland{9}
\def\ucsc{10}
\def\texas{11}
\def\mail{*}

\author{D.~A.~Perley\altaffilmark{\cit,\hubble,\mail},
  A.~J.~Levan\altaffilmark{\warwick},  
  N.~R.~Tanvir\altaffilmark{\leicester},
  S.~B.~Cenko\altaffilmark{\ucb,\goddard},
  J.~S.~Bloom\altaffilmark{\ucb},
  J.~Hjorth\altaffilmark{\dark},
  T.~Kr{\"u}hler\altaffilmark{\dark},
  A.~V.~Filippenko\altaffilmark{\ucb},
  A.~Fruchter\altaffilmark{\stsci},
  J.~P.~U.~Fynbo\altaffilmark{\dark},
  P.~Jakobsson\altaffilmark{\iceland},
  J.~Kalirai\altaffilmark{\stsci},
  B.~Milvang-Jensen\altaffilmark{\dark},
  A.~N.~Morgan\altaffilmark{\ucb},
  J.~X.~Prochaska\altaffilmark{\ucsc}, and
  J.~M.~Silverman\altaffilmark{\texas}
}
\altaffiltext{\cit}{Department of Astronomy, California Institute of Technology,
MC 249-17,
1200 East California Blvd.,
Pasadena, CA 91125}
\altaffiltext{\hubble}{Hubble Fellow}
\altaffiltext{\warwick}{Department of Physics, University of Warwick, Coventry CV4 7AL, UK}
\altaffiltext{\leicester}{Department of Physics and Astronomy, University of Leicester, Leicester LE1 7RH, UK}
\altaffiltext{\ucb}{Department of Astronomy, University of California, Berkeley, CA 94720-3411}
\altaffiltext{\goddard}{NASA/Goddard Space Flight Center, Greenbelt, MD 20771}
\altaffiltext{\dark}{Dark Cosmology Centre, Niels Bohr Institute, Copenhagen, Denmark}
\altaffiltext{\stsci}{Space Telescope Science Institute, Baltimore, MD 21218}
\altaffiltext{\iceland}{Centre for Astrophysics and Cosmology, Science Institute, University of Iceland, Dunhagi 5, 107 Reykjav\'ik, Iceland}
\altaffiltext{\ucsc}{Department of Astronomy and Astrophysics, UCO/Lick Observatory, University of California, Santa Cruz, CA 95064}
\altaffiltext{\texas}{Department of Astronomy, University of Texas, Austin, TX 78712}
\altaffiltext{\mail}{e-mail: dperley@astro.caltech.edu}

\slugcomment{Submitted to ApJ 2013-01-24, accepted 2013-09-27}

\begin{abstract}
We present observations and analysis of the host galaxies of 23 heavily dust-obscured gamma-ray bursts (GRBs) observed by the \Swift\ satellite during the years 2005--2009, representing all GRBs with an unambiguous host-frame extinction of $A_V>1$ mag from this period. Deep observations with Keck, Gemini, VLT, {\it HST}, and {\it Spitzer} successfully detect the host galaxies and establish spectroscopic or photometric redshifts for all 23 events, enabling us to provide measurements of the intrinsic host star-formation rates, stellar masses, and mean extinctions. Compared to the hosts of unobscured GRBs at similar redshifts, we find that the hosts of dust-obscured GRBs are (on average) more massive by about an order of magnitude and also more rapidly star-forming and dust-obscured.  While this demonstrates that GRBs populate all types of star-forming galaxies including the most massive, luminous systems at $z \approx 2$, at redshifts below 1.5 the overall GRB population continues to show a highly significant aversion away from massive galaxies and a preference for low-mass systems relative to what would be expected given a purely SFR-selected galaxy sample. This supports the notion that the GRB rate is strongly dependent on metallicity, and may suggest that the most massive galaxies in the Universe underwent a transition in their chemical properties $\sim 9$ Gyr ago. We also conclude that, based on the absence of unobscured GRBs in massive galaxies and the absence of obscured GRBs in low-mass galaxies, the dust distributions of the lowest-mass and the highest-mass galaxies are relatively homogeneous, while intermediate-mass galaxies ($\sim 10^9$\,M$_\odot$) have diverse internal properties.
\end{abstract}

\bigskip

\clearpage

% S1
\section{Introduction}
\label{sec:intro}

Long-duration gamma-ray bursts (GRBs\footnote{``Long-duration'' GRBs are generally defined as events with durations of $T_{90} \gtrsim 2$\,s.  These are distinguished from short-duration GRBs (typically $T_{90} < 2$\,s), which appear to have a completely different origin, most likely associated with the merger of compact objects \citepeg{Nakar2007,Berger2011}.  Among \Swift\ bursts \citep{Gehrels+2004} with detected afterglows, observed long-duration events outnumber short-duration events by more than 10 to 1, and for simplicity, in the remainder of this paper we will use the term ``GRB'' to refer only to the long-duration class.}) represent a rare, violent endpoint of stellar evolution.  A GRB is generated when a newly formed compact object (a neutron star or black hole\footnote{More exotic compact-object constructs such as quark stars have also been considered; e.g., \cite{Paczynski+2005}.}) within a massive star is able to briefly power a relativistic jet that pushes through the stellar envelope and into the circumstellar medium \citepeg{Usov+1992,Woosley1993}.  Both the radiation associated with the ejecta during the explosion (the prompt gamma-ray and X-ray emission) and the longer-lived multiwavelength afterglow that follows (the emission produced by the relativistic shock wave that results from this explosion; \citealt{Rees+1992,Sari+1998}) are extremely luminous, so they are detectable out to high cosmological redshifts (including a few at redshift $z>8$; e.g., \citealt{Tanvir+2009,Salvaterra+2009,Cucchiara+2011}) even in short observations with mid-sized telescopes.

The association between GRBs and the destruction of massive, short-lived stars \citepeg{Galama+1998,Hjorth+2003,Woosley+2006} predicts that GRBs should form exclusively in star-forming environments.  This prediction generally seems to be upheld---the hosts of GRBs are ubiquitously young and essentially always show evidence of recent star formation, both in an integrated sense \citepeg{Savaglio+2009} and at the precise location of the GRB within the galaxy \citep{Fruchter+2006}.  

A more complicated, still unsettled question is whether GRBs form in \emph{all} star-forming environments---or at least, if they do so in proportion to the star-formation rate (SFR) as one might naively expect given their direct association with young stars.  While individual GRBs are certainly useful probes for studying individual star-forming galaxies, a direct, linear association with cosmic star formation would add tremendous \emph{statistical} power to studies of the GRB host population, directly constraining the relative cosmic SFR in galaxies of different types (i.e., as a function of host mass, luminosity, extinction, morphology, etc.) as well as its evolution with redshift, even in galaxies well below the detection limit of most flux-limited samples \citep{Natarajan+1997,Hogg+1999,Djorgovski+2001,Fynbo+2001,RamirezRuiz+2002,Berger+2003,Jakobsson+2005b,Tanvir+2012}.

Theoretically, there are many reasons to expect that this ideal may not be met in reality.  Metallicity in particular is thought to play an important role in massive-star evolution: metals provide opacity to the stellar envelope, helping an evolved star to expel its diffuse hydrogen and helium layers, which must be removed by some means to enable the jet to escape the star and to be consistent with the observation of hydrogen-free Type Ic supernovae (SNe; see \citealt{Filippenko+1997} for a discussion of SN classification) associated with GRBs\footnote{Alternatively, some models enable the star to evolve homogeneously by continuously mixing the outer layers into the core, thereby converting the entire envelope to heavier elements over the course of its lifetime \citep{Yoon+2005,Woosley+2006}.  Very massive stars may also be able to shed their outer envelopes in massive eruptions having little to do with the envelope's line opacity \citep{Smith+2006}.}. Metals also help strip the star of angular momentum, a process which may inhibit the central engine \citep{Woosley+1999,Woosley+2006}.   Given the large variation in average metallicity between different galaxies, these effects could produce large deviations between the SFR and GRB rate in different galaxies \citep{Hirschi+2005}.  Empirically, metallicity does indeed seem to affect the relative numbers of different types of SNe \citep{Arcavi+2010}, including the Type Ic broad-lined SNe which accompany GRBs.  

Metallicity need not be the only factor:  recent evidence for variation in the initial stellar mass function (IMF; \citealt{vanDokkum+2010,Conroy+2012}) could also produce some variation of the GRB rate in relation to other tracers, since GRBs are probably generated only by extremely massive stars (at least 20--50\,M$_\odot$; \citealt{Mazzali+2003,Ostlin+2008}).  If the distribution of other stellar initial properties beyond mass alone (such as rotation or binary separation) exhibits similar dependencies on environment, these could (in principle) also affect the GRB rate relative to that of overall star formation.

While some studies \citepeg{Jakobsson+2005b,Fynbo+2008,Chen+2009,Mannucci+2011,Michalowski+2012b} do support the notion that the GRB rate is consistent with a model in which it is strictly proportional to the overall SFR in some situations, the notion that the GRB rate shows significant deviations from the prediction of a uniform GRB-to-SFR ratio as a function of environment has received observational support from a number of other studies.  For instance, the number of GRBs reported within spiral galaxies appears to be much lower than predicted given the amount of total star formation (as traced by the rates of Type II SNe) happening in these galaxies \citep{Fruchter+2006,Wainwright+2007}, while the number of GRBs in extremely low-mass, low-metallicity systems appears to significantly exceed predictions \citep{Stanek+2006,Modjaz+2008,Levesque+2010g,Graham+2013}.   At higher redshifts, a significant dearth of GRBs within luminous infrared galaxies (LIRGs; $L_{\rm IR} > 10^{11}$\,L$_\odot$) has also been reported: for example, the work of \cite{LeFloch+2006} found only three LIRG hosts from a sample of 16 GRBs observed, when in reality such galaxies are thought to be responsible for about half of all star formation at $z\gtrsim1$.  Similarly, only a handful of GRBs within ultraluminous infrared galaxies (ULIRGs; $L_{\rm IR} > 10^{12}$\,L$_\odot$) or submillimeter galaxies (SMGs) are known \citep{Berger+2003,Tanvir+2004}, even though these systems contribute substantially to, and may dominate, the cosmic SFR at higher redshifts ($z \gtrsim 1.5$; e.g., \citealt{Smolcic+2009,PerezGonzalez+2005,Michalowski+2010}).

However, most of these studies suffered from a significant limitation.  The ability to search for host galaxies is limited by the need to localize the GRB to subarcsecond precision, and while this is \emph{possible} at many wavelengths (radio, millimeter, near-infrared [NIR], optical, and X-rays are all frequently employed), the majority of afterglow positions before the launch of the \Swift\ satellite were provided optically\footnote{See, for example, the statistics in the table compiled at http://www.mpe.mpg.de/$\sim$jcg/grbgen.html .}.  For events only observed at optical wavelengths, the presence of significant interstellar extinction within a GRB host galaxy could easily conceal the optical afterglow and therefore prevent identification of the host.  If the dust properties of the GRB sightline correlate with those of the host galaxy itself, these dust-obscured GRBs could potentially hide an entire class of hosts with properties quite similar to those that were largely ``missing'' from these pre-\Swift\ works (massive, luminous, and dusty).

Dust-obscured GRBs do exist and even appear to be fairly common, manifesting themselves as so-called ``dark'' GRBs, events with abnormally faint (and as a result usually undetected) optical afterglows\footnote{``Darkness'' can be defined more quantitatively in various ways, and several (conflicting) definitions are employed in the literature (see \S \ref{sec:sample} for an expanded discussion).  Here we use the term more loosely to denote events with atypically faint optical afterglows relative to other wavelengths or to other GRBs at the same epoch.}.  Dark GRBs have been known almost since the beginning of the afterglow era \citep{Groot+1998}, and while dust extinction is not the only possible explanation (absorption by the neutral intergalactic medium at high redshift or intrinsic effects could also produce a faint optical afterglow), it has been favored over alternative interpretations for most well-studied pre-\Swift\ dark GRBs \citep{Taylor+1998,Djorgovski+2001,Klose+2003,Gorosabel+2003b,Jakobsson+2004} and, more recently, for the the large majority of dark GRBs within unbiased samples of \Swift\ events as well \citep{Cenko+2009,Perley+2009b,Greiner+2011}.  Quantitatively, $\sim 25$\% of all \Swift\ GRBs are too faint to detect even if followed up immediately with a 2\,m-class ground-based telescope, most of which (60--80\%, or 15--20\% of all GRBs) are heavily obscured.   Studies of the X-ray attenuation of optically dark GRBs have reached similar conclusions \citep{Fynbo+2009,Melandri+2012,Watson+2012}.  It is of obvious interest to closely examine the host galaxies of these dust-obscured events, both to determine whether they have the potential to alter our conclusions about how the GRB rate is connected to the cosmic SFR, and to extend the detailed analysis permitted by GRB host studies to a wider range of environments than those probed only by unobscured sightlines.

Fortunately, the ability to localize events without a bright optical afterglow has dramatically improved over the past decade.  Ground-based follow-up capabilities have substantially developed since the pre-\Swift\ era, and the early nondetection of an optical afterglow regularly motivates deeper follow-up studies from 8\,m-class telescopes (frequently at NIR wavelengths) or observations in unobscured parts of the electromagnetic spectrum (X-ray, submillimeter, or radio) that do successfully detect a counterpart.  But even if no additional detections are secured, \Swift's onboard X-ray Telescope (XRT; \citealt{Burrows+2005}) effectively guarantees a $\sim 2$\arcsec\ position for every long-duration burst \citep{Butler2007,Goad+2007,Evans+2009}, sufficient to localize a host galaxy with reasonable confidence a large majority of the time.  Earlier X-ray cameras on GRB-detecting satellites had much poorer angular resolution, and were not sufficient for uniquely identifying the host galaxy.

Studies to date have provided mixed results regarding the extent to which the host population unveiled by dark GRBs actually differs from the host population identified by optically bright GRBs.  The pre-\Swift\ sample of \cite{LeFloch+2006} did include a few dark GRBs, only one of which had a sufficiently large stellar mass or mid-IR dust emission to be detected in {\it Spitzer Space Telescope (Spitzer)} observations.   The darkest GRB hosts within the uniform \Swift\ sample of \cite{Cenko+2009} did not have markedly unusual optical properties distinguishing them from other GRB hosts \citep{Perley+2009b}, and the late-time observations of the host of GRB 060923A by \cite{Tanvir+2008} identified only a faint host with fairly ordinary color.   However, the hosts of several other individual dark GRBs have now shown properties that are actually quite unlike those of the supposedly ``typical'' low-mass, low-$A_V$ host galaxy.  Specifically, GRBs 051022 \citep{CastroTirado+2007,Rol+2007}, 080207 \citep{Svensson+2012,Hunt+2011}, 080325 \citep{Hashimoto+2010}, and 080607 \citep{Chen+2010} have now all been associated with quite luminous and massive host galaxies, and the host of GRB 020819 has recently been shown to be a high-metallicity spiral \citep{Levesque+2010f}.  While the hosts of optically bright bursts are occasionally quite luminous in the ultraviolet (UV) and can also be moderately massive and occasionally metal rich \citepeg{Levesque+2010g,Kruehler+2012a}, substantial \Swift\ and pre-\Swift\ host surveys and compilations \citep{Chen+2009,Savaglio+2009} have turned up only a few host galaxies with stellar masses or average dust extinctions even approaching those of the dark GRB hosts above.  Still, given that the study of these individual objects and publication of the resulting discovery was in some cases surely motivated by the properties of the host itself, it is difficult to determine whether these events represent the ``typical'' dark-burst host or are rare exceptions to the blue-and-faint rule.

The study of \cite{Kruehler+2011} was the first to extend these investigations of individual objects to the broader population, using a sample of eight dark GRBs from the literature and from the Gamma-Ray Burst Optical Near-IR Detector (GROND), supplemented by New Technology Telescope (NTT) and Very Large Telescope (VLT) observations.  Mirroring the results for individual dark GRBs, they measure a wide range of properties among the hosts in this sample, ranging from relatively small, minimally obscured galaxies typical of nondark hosts up to very luminous, massive, and dusty galaxies.  While their sample size is too small to make strong statistical statements, these results suggest that massive hosts are indeed reasonably common (if not ubiquitous) among dark GRBs.  Similar results (i.e., a substantial fraction of very red galaxies) were seen in the study of \cite{Rossi+2012}, which targeted 17 GRBs with no detected optical afterglow.   A much larger sample of 69 uniformly selected hosts observed with the VLT was presented by \cite{Hjorth+2012}.  While only two-color ($R$ and $K$) photometry of each host is available and deep constraints on the presence or absence of an optical afterglow are not always available, bursts with no detection of an optical afterglow do seem to have substantially redder hosts on average (Malesani et al. 2013, in prep.), suggesting a trend toward dustier and more massive hosts.

It is necessary to continue moving from a regime dominated by studies of individual objects toward the statistical examination of large samples based on understandable selection effects.   To this end, over the past several years we have been conducting a comprehensive, multiwavelength campaign devoted to the observation and characterization of optically dark GRBs and their hosts. Here we present the first results of our campaign---confirming that massive, dusty, luminous hosts are in fact typical (but not ubiquitous) among the population of ``dark'' bursts, and examining in detail the implications of this discovery for the overall population of GRB hosts and for the connection between the GRB rate and SFR.

An outline of this paper is as follows.  In \S \ref{sec:sample} we briefly describe our selection of the sample.  In \S \ref{sec:obs} we present our observations of the hosts with Keck, Gemini, {\it Spitzer}, and the {\it Hubble Space Telescope (HST)}.  Section \ref{sec:modeling} describes our method for estimation of stellar masses, SFRs, and minimum bolometric luminosities from these measurements and other observations in the literature. In \S \ref{sec:hosts} we outline each burst individually; we describe the characteristics motivating its inclusion in the sample, identify its host galaxy and discuss observations thereof, and summarize the host's properties.  We examine in \S \ref{sec:properties} the properties of the entire sample as a population in comparison to other, previously published samples of GRBs and other populations of high-$z$ galaxies.  In \S \ref{sec:discussion} we discuss the implication of our results for the origins of dark bursts, the distribution of dust in high-$z$ galaxies, and the ability of GRBs to serve as unbiased probes of star formation. Our conclusions are summarized in \S \ref{sec:conclusions}.

% Figure 1
\begin{figure}
\centerline{
\includegraphics[scale=0.75,angle=0]{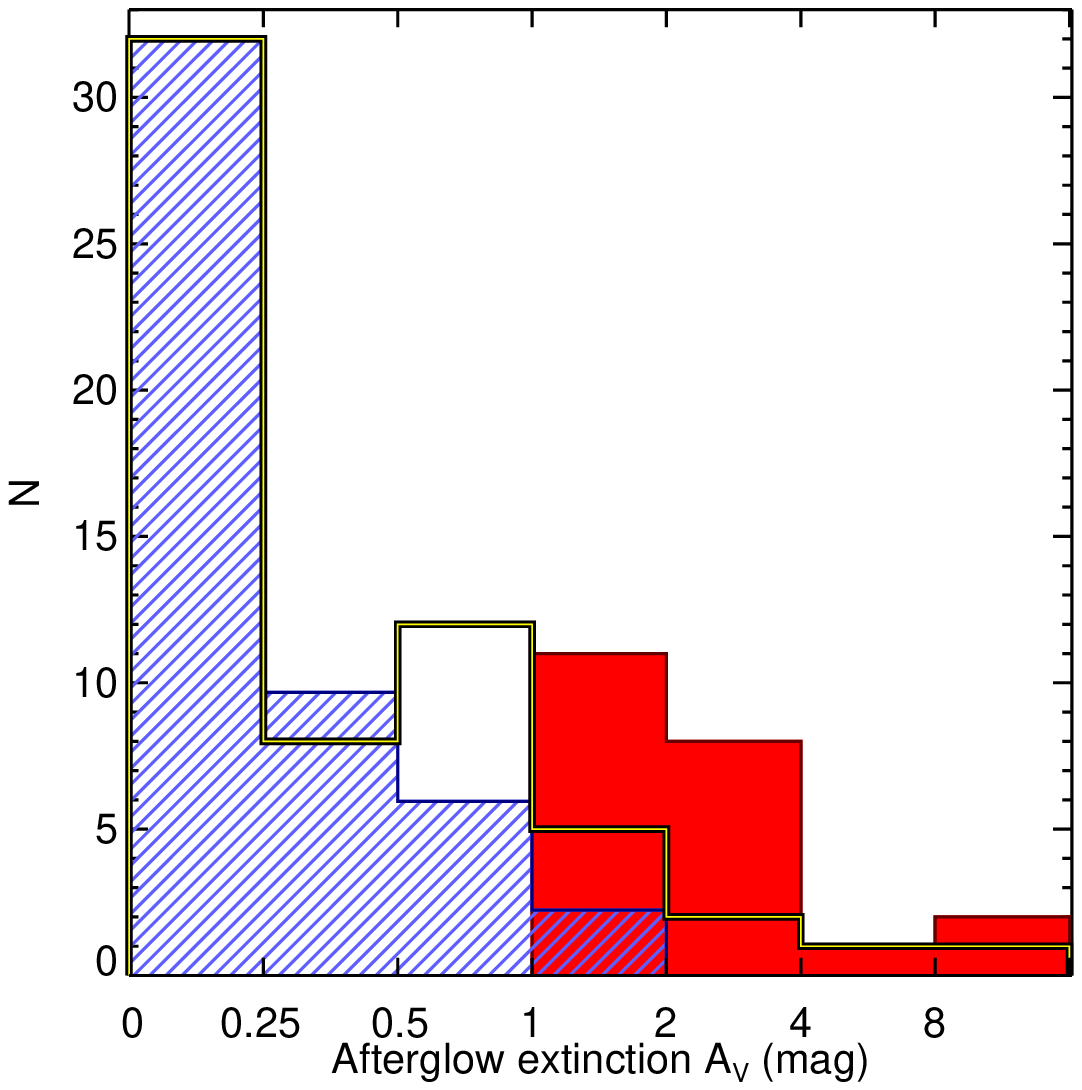}} 
\caption{Histogram of measured or limiting afterglow extinction columns (in terms of rest-frame $A_V$) of several samples of \Swift\ GRBs, including events presented in this paper.  The unfilled yellow-and-black line shows an estimate of the intrinsic, unbiased distribution based on a nearly complete sample of events combined from \cite{Cenko+2009} and \cite{Greiner+2011}.  The hatched blue histogram shows measured values of $A_V$ for a sample of well-observed, optically bright events from \cite{Kann+2006} and \cite{Kann+2010}, rescaled to match the $<0.25$\,mag bin for the complete sample; the $\sim 20$\% dustiest events are systematically missed in these optically selected studies including effectively all GRBs with $A_V > 2$\,mag.  The solid red histogram shows the $A_V$ measurements (or lower limits) for the 23 events analyzed in this sample.}
\label{fig:avhist}
\end{figure}

% S2
\section{Sample Selection}
\label{sec:sample}

\subsection{Motivation}

Our primary goals in this paper are to \emph{characterize the GRB host population missed in earlier work due to dust extinction of the optical afterglow} and to \emph{examine the implications for inclusion of this missing population} on larger questions of the coupling between the GRB rate and SFR, and on general questions relating to GRB hosts and high-redshift galaxies.  Approaching this task is not straightforward, since optical coverage of GRBs is spotty because of the unpredictable times and locations of these events---GRBs often occur too close to the Sun or Moon to observe, and ground-based observations may by stymied by weather and other conditions.  Many optically bright bursts are missed owing to these prosaic reasons, and without careful attention to selection (for example, had we merely chosen a sample of events with no reported afterglow) our sample may be ``polluted'' by ordinary events and lead us to incorrect conclusions about the types of events that are \emph{systematically} missed.

In addition, many factors other than dust extinction can influence the optical brightness (and therefore detectability) of a particular GRB at a given time:  the burst's energetics, its circumstellar density, the temporal evolution of its light curve, as well as a variety of microphysical parameters and of course the burst's distance and redshift \citepeg{Sari+1998,Groot+1998,Djorgovski+2001,Fynbo+2001,Lazzati+2002,Nysewander+2009}.  These causes are all physically distinct, and most of these factors probably have little to do with the large-scale properties of the host galaxy the burst inhabits.  On the other hand, given the strong correlations observed between the mean obscuration of a galaxy's stellar population and its fundamental properties such as mass and SFR \citepeg{Meurer+1999,Shapley+2001}, there is significant reason to expect that dust-obscured GRBs may indeed reveal a different host population.  Dust obscuration also is the predominant cause of optical nondetection within samples where GRBs \emph{are} well observed at early times \citep{Cenko+2009,Perley+2009b,Greiner+2011}, especially among luminous and well-observed populations.

For these reasons, in this study we focus \emph{only} on GRBs whose afterglows have been heavily absorbed by dust in the interstellar medium (ISM) of their host galaxy.  We will generally refer to this class as \emph{dust-obscured} GRBs to distinguish them from the more general class of optically ``dark'' bursts, which can result from numerous causes (or combinations of causes), although we will occasionally continue refer to dust-obscured events using simply ``dark'' as a shorthand.

\subsection{Selection Criteria and Implementation}
\label{sec:select}

Since we desire a quasi-complete sample of all known dust-obscured events during a given time period to avoid biasing our host sample toward galaxies that are particularly bright or have other conspicuous properties that may have attracted our (or others') attention first, we restrict our search at the outset to events observed by \Swift\ that occurred during the five-year period of 2005--2009.

Within this temporal window, our primary condition for the inclusion of a given GRB in our sample is direct evidence (from observations of the X-ray, optical, and NIR afterglow) for \emph{extinction of at least $A_V = 1$\,mag} in the host rest frame\footnote{Throughout this section we remove Galactic foreground extinction at the outset using the dust maps of \citet{Schlegel+1998} and neglect intergalactic extinction.}.  We select this threshold on the basis of the fact that that very few optically bright GRBs exceed it \citep{Kann+2006,Kann+2010,Schady+2007,Schady+2010}\footnote{Although 1\,mag of attenuation is a relatively small amount and GRBs are very luminous, since the typical GRB lies at $z = 1$--3 the actual observed attenuation in the observer-frame optical is much larger than this, since optical filters correspond to the rest-frame UV at these redshifts.}, yet events of higher $A_V$ do represent a significant contribution to the overall GRB rate ($\sim 15$\%; Figure \ref{fig:avhist}).  GRBs with $A_V > 1$\,mag therefore represent a population that is intrinsically common yet highly underrepresented in previous host-galaxy work.   This magnitude threshold is also achievable in practice: as long as a burst is followed up rapidly with a moderate-size telescope (or within the first day by a large-aperture telescope) it is usually possible to determine whether it has $A_V < 1$\,mag or $A_V > 1$\,mag.

Since for heavily obscured events the optical or NIR afterglow is often not detected at all, making this determination requires a few assumptions about the intrinsic spectrum.   Following \cite{Jakobsson+2004}, we assume that based on simple synchrotron models \citepeg{Sari+1998} a fading GRB afterglow must have an intrinsic spectral index (defined as $F_\nu \propto \nu^{-\beta}$) of $\beta_{\rm OX} \geq 0.5$ in all circumstances, or (more stringently) an intrinsic $\beta_{\rm OX} \geq \beta_{\rm X,min}-0.5$ \emph{if} late-time XRT observations \citep{Butler+2007b} indicate an X-ray spectral slope steeper than $\beta_{X} = 1.0$ at 95\% confidence (following \citealt{vanderHorst+2009}, but we use the 95\% confidence lower limit on $\beta_{X}$ rather than the best-fit value). 

For events where there is no detection of an optical/NIR counterpart (or there is one detection in only a single NIR filter, typically the $K$ band), we simply take this baseline minimum intrinsic flux and determine the minimum extinction needed \emph{beyond} this to explain the optical/NIR nondetections.  The \Swift\ XRT X-ray flux is determined from the \Swift\ data pages \citep{Evans+2007,Evans+2009}, converted to a 1\,keV flux density using the time-averaged spectral index and absorption correction, and interpolated to the exact time of observation based on a power-law fit to neighboring data points.  An extinction curve like that of the Small Magellanic Cloud (SMC)\footnote{The SMC curve is employed because it produces the largest rest-frame UV (observer-frame optical for typical \Swift\ GRB redshifts) extinction for a given $A_V$ among the Local Group extinction curves, and therefore is the most appropriate choice for evaluating the minimum $A_{V}$.  Of course, even steeper extinction curves, such as those implied from populations of Type Ia SNe and perhaps some GRBs \citep{Poznanski+2009,Butler+2006,Zafar+2011}, could produce (slightly) lower values of $A_V$ for a given $A_{\rm UV}$, so in this sense our $A_V$ limits are not strict.  Nevertheless, for consistency and simplicity we use the SMC curve in this work in cases where the extinction curve cannot be inferred directly from multicolor detections of the afterglow.}, with redshift fixed to the host value (if known) or to a fiducial $z=2$ (otherwise), is then applied to determine how much dust is necessary to interpret the upper limit or measurement.

Even if analysis of the optical/NIR points individually versus the X-ray flux does not unambiguously establish a large $A_V$, a separate constraint can be established based on the optical or NIR \emph{color}---an extremely red afterglow can potentially provide much stronger (and physically definitive) constraints on extinction than even a quite deep nondetection.  If the afterglow is detected in at least one filter, all UV/optical/IR points and limits are scaled to a single epoch.  We use the optical light curve if there are enough observations in the same filter to establish it; otherwise we assume an optical decay rate in the range $0 < \alpha < 2$, conservatively selecting the value that produces the bluest spectral energy distribution (SED).  We then again calculate the minimum $A_V$ that can simultaneously explain these observations (and the X-ray observations) within the synchrotron model.  Both an unbroken intrinsic optical-to-X-ray SED and an SED with a break between the bands ($\beta_O = \beta_X - 0.5$) are attempted.

In principle, we must take into account the possibility of Lyman-$\alpha$ or Lyman-break absorption from the intergalactic medium in this analysis, since these can also significantly attenuate the flux in blue or UV filters or at high redshift.  At most redshifts typical of GRBs this is not a concern, since in practice the redder filters ($R$ band and redward) always provide the strongest constraints on extinction anyway, and the $R$ band is not significantly suppressed at $z \lesssim 5$.  However, we do have to consider the possibly that a given GRB is at $z>5$ as an alternative hypothesis; higher-$z$ GRBs represent a small but nontrivial fraction of the GRB sample (about 5--10\%; e.g., \citealt{Perley+2009b,Fynbo+2009,Greiner+2011,Jakobsson+2012}).  We can exclude the high-redshift hypothesis in any of the following ways.  

\begin{enumerate}
  \item From detection of the transient in any optical filter bluer than the filter(s) used to determine the presence of significant $A_V$.  (The combined effects of Lyman-break and Lyman-$\alpha$ absorption produce a sharp cutoff at $z>5$ and partial attenuation is only possible in one broad-band filter at a time.)
  \item From a measured $\beta_{\rm OX} < 0.5$ in the $K$ or $H$ bands, which are not absorbed by neutral gas except at extremely high redshifts ($z>11$).
  \item From detection of significant X-ray absorption in excess of the expected Galactic value.  Following \cite{Grupe+2007}, we determine the $N_{\rm H}$ excess (fit from the XRT data at $z=0$ with the known Galactic column subsequently subtracted) to rule out very high redshifts.  Specifically, we employ the automatic online tables of \cite{Butler+2007b}\footnote{\url{http://butler.lab.asu.edu/Swift/xrt\_spec\_table.html}} and apply a minimum threshold of $N_{\rm H} > 10^{21}~ {\rm cm}^{-2}$ to the lower limit on $N_{\rm H}$ in that work. (The value of $10^{21}~{\rm cm}^{-2}$ corresponds to a limit of $z_{\rm max}=5$ in the formula given by \cite{Grupe+2007}. While this criterion was developed based only on events through 2007, we empirically verified that it still comfortably excludes all known $z>5$ events among \Swift\ GRBs through to the present time.)
\end{enumerate}

In all cases, the lower-redshift association was independently verified by the detection of a host galaxy underlying the afterglow position and direct measurement of its redshift spectroscopically or photometrically.  (However, host detection and redshift measurement was \emph{not} a criterion for inclusion.)

The implementation of this procedure in practice will be described in detail in a separate work (Perley et al. 2014, in prep.)  In brief, we downloaded all photometric observations (including upper limits) from the GCN Circulars\footnote{GCN Circulars are, by their nature, preliminary reports and may contain additional, unreported systematic errors due to calibration uncertainties or mistakes.  In cases where we have access to the original data we reanalyzed and recalibrated these observations to confirm or update the GCN results.  In general, however, the inclusion of an event in this sample would not be affected by even fairly large errors in calibration or photometry unless an event is very close to the $A_V$ threshold.} in 2005--2009 as well as the online library of XRT light curves provided by \citet{Evans+2007,Evans+2009}, and automatically calculated $\beta_{\rm OX}$ and $A_{V,{\rm min}}$ for each point, as well as $A_{V, {\rm fit}}$ if possible.  Events close to satisfying the $A_V > 1$\,mag criterion were followed up in more detail with observations from published or unpublished sources as necessary (where possible) in order to verify this association and provide the deepest possible limit.
 
A few targets were excluded (specifically GRBs 050716, 060923C, and 080229A) because of the presence of a bright Galactic star within 2\arcsec\ of the GRB afterglow position.  We also excluded GRB 070412, which is in the outer halo of a bright foreground galaxy.  Moreover, we elected not to include GRBs 060807, 080605, 080805, and 090926B, for which certain estimates indicate a best-fit value slightly above $A_V \approx 1.0$\,mag but with a range of uncertainty (or alternate solutions) permitting a lower value \citep{Kruehler+2011,Zafar+2012}.  After these exclusions, the final $A_V > 1$\,mag sample contains a total of 23 GRBs as summarized in Table \ref{tab:grbprop}.  In brief, the sample has the following properties.

\begin{itemize}
\item All but one event (GRB 051022) were discovered by the \Swift Burst Alert Telescope (BAT).  While we do not restrict our search to \Swift\ bursts, our procedure requires XRT follow-up observations nearly simultaneous with the optical measurements in order to evaluate the spectral index.

\item Eight events were detected in multiple NIR/optical filters and can provide a direct estimate (that is, well-constrained lower \emph{and} upper limits) on the rest-frame extinction.  Six were detected in only one filter (always only the $K$ band), and while an upper limit on $A_V$ could be placed in principle by assuming a minimum redshift and maximum $\beta_{OX}$, it would be very large (typically $\sim 10$--30\,mag) and not particularly useful.  The remaining ten were not detected in any NIR/optical filter and have only lower limits on $A_V$.

\item All but four events (GRBs 060923A, 070802, 071021, and 080325) have detection of significant $N_{\rm H}$ excess.  Among the exceptions, GRB 070802 has a spectroscopic afterglow redshift and the remaining three had $K$-band afterglow detections below $\beta_{\rm OX}=0.5$, so they are also extinguished (as opposed to being at high redshift).  Indeed, additional observations confirmed the presence of a host galaxy underlying all four positions.
\end{itemize}

\begin{figure*}
\centerline{
\includegraphics[scale=0.85,angle=0]{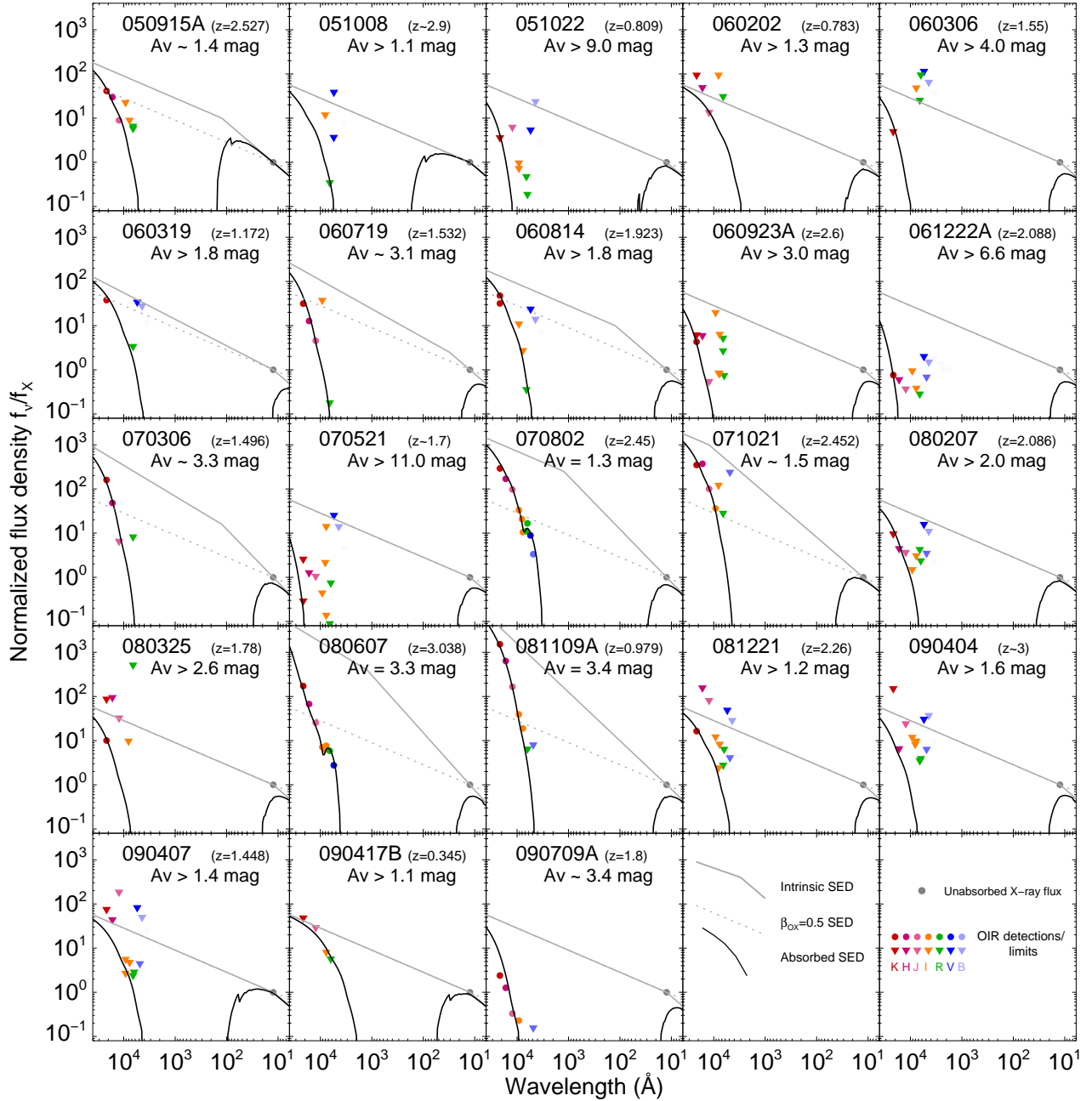}} 
\caption{Equivalent coeval \emph{afterglow} SEDs for the sample of 23 \Swift-observed GRBs whose host galaxies represent the subject of this paper.  The gray line shows the intrinsic afterglow flux corresponding to the model with minimum $A_V$ that is consistent with the observations; the black curve indicates the observable flux for this model including extinction and X-ray absorption.  The dashed line indicates $\beta_{\rm OX} = 0.5$ unless the minimum-flux model itself assumed this value.   SEDs are normalized to the observed (absorption-corrected) X-ray flux at 1\,keV.}
\label{fig:minextseds}
\end{figure*}

\noindent
The prompt-emission and afterglow properties of our sample are also summarized in Table \ref{tab:grbprop} and in Figure \ref{fig:minextseds}.

% Figure 2

\subsection{Impacts of Possible Biases in Sample Selection}

Our sample is necessarily not complete, constituting 23 events out of $\sim$400 long-duration GRBs localized by the XRT during the five-year window (among which $\sim 40$--80 probably had a ``true'' $A_V > 1$\,mag).  The majority of dust-obscured bursts are inevitably missed, because the rapid or deep follow-up observations necessary to identify them conclusively are usually not conducted.  In addition, the inclusion or exclusion of a few events close to the $A_V \approx 1$\,mag threshold could be debated (i.e., varying assumptions about $\beta_{\rm OX}$, the choice of extinction law, the X-ray fitting procedure, etc.\ would alter a few events in the sample).  Neither of these points is problematic for our sample, which aims only to gather a set of events that is representative of the population missed by optical-afterglow searches without introducing a dependence on the characteristics of our host galaxies.  (Such a bias would, for example, be present if we had required a reported host redshift, which would necessarily disfavor faint galaxies.)

As the vast majority of afterglow observations were carried out (and reported in the GCN Circulars) before the host galaxy itself was identified, our procedure above should be almost completely independent of the properties of the host.  A weak bias in favor of brighter hosts could in principle be present only as a result of the fact that in a some cases host-redshift measurement preceded selection, and in principle some events would not have made the $A_V > 1$\,mag cut if the fiducial $z \approx 2$ was assumed at the outset instead of the actual redshift.  In practice, only GRBs 060202 and 090417B would have failed our cut at $z \approx 2$ while passing it at the host-measured redshift.  

The sample is clearly \emph{not} unbiased in terms of the intrinsic afterglow properties---in particular, events with more luminous X-ray afterglows will naturally be favored as optical or NIR follow-up observations to a given depth are increasingly likely to be constraining if the X-ray afterglow flux is higher.  For a given X-ray luminosity, events with relatively flat intrinsic spectral indices ($\beta \approx 0.5$) will be weakly favored if the extinction is very large since relatively less extinction is needed to suppress the afterglow, but events with steep intrinsic indices ($\beta \approx 1$) can be favored if the extinction is more modest, since optical detection of a reddened afterglow becomes feasible.  However, as these properties are intrinsic to the GRB itself (the spectral index is also sensitive to the immediate circumburst density), we do not expect strong correlations with the large-scale host-galaxy environment.  We therefore anticipate that our sample should be reasonably representative of the hosts of ``all'' $A_V > 1$\,mag bursts, including those with fainter afterglows.

We also expect some biases with redshift, since it is easier to place constraining limits on the afterglow of an event that is nearby relative to an event with similar luminosity that is far away\footnote{On the other hand, lower-luminosity GRBs may not be detected by the satellite in the first place at greater distances, and an afterglow will stay bright for longer due to time dilation, at least partially compensating for this bias.}.  Because rest-frame optical/NIR observations correspond to bluer light at higher redshifts where even relatively small dust columns will absorb a large amount of light, very large extinction values also become harder to recognize for this reason (especially at $z>4$).  For these reasons we certainly cannot expect our sample to provide a representative redshift distribution of all dust-obscured GRBs.  Nevertheless, as long as comparisons are restricted to objects at similar redshifts, this possible bias should not affect any conclusions drawn by the set of hosts probed by our sample.

\section{Observations}
\label{sec:obs}

\begin{figure*}
\centerline{
\includegraphics[scale=0.85,angle=0]{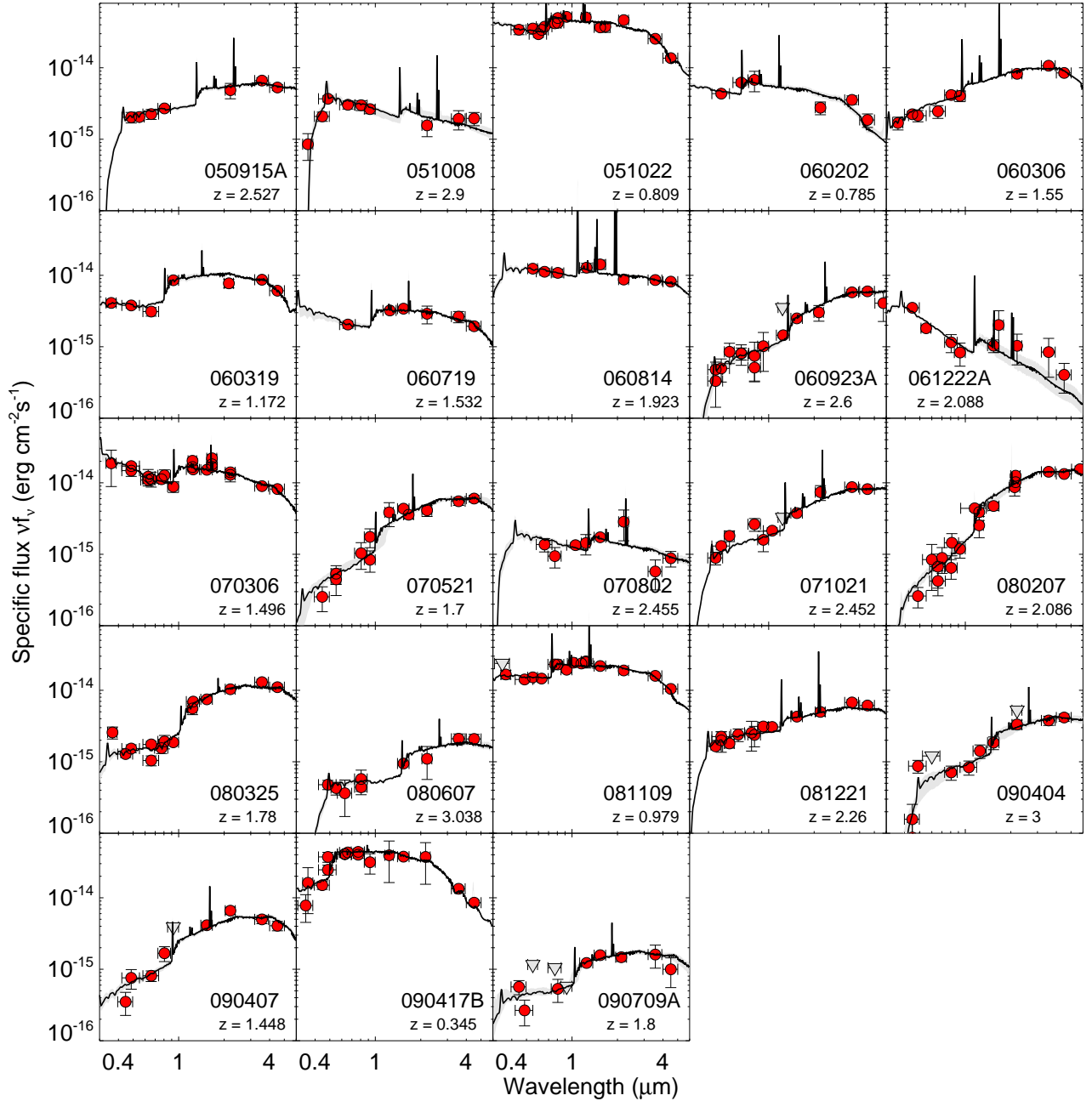}} 
\caption{Broad-band photometry of our sample of dark GRB host galaxies fitted by our population-synthesis SED modeling procedure.  Red points are photometry (2$\sigma$ uncertainty error bars), with the black curve indicating the best fit and the gray envelope the 1$\sigma$ model uncertainty region.  Gray triangles indicate upper limits (2$\sigma$).}
\label{fig:hostsedtile}
\end{figure*}

\begin{figure*}
\centerline{
\includegraphics[scale=0.333,angle=0]{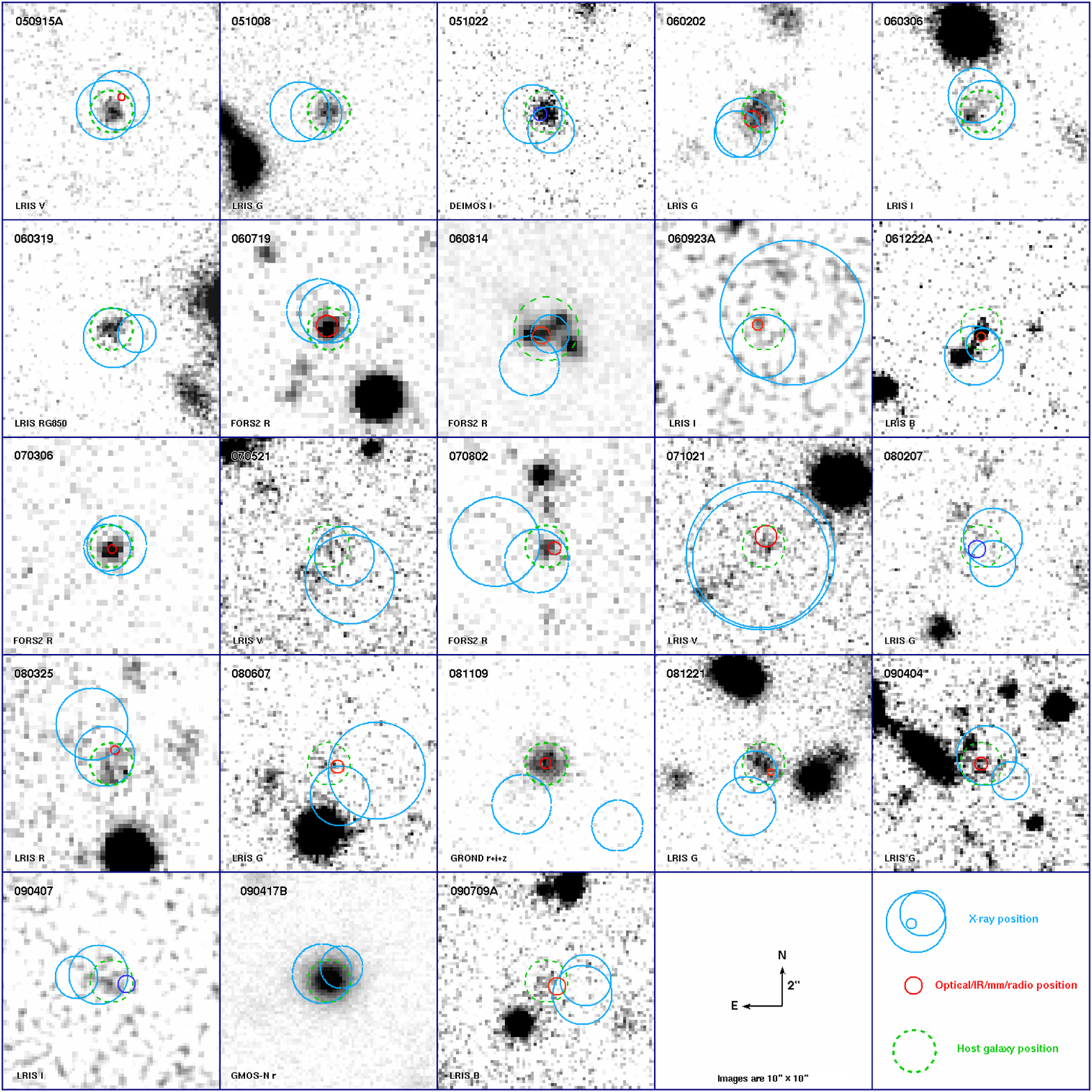}} 
\caption{Optical mosaic, showing a $10'' \times 10''$ cutout of an image chosen from our optical imaging of each source (generally, the filter showing the clearest detection was chosen).  Blue circles indicate X-ray positions from the \Swift\ XRT (from \citealt{Butler2007} and \citealt{Evans+2009}) or from {\it CXO}, and red circles indicate a position at a longer wavelength (usually NIR, but in a few cases radio or optical).  The host itself is centered in each tile and encircled with a dotted green line.}
\label{fig:optmosaic}
\end{figure*}

We observed the fields of all 23 targets satisfying the above criteria using a variety of resources from both the ground and space.  In the following sections we briefly summarize these observations, as well as the reduction, calibration, and analysis of the data.  SEDs showing the broadband photometry of all host galaxies are presented in Figure \ref{fig:hostsedtile}; imaging of the fields is presented in Figures \ref{fig:optmosaic}, \ref{fig:irmosaic}, \ref{fig:iracmosaic}.

\subsection{Keck/LRIS}

The Low Resolution Imaging Spectrometer (LRIS; \citealt{Oke+1995}) on the Keck I telescope is an optical imager and spectrograph equipped with both blue- and red-optimized cameras split by a dichroic.  Imaging observations of the galaxies were acquired mostly as part of our multi-year GRB host follow-up campaign and were reduced via standard procedures (Perley et al. 2014, in prep.).  Photometric calibration was performed relative to the Sloan Digital Sky Survey \citep[SDSS;][]{DR8} in cases where SDSS covered the field in question; otherwise, we used observations of \cite{Landolt+2009} standards (on photometric nights) or our own secondary standards obtained with either the 1\,m Nickel telescope at Lick Observatory or the roboticized 60-inch telescope at Palomar Observatory \citep[P60;][]{Cenko+2006}.  Magnitudes\footnote{Except where specified, when reporting apparent magnitudes or colors, we use the Vega system (for non-SDSS filters) or the SDSS system (for SDSS filters; \citealt{Fukugita+1996}).}
are determined via aperture photometry, using a custom wrapper around the \texttt{aper} task of the IDL Astronomy User's Library\footnote{\url{http://idlastro.gsfc.nasa.gov/ .}}.  In cases where the host is seen to be resolved in any filter we use a consistent choice of aperture for all filters, with exceptions for images having particularly bad seeing ($>1.0 \arcsec$) where a larger aperture is employed. Photometry (from LRIS and all other instruments, below) is presented in Table \ref{tab:hostphotometry}.

A small number of host galaxies were also observed with LRIS in long-slit spectroscopic mode.  These data will not be discussed in detail here, except as they pertain to establishing or constraining the host-galaxy redshift.  A more comprehensive spectroscopic study of dark GRB hosts will be presented in future work.

% Figure 3

\subsection{Keck/NIRC and Keck/MOSFIRE}

We employed the Near-Infrared Camera (NIRC; \citealt{Matthews+1994}) on Keck I during a one-night classical run on 2009 May 31 (UT is used throughout this paper) to observe several objects.  These images were reduced using standard NIR techniques within a custom Python pipeline.  Seeing conditions were good and the night was photometric, so we used observations of the standard stars FS23/M3-193, FS33/GD153, and FS29/G93-48 to establish the photometric calibration for most fields.  We checked these for consistency using stars present in our science fields with magnitudes from the Two-Micron All Sky Survey (2MASS; \citealt{Skrutskie+2006}), or with PAIRITEL (the Peters Automated Infrared Telescope; \citealt{Bloom+2006c}) calibration observations which we used to extend the 2MASS calibration down to fainter stars, when possible.  PAIRITEL employs the same camera, telescope, and filters as the northern 2MASS survey.  We also include a $J$-band observation of GRB 070521 from MOSFIRE taken on 2013 June 20, reduced and analyzed using similar techniques and calibrated relative to 2MASS.

\subsection{Gemini/NIRI}

Deep imaging observations from the Near-Infrared Imager (NIRI; \citealt{Hodapp+2003}) on Gemini-North were used to constrain the properties of host galaxies in the sample.   Data came both from our target-of-opportunity (ToO) program (in cases where no transient behavior was observed at early times, or when a late-time image was obtained to confirm suspected early variability) and from two classical nights in 2010 obtained as part of the Keck-Gemini exchange program.  Data from both runs were reduced using the NIRI reduction utilities in the Gemini IRAF package, automated using a custom Python script.  We calibrated the observations using 2MASS standards in the field or against our PAIRITEL calibrations.

\subsection{Hubble Space Telescope}

We have obtained {\it HST} images of several of our targets, taken from a variety of programs during Cycles 16--20.  Different programs employed different strategies.    In GO programs 11343, 11840, 12378, and 12764 (PI Levan) we obtained observations in one optical and one NIR band, for which we chose F606W (broad $V$ to $R$) and F160W (broad $H$) to provide a good combination of sensitivity and wavelength range.  Sometimes we acquired two orbits of observations (one per filter) to obtain these data, in which case we use ACS and WFC3 for the optical and NIR, respectively, with a standard 4-point box dither patten. In other cases we used a single WFC3 orbit, split between F606W and F160W, with a 3-point line dither in each filter. The resulting data were combined within multidrizzle, with {\tt pixfrac=1} and the {\tt scale} left at the native value for these moderately dithered images.  Many of our targets were also observed as part of GO program 12949 (PI Perley), this time using WFC-IR exclusively in two filters (F160W plus a broad $Y$ or $J$ filter, depending on the redshift) using a 3-point line dither.  Additional images were taken from our Snapshot program (using WFC3-IR in just the F160W band; GO-12307) for GRB 070521.  Images of GRBs 080207 and 080607 were taken from previous studies of these objects (see \citealt{Chen+2010}; \citealt{Svensson+2012}).

% Figure 4

We obtain magnitudes by aperture photometry via {\tt mag-auto} within SExtractor \citep{Bertin+1996}; results are presented in Table \ref{tab:hostphotometry}. In cases of observations in two filters we use the same physical aperture for consistency, and in most cases we employ a similar aperture as that used in ground-based imaging as well.  

\subsection{Spitzer Space Telescope}
\label{sec:spitzer}

Observations of all of our targets were carried out with the InfraRed Array Camera (IRAC; \citealt{Fazio+2004}) on {\it Spitzer} \citep{Werner+2004}, mostly as part of our programs (GO 70036 and 90062, PI Perley) during Warm Mission Cycles 7 and 9.  In all cases we employed dithered 100\,s observations of the field, usually totaling 1500\,s (15 dither positions) per filter, but occasionally more or less depending on the anticipated magnitude from previous ground-based observations of the target.  Targets were observed in Channels 1 and 2 (3.6 and 4.5\,$\mu$m, respectively.)  GRBs 060923A and GRB 080207 were previously observed during the cold mission, in each case in all four IRAC filters.

We downloaded the PBCD images from the {\it Spitzer} archive to disk and identified the host galaxy by reference to our ground-based or {\it HST} observations.  For several fields, the host galaxy is blended with one or more nearby sources in the field; the diameter of the point-spread function (PSF) of IRAC is $\sim 2$\arcsec.  In these cases, we used the GALFIT package \citep{Peng+2002} to subtract the nearby objects based on a model of the PSF measured from a bright, isolated star elsewhere in the IRAC image.  Photometry of the host was then performed using IRAF, employing an aperture radius of 2 native pixels (4 resampled pixels in the PBCD observations, or 2.4\arcsec) with a sky annulus of inner and outer radii 8 and 12 pixels, respectively, and calibrated via the zeropoint values in the {\it Spitzer} IRAC handbook.  In all cases, the angular size of the host galaxy was sufficiently small that aperture effects are not significant.

% Figure 5
\begin{figure*}
\centerline{
\includegraphics[scale=0.333,angle=0]{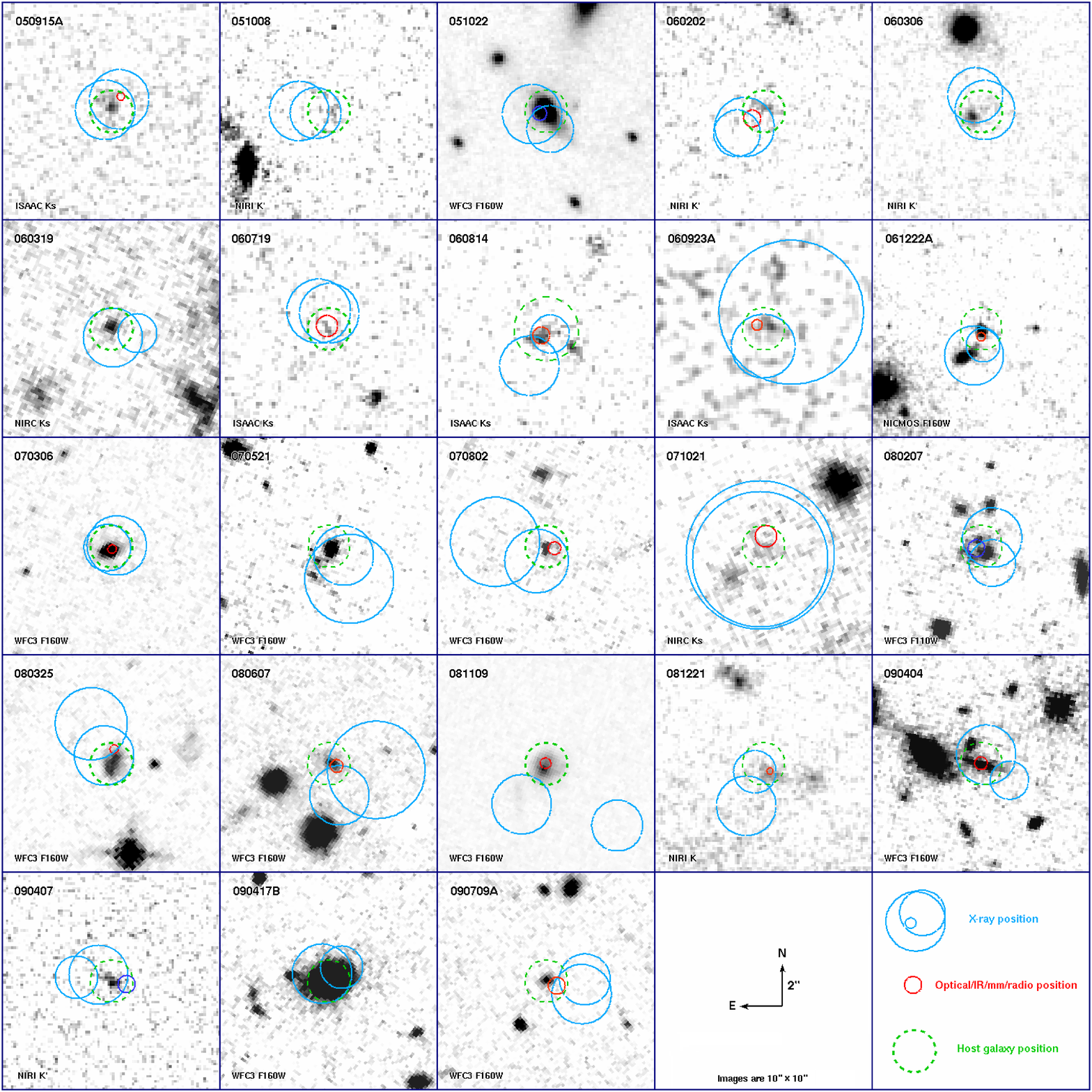}} 
\caption{Near-infrared mosaic of our GRB host targets.  Image sizes and overplotted information are the same as in Figure \ref{fig:optmosaic}.}
\label{fig:irmosaic}
\end{figure*}

% Figure 6
\begin{figure*}
\centerline{
\includegraphics[scale=0.333,angle=0]{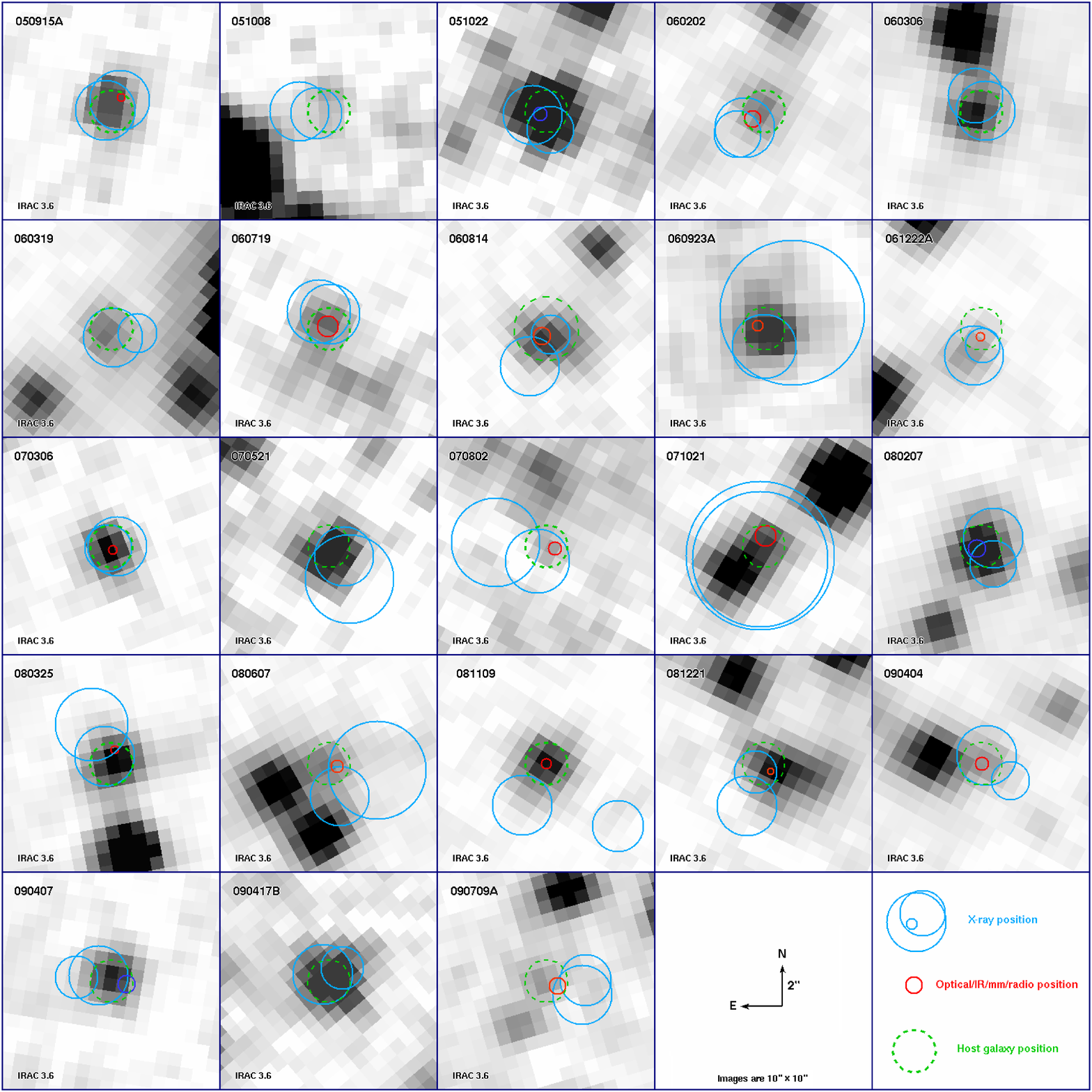}} 
\caption{IRAC mosaic showing 3.6\,$\mu$m imaging with {\it Spitzer}.  Despite our relatively short integrations (typically 1500\,s), we detect every object in our sample, indicating a significantly higher average stellar mass than in previous, optically selected GRB host samples.  In a few cases the host is blended with a nearby object; the host flux is recovered by fitting and removing the flux of any possible nearby contaminants before performing aperture photometry.}
\label{fig:iracmosaic}
\end{figure*}

\subsection{Very Large Telescope}

Many hosts in this sample were observed with the Very Large Telescope (VLT).  Most of these data were previously published; in particular, we rely heavily on the FORS2 $R$-band and ISAAC $K$-band photometry products from the TOUGH survey \citep{Hjorth+2012} and several X-shooter redshifts reported in the work of \cite{Kruehler+2012b}.  We also use our independent rereduction of the X-shooter observations reported by \cite{Salvaterra+2012}, retrieved from the ESO archive.

\subsection{Previous Work}

We acquired additional photometry from a variety of other literature sources, usually burst-specific papers including photometry of the host galaxy under study.  These sources are cited where appropriate.

\subsection{Host Identification}
\label{sec:hostid}

Historically, the most significant challenge affecting the identification and characterization of dark GRB hosts has been the difficulty in localizing the afterglows with sufficient accuracy ($\sim 2$--3\arcsec) to uniquely identify the host.  As we have mentioned, \Swift\ now provides positions accurate to this level for essentially all (long-duration) GRBs observed with the XRT, based on the optical/UV registration techniques of \cite{Butler2007} and \cite{Evans+2009}.

Nevertheless, XRT positions are never better than $\sim 1$\arcsec\ accuracy and occasionally can leave some ambiguity about the host identification (e.g., see \citealt{Rossi+2012}).  In most cases, the afterglow was also detected at optical or NIR wavelengths, providing (in principle) a position accurate to 0.5\arcsec\ or better.  These positions were obtained either by our own observations, from published work, or from the GCN Circulars. (In cases where the accuracy of a reported position in the GCN Circulars was in any doubt, we reobtained the afterglow image and recalculated the position internally.)

We have also been acquiring ToO observations of optically undetected GRB afterglows using the {\it Chandra X-ray Observatory} ({\it CXO}).  The ToO observations of $\sim 15$\,ks per exposure were typically conducted with ACIS-S in imaging mode within a few days of the occurrence of the GRB.  Coordinates from these observations (placed in an absolute frame using the 2MASS catalog) are presented in Table \ref{tab:positions}, along with our optical/NIR or XRT positions where appropriate.

The host galaxy was identified in the standard way by finding the brightest source in the image that is consistent with the afterglow position.  In nearly all cases this is unambiguous: a single, moderately bright, well-detected object is seen directly at the afterglow position.  The significance of the association (following the standard $P_{\rm chance}$ metric for estimating the probability that a circle of a given size placed randomly in the sky encloses a galaxy brighter than a certain flux level; e.g.,\ \citealt{Bloom+2002}) depends on the filter chosen to make the comparison---we adopt the $K$ band for our targets since it does not penalize very red, dust-attenuated galaxies, although the results would not be qualitatively different if the more commonly employed $R$ band were chosen.  (In cases where $K$ photometry was not available, magnitudes are calculated by interpolation from other filters.)  In every case, the probability of false association is low (always $P_{\rm chance} < 0.1$, and in most cases $P_{\rm chance} < 0.01$).  Because of the large size of the sample, the cumulative probability of having a small number of misidentifications present is significant despite these low \emph{individual} probabilities:  the probability of at least one misidentified host is 39\% and the probability of two or more misidentified hosts is 8\%.   However, as our study focuses on the aggregate properties of the sample and is not particularly concerned with the properties of any individual case, we do not expect this to be a significant limitation.

% S3
\section{Modeling}
\label{sec:modeling}

A primary goal of this work is to infer the properties of the host galaxy (in particular the UV-based SFR, the average extinction $A_V$, and the total stellar mass $M_*$) for direct comparison to previous samples.  While these parameters can be crudely estimated from individual photometric points independently using various prescriptions (e.g., using the optical slope $\beta$ to measure $A_V$ and correcting the rest-frame UV luminosity to measure the SFR; or using the rest-frame $K$-band luminosity as an indicator of stellar mass), given the complex interplay between the parameters and the large variation of redshifts and photometric completeness across our sample, we instead adopt a more general population-synthesis-based SED-fitting procedure to fit all parameters to all data points simultaneously.

The fitting of the host SEDs is implemented with our own software written in IDL using the population-synthesis libraries of \citet{bc03} (hereafter BC03), assuming a \cite{Chabrier2003} IMF and the Padova (1994) stellar evolution tracks \citep{Bertelli+1994}.  Taking as inputs the galaxy's total stellar mass, average metallicity, current SFR, and an analytic expression for the time-dependent star-formation history, the code combines the appropriate BC03 templates and extinguishes them using the \cite{Calzetti+1994} attenuation law to produce the galaxy-integrated spectral luminosity distribution.  Major UV/optical nebular emission lines are included, calculated from the SFR and metallicity following the prescriptions of \citet{Kennicutt1998}, \citet{KD02}, and \citet{Kewley+2004}, since in some cases these can contribute substantial flux even to broad-band filters.  These are then redshifted appropriately to produce the observed SED, and an intergalactic neutral-gas absorption prescription (a semiempirical sightline-averaged Lyman-$\alpha$ forest model as a function of wavelength and redshift following \citealt{Madau1995}) is also applied.

The Calzetti law---a featureless curve similar to a power law---is used to model the host-galaxy SED even in cases where direct line-of-sight measurements through the host galaxy provided by the GRB afterglow favor a different extinction law, for three reasons.  First, the Calzetti curve is the standard one employed in most current high-$z$ galaxy work (where it has generally been successful at modeling the UV SEDs of distant galaxies in current surveys; e.g., \citealt{Calzetti+1999,Reddy+2010}), and this convention facilitates direct comparison of the $A_V$ and SFR to other work.  Second, the Calzetti law is explicitly an \emph{attenuation} law rather than an \emph{extinction} law.  An extinction law refers to a point source shining through a single sightline (with a single optical depth and in which photons are lost to scattering), whereas an attenuation law refers to the ratio of total emitted energy to total energy escaping (the exact situation desired here, in which stars can exist at varying optical depths in the cloud and in which scattering photons are not lost).  Third, in cases in our sample where sufficient rest-frame UV data were available to attempt multiple extinction laws, the Calzetti extinction law was favored (e.g., there was no evidence of a 2175\,\AA\ absorption feature or strong UV curvature in the host SED).

The SED is then converted to an observable spectral flux distribution (SFD) using a basic correction for distance (we employ $\Omega_\Lambda=0.7$, $\Omega_{\rm m}=0.3$, H$_0 = 70$\,km\,s$^{-1}$\,Mpc$^{-1}$) and redshift.  Finally, synthetic photometry is performed (using a library of standard filter transmission curves) to produce the predicted fluxes in various standard filters as observed at a given redshift.  The broad-band observations can then be fitted against our model using the Levenberg-Marquardt least-squares fitting procedure as implemented in the {\tt mpfit} package \citep{More1978,Markwardt+2009}.  Upper limits are treated as a flux measurement of zero with an $N\sigma$ uncertainty equal to the uncertainty level of the limit.  (There are only a few constraining upper limits in the sample.)

Our set of broad-band observations is generally too limited to actually constrain the star-formation history or metallicity independent of other parameters, even for the best-studied galaxies in our sample.  To help standardize our results against each other and against other studies, we apply a few simplifying assumptions.  The stellar metallicity for all galaxies is assumed to be 0.5\,Z$_{\odot}$ (more precisely, $Z = 0.01$ in the BC03 model), with two exceptions.  For the host of GRB 061222A, the UV spectral index is too blue for any solar-metallicity template and we instead use 0.1\,Z$_{\odot}$.  For the host of GRB 051022, a value close to Solar has been measured spectroscopically \citep{Graham+2009} and so we assume 1.0\,Z$_\odot$.  (In general, however, the assumed stellar metallicity does not make a significant difference on our results relative to other sources of uncertainty.)  Second, we apply a simple prescription for the star-formation history as follows. The system is first fit with a constant star-formation history.  If the effective age ($M_*/$SFR) converges to a value in excess of 1\,Gyr or if the fit is poor, we allow the young and old populations to be fit separately by permitting a step-function change of the SFR at $5 \times 10^7$\,yr before present.  The age of the older population ($t_{\rm max}$) is a free parameter in the fit in this model, but is constrained to be less than the age of the Universe at that redshift.  (In one case, GRB 080325, the age was forced to this maximum limit, and for this case only we modified the star-formation model to be exponentially decaying with time for the old population.)  If the effective age converges to less than 1\,Gyr, we retain the assumption of a constant star-formation history (we are not sensitive to variations) but require a minimum value for the effective age of $>2 \times 10^7$\,yr.

To estimate the statistical uncertainties in each parameter, we run a series of 100 Monte-Carlo trials for each galaxy in which each flux measurement is modified by a random amount (following a Gaussian distribution with $\sigma$ equal to its observational uncertainty).  Reported uncertainties in the output parameters correspond to the 15th-to-85th percentile (that is, 1$\sigma$) of the results.  These results are given in Table \ref{tab:hostmodelpar}.  Note that these values do not include the systematic uncertainties due to assumptions about the extinction, star-formation history, and IMF, which can be significant \citep{Michalowski+2012a} but should affect all galaxies (and comparison samples) in a similar way.  Our primary interest is in the aggregate properties of the various samples in comparison to each other (rather than in the absolute values of individual objects), and we do not expect large systematics that would endanger these comparisons.

% S4
\section{GRBs and Host Galaxies}
\label{sec:hosts}

\subsection{GRB 050915A}
GRB 050915A was a relatively faint and fast-fading event with minimal late-time follow-up.  However, it was observed rapidly by a number of telescopes including (critically) the Palomar P60 and PAIRITEL. The P60 did not detect the transient in any exposure \citep{Cenko+2009}.  PAIRITEL reported a weak detection of a transient source in an early $H$-band stack \citep{GCN3984}, although no detection was reported in simultaneous $J$ or $K_s$ imaging.   We reanalyzed these frames and performed photometry on all three filters by forcing an aperture at the location of the $H$-band detection, and we do detect (at about 4--5$\sigma$) a weak source in the other filters as well.  The NIR colors and optical nondetection are consistent with a dust-reddened afterglow with an inferred extinction of $A_V \approx 1.5$\,mag.

Previous imaging of this field was reported by \cite{Perley+2009b}, including the identification of a host galaxy offset by 0.85\arcsec\ from the transient location.  The photometry presented in Table \ref{tab:hostphotometry} and used in modeling has been reanalyzed since that time using our Lick Nickel field calibration.  Combining these observations with $R$- and $K$-band photometry from the VLT clearly shows that the putative host is quite red; $R-K = 3.9$\,mag (Vega).  VLT X-shooter observations \citep{Kruehler+2012b} have also recently established the redshift of this source based on several strong nebular emission lines at $z=2.527$.

The fit to the host SED converges to a young, strongly obscured stellar population with a very high current star-formation rate of $\sim 140$\,M$_\odot$\,yr$^{-1}$.   The relative optical/NIR faintness of this source (given the bright, steep UV continuum) indicates a rather young age and modest stellar mass  ($\sim 4 \times 10^{9}$\,M$_\odot$)  The galaxy has characteristics of a luminous, dusty starburst.

\subsection{GRB 051008}

GRB 051008 is the subject of the detailed study by Volnova et al. (2013, in prep.; see also \citealt{Volnova+2010}), and most of the observations reported here will also appear there, although the analysis reported in this paper is independent of that work.  The source was observed within 1\,hr by the 2.6\,m Shajn telescope at the Crimean Astrophysical Observatory and by the Tautenburg 1.34\,m telescope; although early reports suggested some possible optical variation of a galaxy pair inside the original XRT error circle \citep{GCN4081,GCN4087}, this was not confirmed by later analysis, and the most recent XRT error circle excludes these galaxies.  The large $N_{\rm H}$ excess inferred from the XRT spectral analysis rules out a high-redshift origin (see \S \ref{sec:select}).

We have imaged the field using LRIS on Keck in numerous optical filters ($U$, $B$, $g$, $R$, $I$, $RG_{850}$) as well as in the $K$ band with NIRI and with \emph{Spitzer}.  A single, optically bright galaxy is detected consistent with the XRT position in these data.  The SED of this object shows the features of a classic Lyman-break galaxy:  a flat, steadily rising spectrum through most of the optical with a sharp cutoff in the $B$ and $U$ bands.  Volnova et al.\ estimate a photo-$z$ of $2.85^{+0.03}_{-0.05}$ from these data; our own estimate using EaZy \citep{Brammer+2008} results in $z = 2.96 \pm 0.11$.   We also acquired spectra of this source and do not detect any emission lines at the location of the host.  At the putative redshift only Lyman-$\alpha$ is expected to lie within our spectral range, but because of the resonant scattering properties of the Lyman-$\alpha$ line, the nondetection is not a particularly strong constraint on the properties of the galaxies, nor is it unusual for LBGs \citepeg{Stark+2010} or for GRB hosts \citep{MilvangJensen+2012}.

The strong detections in many optical filters allow us to constrain the host SFR fairly securely ($\sim 70$\,M$_\odot$\,yr$^{-1}$ at $z=2.9$).  The galaxy is young and has a low mass ($\sim 5 \times 10^9$\,M$_\odot$), and it is moderately dust obscured ($A_V = 0.8$\,mag), typical properties for luminous LBGs.

\subsection{GRB 051022}

GRB 051022 was a High-Energy Transient Explorer 2 (HETE-2) burst \citep{Nakagawa+2006} rapidly followed by a range of other instruments, including \Swift.  Early nondetections at optical wavelengths quickly motivated deeper searches and NIR observations, all of which reported only upper limits on an optical afterglow; comparison of these limits to the X-ray flux unambiguously indicates a very large dust attenuation column ($A_V > 9.3$\,mag).  However, a bright host-galaxy candidate was immediately evident even in small-telescope images.  This GRB and its host have already been the subject of intensive studies by several authors \citep{CastroTirado+2007,Rol+2007}, which clearly characterize it as a young, high SFR ongoing merger at $z=0.809$.  It seems to have a relatively high metallicity \citep{Graham+2009}.

Our observations of this source include {\it HST} imaging as well as a single Keck/DEIMOS $I$-band image from 2005.  In addition, the two published papers on this object give an extensive library of ground-based optical/NIR photometry.  Unfortunately, there is a large systematic discrepancy between the values in two published works: most photometric points in \cite{CastroTirado+2007} are on average 0.3\,mag brighter than the corresponding values in \cite{Rol+2007}.  Since both papers use the same calibration \citep{GCN4184}, and the host is fairly extended ($\sim 1'' \times 2''$ in the {\it HST} imaging), we suspect that the discrepancy originates from use of different apertures under different seeing conditions (neither paper states what aperture size is used for the host-galaxy photometry, and the extension is significant even in ground-based images) or possibly from the use of extended objects from the Henden catalog as calibrators (a significant fraction of the objects near the GRB position in the Henden catalog are themselves significantly extended galaxies).  

We downloaded the $riz$ GMOS-S images from the Gemini Science archive and performed photometry with a 2.0\arcsec\ radius aperture using only objects with stellar PSFs from the Henden catalog, transformed to the SDSS system using the equations of \cite{Jester+2005}.  We also acquired the images from the Danish 1.54\,m telescope and UKIRT (published by \citealt{Rol+2007}) and redid the photometry using a similarly large aperture and the Henden calibration (or 2MASS for the NIR bands).   These reanalyzed points give much better self-consistency and are used in the model in place of the numbers in the literature.  The photometry from \cite{CastroTirado+2007} is not used in this analysis.

Consistent with previous work, we find that the host of GRB 051022 is a relatively young ($\sim 700$\,Myr), massive ($2 \times 10^{10}$\,M$_\odot$\,yr$^{-1}$), rapidly star-forming ($\sim 30$\,M$_\odot$\,yr$^{-1}$) galaxy with moderate dust obscuration ($\sim 0.7$\,mag).

\subsection{GRB 060202}

Early, moderately deep nondetections of this source by the Faulkes Telescope North \citep{GCN4630} motivated the initial identification of this GRB as a dark burst.  The host-galaxy candidate was first reported in twilight Keck imaging by \cite{GCN4636}; a faint source at this position was also seen with Gemini-NIRI \citep{GCN4639} and UKIRT \citep{GCN4653}.  While the comparison of UKIRT and Gemini-NIRI $K$-band images \citep{GCN4653} showed no obvious fading, a comparison between the $K_s$-band imaging from the ToO observation and our significantly deeper late-time imaging of the field (also with NIRI using the $K'$ filter) reveals significant fading and an astrometric shift of about 0.5\arcsec\ between the object centroids, clearly indicating that the early-time observations were afterglow-dominated.  Since we do not have late-time observations in the $J$ or $H$ filters, the relative contributions of afterglow and host in these bands are ambiguous, but even if the flux in these bands was entirely afterglow, the color was much redder than a typical afterglow or than comparison to contemporaneous X-ray measurements would suggest, supporting the notion that this source was reddened by host-galaxy dust.

Our late-time imaging of the field includes only Keck $g$- and $R$-band images and the Gemini-NIRI $K'$-band image mentioned above. (We also cautiously use the $I$-band photometry from \citealt{GCN4636} pending independent calibration of the field.)  In addition, we acquired spectra with LRIS on 2006 Sep.\ 21 and detect a bright, relatively broad emission line at a wavelength of 6660\,\AA.  No other lines are evident in the spectrum.  We associate this line with \OII\ at $z=0.78$:  identification with any other common nebular line (\OIII, H$\beta$, H$\alpha$) would place other lines further to the blue where they would almost certainly be detected.  While the \OII\ association would put \OIII\ and H$\beta$ in the detectable spectral range, at this redshift they land on strong night-sky lines that are subject to heavy fringing in the original LRIS-R CCD.

We have only one measurement blueward of the rest-frame Balmer break; consequently, we can only crudely constrain the extinction and SFR, but both are relatively modest ($\sim 1$\,mag and $\sim 6$\,M$_\odot$\,yr$^{-1}$, respectively).  The faint $K$-band measurement implies a low mass of $\sim 10^9$\,M$_\odot$.

\subsection{GRB 060306}
GRB 060306 was rapidly followed by several small telescopes and no afterglow was detected, although most of these limits are not particularly constraining.  However, the NIC-FPS image \citep{GCN4861} is quite deep.  The faint source reported as a possible afterglow candidate in this GCN Circular is no longer inside the most recent XRT error circle and is most likely a foreground galaxy, but we downloaded the reduced image and calculated a point-source limit on any afterglow (or host) of $K > 18.2$\,mag at the current GRB position relative to 2MASS standards.  This is several magnitudes below even the minimum ($\beta_{OX}=0.5$) expectation from the X-ray light curve and indicates substantial extinction.
  
We acquired late-time imaging of this position in many different optical filters at Keck as well as in the $K$ band with NIRI; it was also observed as part of the TOUGH project.  A source is detected in the XRT error circle in all of these observations.  In imaging conducted on nights with the best seeing, the source resolves into two components: a bright southeastern part and a much fainter northwestern source. 

A redshift for the host galaxy of 3.5 was reported by \cite{Salvaterra+2012} based on an NIR emission line at 16800\,\AA, which was claimed to be associated with the \OII\ doublet.  Our reanalysis of the spectrum shows this redshift to be incorrect:  the apparent separation of the line into a doublet is actually caused by the velocity offset of the two components, an effect that can be clearly seen in the two-dimensional spectrum as spatial variation in the line center between the two.  Extracting the components individually, the line is too narrow to be \OII.

No other unambiguous lines are detected in the X-shooter spectrum, so in principle this line could correspond to any major emission feature--- most likely H$\alpha$ at $z=1.55$ or \OIII\ at $z=2.35$, as suggested by \cite{Jakobsson+2012} and \cite{Kruehler+2012b}.  Photometrically, the lower redshift is favored, as we see no evidence of any spectral break due to Lyman-$\alpha$ affecting the $U$ band, and a fit to the photometry with EaZy provides a consistent $z_{\rm phot} = 1.52^{+0.30}_{-0.36}$ (1$\sigma$).  Furthermore, a weak resolved emission feature is detected in the X-shooter spectrum at the expected location of \OII\ for an assumed redshift of 1.55.  Together with the photometric constraints, we therefore consider the association with this GRB and its host system at $z=1.55$ reasonably secure.   We report photometry of the entire host complex (based on a single aperture enclosing both objects) in Table \ref{tab:hostphotometry}.

The host colors are fairly red with no indication of a strong Balmer break, and as with several other galaxies discussed previously we infer a dusty, young, rapidly star-forming host, with values (at $z=1.55$) of SFR $\approx 250$\,M$_\odot\,$yr$^{-1}$, $A_V \approx 2.2$\,mag, and $M_* \approx 8 \times 10^9$\,M$_\odot$.

\subsection{GRB 060319}

The identification of GRB 060319 as a dark burst is based almost entirely on the early WHT $K$-band observation of \cite{GCN4897}, who reported a $K$-band source inside the XRT error circle.  While the coordinates given in that GCN Circular are actually not in the most recent XRT error circle, reanalysis of the WHT image with improved astrometry shows this source to in fact be coincident with the most recent XRT position, as well as with an underlying object detected in late-time Keck imaging with LRIS and NIRC.  Comparison to our NIRC $K_s$-band imaging in particular shows obvious fading of this source by almost 2\,mag since the WHT observation, confirming it as the NIR transient.  The detected flux is already below the $\beta_{\rm OX}=0.5$ line and the redness of the transient is confirmed by deep $R$-band nondetections at earlier times.

We acquired spectra with LRIS on Keck I on 2007 July 18.  A single, relatively broad emission line is detected in the red part of the spectrum at 8096\,\AA\ that we associate with the \OII\ doublet at $z=1.172$.  No other lines are expected to be detected in our spectral range at this redshift.

A break is evident in the broad-band SED between the $R$ and $z$ bands, consistent with a strong Balmer break at this redshift and the presence of a relatively old, evolved population ($M_* \approx 2 \times 10^{10}$\,M$_\odot$ with an age of $\sim$2 Gyr) in addition to a modest amount of more recent star formation (SFR $\approx 8\,{\rm yr}^{-1}$).

\subsection{GRB 060719}

The position of this GRB was observed rapidly (within 2\,min) by several robotic telescopes, none of which reported a detection \citep{GCN5343,GCN5344,GCN5345}.  Deep optical imaging was conducted with the VLT (using FORS2) after only 38\,min, and deep VLT NIR imaging was acquired several hours later (using ISAAC).  Only the NIR observations resulted in the detection of an afterglow \citep{GCN5347,GCN5350}.   Provisional photometry (Malesani et al. 2013, private communication) plotted in Figure \ref{fig:minextseds} shows that the NIR colors of the transient are quite red; our fit indicates a total extinction of $A_V \approx 3$\,mag.

Photometry of the host galaxy comes from the TOUGH project at the VLT, and from our {\it HST} and {\it Spitzer} observations.  A faint, red object is detected at the afterglow position in all of these images.  X-shooter spectroscopy identified an emission line corresponding to H$\alpha$ at $z=1.532$ \citep{Kruehler+2012b}, consistent with the blue color in the two IRAC channels.  The stellar mass is well constrained by the abundant NIR photometry; we estimate  $M_* = 1.3 \times 10^{10}$\,M$_{\odot}$.  Since only a single photometric measurement blueward of the Balmer break is available, the properties of more recent star formation are constrained less well, but we infer $A_V \approx 0.4$ mag and SFR $\approx 5 M_\odot {\rm yr}^{-1}$.

\subsection{GRB 060814}

Bright GRB 060814 was observed at early times by a number of telescopes, including the small robotic TAROT within a few minutes \citep{GCN5448} and the VLT in the $R$ band after one hour \citep{GCN5450}.  No counterpart was reported, although a complex extended source near the XRT error circle was noted by \cite{GCN5456}.  This source is also detected by SDSS \citep{GCN5458} and no evidence of variability was reported.  Additional observations several hours later in the NIR with UKIRT and the Palomar 200-inch Hale telescope (P200) detected a $K \approx 18$\,mag source consistent with this location which faded in subsequent exposures \citep{GCN5461}, confirming it as the afterglow of the GRB.

Our first late-time imaging of this position was carried out with LRIS during a night of poor seeing (1.8\arcsec).  The extended source reported in the GCN Circulars was still easily detectable (and still obviously extended despite the poor seeing); we obtained spectra of it several months later and reported a redshift of 0.84 based on the detection of several nebular lines \citep{GCN6663}.

We reobserved the source in imaging mode with NIRC two years later (in the $J$ and $K$ bands), and it was also included as part of the TOUGH project ($R$ and $K$ bands) and in our {\it HST} campaign.  The seeing conditions for all of these observations were much better ($<0.6\arcsec$), and these images show that the extended object has significant substructure.  In particular, it is resolved into two distinct subcomponents separated by about 1\arcsec: a point-like object in the southwest, and an extended source (consisting of two connected blobs in the north and east).

Additional spectroscopy of this object with FORS-2 and X-shooter on the VLT was conducted under the TOUGH program \citep{Hjorth+2012,Jakobsson+2012,Kruehler+2012b}.  These observations clearly show that the two objects are at different redshifts: the $z=0.84$ emission lines correspond only to the southwest object, while the northern/eastern components are at a common redshift of 1.923 based on the detection of several emission lines in the NIR.  We reanalyzed our LRIS spectra and confirm that the $z=0.84$ emission lines correspond only to the western source, with no optical emission lines detectable from the eastern source.  Relative astrometry between the UKIRT images and the late-time VLT images shows that the GRB afterglow is associated with the $z=1.92$ object, indicating that the $z=0.84$ object is an unrelated foreground system \citep{Jakobsson+2012}.

The two unrelated sources are heavily blended in our Keck/LRIS and {\it Spitzer} images. However, given the well-determined positions and morphologies of the two objects from the VLT and {\it HST} imaging and the brightness of both objects in all bands, it is not difficult to isolate the flux of the true host from the foreground object using the same basic technique as for other blended {\it Spitzer} images (\S \ref{sec:spitzer}) using a GALFIT model.  The photometry reported for this object in Table \ref{tab:hostphotometry} is for the host alone.

Given the galaxy's remarkable brightness despite its redshift of 1.92, it is no surprise that our models indicate that the host is extremely luminous and rapidly forming stars.  Our best-fit model converges to an SFR of $\sim 240$\,M$_\odot$\,yr$^{-1}$ with an attenuation of $A_V = 1.2$\,mag; the mass is moderate ($10^{10}$\,M$_\odot$) and the inferred age very young (30\,Myr), indicating an extremely rapid and powerful starburst.  This starburst could be coupled to ongoing merger activity in light of the apparent binary morphology of the host system.

\subsection{GRB 060923A}

GRB 060923A was the subject of the study by \cite{Tanvir+2008}, who report deep optical limits and a faint $K$-band detection of the afterglow (establishing the afterglow as dust obscured) and a basic characterization of the host galaxy.  We also previously discussed this object as part of the study of \cite{Perley+2009b}.  Our reanalysis here includes all of these data in addition to further, unpublished photometry from Gemini-N/NIRI and {\it Spitzer} (cold mission; PI Fox) as well as recent {\it HST} observations.

While the host has been observed in a large number of (mostly optical) filters, the detection is marginal in all of them.  The clearest detections come from the Keck $B$-band and the VLT $R$-band observations; inspection of these images hints at a binary morphology with distinct components in the northwest (where the afterglow is located) and southeast, structure that is confirmed in the higher signal-to-noise ratio {\it HST} imaging.  We have therefore redone the aperture photometry on the VLT images to ensure that the same aperture is used as for the other optical filters.

\citet{Tanvir+2008} noted that the host galaxy has an ordinary $R-K$ color, although relative to most GRB hosts it is actually fairly red (especially when the {\it Spitzer} data and measurements in bluer optical filters are considered).  The redshift of this source is not known spectroscopically:  a deep FORS2 integration resulted in detection of continuum down to 4600\,\AA\ but no emission lines \citep{Tanvir+2012, Jakobsson+2012}.  Nevertheless, the redshift can be constrained by our broad-band data; the 95\% confidence redshift range from EaZy is $1.98 < z < 3.08$ with a best-fit value of $2.47$.

Regardless of the exact redshift, the bright {\it Spitzer} detections point clearly toward a large stellar mass ($M \approx 10^{11}$\,M$_\odot$ at $z=2.5$).  The derived optical SFR is not strongly dependent on the redshift and is high but not remarkably so ($\sim 90$\,M$_\odot$\,yr$^{-1}$); given its mass this value is in fact quite modest, indicative of an ordinary but massive high-$z$ galaxy not undergoing a starburst episode.

\subsection{GRB 061222A}

GRB 061222A is one of the most obscured afterglows in the sample.  The only reported detection in the optical/NIR bands is an early-time Gemini/NIRI $K$-band point, even though this GRB was among the brightest events of that year in X-rays and gamma-rays.  Deep observations in all other filters resulted only in upper limits, although these are not as constraining as the $K$-band point.  As reported by \cite{Perley+2009b}, the host is quite blue, and in fact shows a strong Lyman-$\alpha$ emission feature.  It is the only dark GRB host known to exhibit this feature \citep{MilvangJensen+2012}.

In addition to a reanalysis of the photometry presented in that work, here we include several new observations: specifically, {\it Spitzer}/IRAC observations and an archival {\it HST}/NICMOS F160W measurement (PI: Berger). 
Consistent with the very blue color of this object reported by \cite{Perley+2009b}, the {\it Spitzer} data only marginally detect the host (it is one of only a few sources in the sample which do not have a strong {\it Spitzer} detection), but the source is well detected by {\it HST}.

Given the very blue rest-frame UV color (which may indicate an even younger starburst than our $5\times10^7$ yr minimum value) the extinction is extremely low, converging to the physical limit of $A_V=0$ for nearly all trials, even if a significantly lower metallicity is assumed (the extremely blue color and low mean extinction mark the host of GRB 061222A as a dramatic outlier compared to most other dark GRB hosts in our sample).  This is particularly curious given the tremendous extinction evident in the afterglow SED, and suggests extreme heterogeneity in the galaxy's internal dust distribution:  either the dust is localized to a small number of very optically thick clouds and the GRB sightline happened to intersect one, \emph{or} the bulk of the star formation is actually heavily obscured.  The small stellar mass of this object probably favors the former interpretation (the estimated stellar mass of $\sim 10^9$\,M$_\odot$ is much lower than that of typical submillimeter galaxies), but deep long-wavelength observations (e.g., with ALMA or EVLA) will be needed to resolve the question unambiguously.

Interestingly, the nearby galaxy seen in our images at an offset whose relative orientation and proximity might suggest a merging companion is not at the same redshift \citep{Perley+2009b}.

\subsection{GRB 070306}

GRB 070306 was the subject of \cite{Jaunsen+2008}, who identify it as a highly obscured afterglow based on deep, early-time WHT and VLT follow-up observations:  the afterglow is detected only in the $K$ and $H$ bands in these data, and it is very red.  Data points for our modeling (and the host redshift of 1.496) are derived from this work (Table 2) and from the recent study of \cite{Kruehler+2011}, supplemented by our {\it Spitzer} photometry.  This combined dataset is one of the most complete of any object in the sample.

The host galaxy is relatively bright and well detected in most filters, and it has a relatively blue SED showing a clear Balmer break that matches the emission-line redshift.  The stellar mass and SFR are both substantial but the mean extinction is relatively low.  The fit quality is actually improved significantly by allowing a nonconstant star-formation history that is elevated in the past relative to the present; even with this flexibility added, our model converges to a well-determined SFR (a relatively modest $\sim 13$\,M$_\odot$\,yr$^{-1}$) and significant stellar mass ($\sim 5 \times 10^{10}$\,M$_\odot$).  Given the blue rest-frame UV slope the extinction is very low, $A_V < 0.4$\,mag.

As with GRB 061222A (and the recent GRB 100621A; \citealt{Kruehler+2011}), the combination of blue host and highly reddened afterglow implies either a very heterogeneous ISM or a highly obscured super-starburst occurring inside the galaxy core.  The much larger mass of this galaxy is consistent with either picture.  In fact, we have recently detected this object at 2\,GHz with the EVLA (Perley et al., in prep.), suggesting that, if the radio emission is not due to afterglow, this object is a blue submillimeter galaxy (the inferred SFR from the radio continuum is $\sim 200$\,M$_\odot$\,yr$^{-1}$.)

\subsection{GRB 070521}

GRB 070521 is an extremely dark burst; it occurred during dark time and triggered both the Faulkes Telescope North and P60 for rapid observations, neither of which detected a transient \citep{GCN6435,Cenko+2009}.  Follow-up observations within 24\,hr were conducted with P200, Gemini, Keck, and Subaru, none of which reported any variation.  (Some possible, very marginal evidence of optical variation was suggested by \citealt{GCN6457} and \citealt{GCN6936}, but neither of the reported positions are in or near the final XRT error circle, nor would variation in these optical bands be expected given the much deeper NIR limits available from Gemini.)  Given the deep, early NIR limit and bright X-ray afterglow, the inferred limit on the extinction column is extremely large ($A_V \gtrsim 12$\,mag).

The best position available for this source is the UVOT-enhanced XRT error circle.  There is only one source that is clearly inside this circle: a very red, extended object toward its eastern side.  A fainter (in most bands), bluer, morphologically complex source is located immediately southeast of this galaxy, just outside the XRT error circle.  It is not completely unambiguous which of the two sources is the true host galaxy (or if they are related), but given the first object's brightness and closer proximity to the error circle we identify it as the host galaxy.  This is the same object we previously suggested as the host \citep{Perley+2009b}; our imaging was not deep enough to identify the fainter galaxy in that work.

We previously estimated \citep{Perley+2009b} a photometric redshift of 1.35 for this host.  Our additional $z$-band images and a spectrum of the host which identified no sign of \OII\ emission out to $\sim$\,10000\,\AA\ now lead us to favor a slightly higher redshift of about 1.7 (the 95\% confidence range from the EaZy fit is 1.37--2.20).  We continue to see evidence of a strong Balmer break blueward of the $J$ band that suggests an evolved stellar population and significant mass---in this case about $3 \times 10^{10}$\,M$_\odot$.  The SFR is modest at $\sim 40$\,M$_\odot$ yr$^{-1}$, and the galaxy is also quite dust obscured ($A_V \approx 2.2$\,mag).

\subsection{GRB 070802}

The relatively faint, red afterglow of this event was observed by several telescopes including GROND on the ESO/MPG 2.2\,m, which detected the afterglow in all seven filters from $g$ though $K$ \citep{Kruehler+2008}.  (GRBs 080607 and 081109 are the only other events in this sample to be detected in the $R$ band or bluer filters.)  The optical afterglow was also observed spectroscopically with the VLT, establishing an absorption redshift of 2.454 \citep{Eliasdottir+2009}.  Both the GROND SED and the VLT spectrum show a clearly reddened continuum with a significant detection of the 2175\,\AA\ dust bump.  The actual degree of dust attenuation is borderline for inclusion: \cite{Kruehler+2011} estimate $A_V = 1.23 \pm 0.17$\,mag, only marginally establishing $A_V \geq 1$\,mag.

Most of our photometric data points are taken from the VLT observations of \cite{Eliasdottir+2009} and the GROND observations of \cite{Kruehler+2008}.  We add to this {\it HST} and {\it Spitzer} observations of the source from our programs.  The host position in the 3.6\,$\mu$m IRAC image was affected by column pull-down originating from a bright star elsewhere in the image; we removed this effect by adding flux to this column in the vicinity of the host until the artifact disappeared and then performed photometry in the usual way, but this point should nevertheless be used with caution.

While well detected in the optical bands, the host of GRB 070802 is only weakly detected by {\it Spitzer}, indicating a low stellar mass and relatively young population.  We estimate an SFR of $\sim 16$\,$M_\odot$\,yr$^{-1}$ and a mass of $\sim 5 \times 10^9$\,M$_\odot$.

\subsection{GRB 071021}

GRB 071021 was a very long burst initially flagged as a high-redshift candidate based on its prompt emission properties  \citep{GCN6967}, a possibility seemingly encouraged by the lack of detection of an optical counterpart in early imaging \citepeg{GCN6961,GCN6964}.  Subsequent imaging did detect a faint, fading NIR counterpart with red $KJz$ colors consistent with a dust-reddened afterglow \citep{GCN6968}.  

We have imaged this field in several different filters with LRIS and NIRC on Keck, as well as with NIRI ($J$ band during partly cloudy conditions).  A faint galaxy is detected at the afterglow position in all of these images except $J$.  The surrounding environment is complex; the host galaxy is part of a ``chain'' of four different sources along a line separated by a few arcseconds.  The association of these galaxies with each other is not clear (the nearest source, about 2\arcsec\ southeast of the host, has a similar, red color), but fortunately the host is well separated from the other objects and is relatively bright, allowing construction of a good SED.  This source was also included in the X-shooter spectroscopic campaign of \cite{Kruehler+2012b} and has a measured redshift of 2.452.

The combination of bright {\it Spitzer} fluxes and prominent optical detections indicate a massive and rapidly star-forming galaxy.  The SFR is very high (SFR $\approx 200$\,M$_\odot$\,yr$^{-1}$), as are the stellar mass ($\sim 10^{10}$\,M$_\odot$) and the extinction ($A_V \approx 1.9$\,mag).

\subsection{GRB 080207}

GRB 080207 was followed quickly with deep imaging at the VLT and not detected in any filter, including in the NIR.   The corresponding upper limit on the extinction ($A_V > 2.0$\,mag) is actually not as strong as that of most other sources in our sample since the X-ray afterglow faded quite rapidly, but the depth of the nondetections (coupled with a very high X-ray $N_{\rm H}$) clearly identify this event as a dust-extinguished dark burst.

The host galaxy of GRB 080207 has been studied in detail by \cite{Rossi+2012}, \cite{Svensson+2012} and \cite{Hunt+2011}.  The photometric redshifts presented in the latter two papers ($z=1.74^{+0.34}_{-0.18}$ and $z=2.2^{+0.2}_{-0.3}$, respectively) have also recently been confirmed by X-shooter spectroscopy that establishes the redshift of 2.086 \citep{Kruehler+2012b}.  Our analysis here combines photometry from these two sources and fixes the redshift to the spectroscopic value. (Except in the case of the NIRC $K_s$-band observation, we have not reanalyzed the photometry independently.)

Because the source is so red (in terms of $R-K$ it is the reddest object in the sample), and the optical detections are all at relatively low significance, there is significant degeneracy between a mature population and a highly dust-obscured, rapidly star-forming one.  This results in a large range of uncertainty in the SFR and $A_V$, and even an SFR of zero is only ruled out at about $1\sigma$, although the clear detection of several emission lines in the NIR with X-shooter suggests that the actual value is probably not far from the best fit, $\sim 50$\,M$_\odot$\,yr$^{-1}$ (the dust-uncorrected H$\alpha$ SFR from the X-shooter provides a lower limit of 10--15\,M$_\odot$\,yr$^{-1}$).  In addition, the {\it HST} magnitudes are somewhat discrepant (by about $\sim 0.5$\,mag, many times the photometric error) from ground-based photometry in the nearby filters, leading to an unavoidably poor fit.  The mass is not affected by these issues and is well constrained at $\sim 1.2 \times 10^{11}$\,M$_\odot$, a large value intermediate between (but consistent with) the estimates provided by \cite{Svensson+2012} and \cite{Hunt+2011}.

\subsection{GRB 080325}

This event and its host galaxy were the subject of the study of \cite{Hashimoto+2010}.  A faint, red, fading NIR transient was detected by Subaru at a position offset slightly from a fainter, extended source that represents the presumptive host of the GRB.  The photometric redshift of the galaxy from that work is $z=1.9_{-0.1}^{+0.3}$.  Further Subaru observations since that time using the MOIRCS spectrograph have confirmed the redshift of this host to be 1.78 (Hashimoto et al. 2013, private communication).  Our own EaZy photometric redshift estimate ($z_{\rm phot} = 1.57 ^{+0.71}_{-0.15}$) is also consistent with this value.  

We have observed this field with LRIS in several optical filters and with {\it Spitzer}.  These observations confirm the characterization of this source as a faint, red galaxy (although we measure somewhat different optical magnitudes than reported by \citealt{Hashimoto+2010}).

Given the very bright $K$-band and {\it Spitzer} fluxes this galaxy is quite massive.  Our step-function star-formation history initially converged to a maximum stellar age older than the Universe at this redshift; to prevent this unrealistic behavior for this event we modified the model to an exponentially falling star-formation history, which produces a cosmologically consistent answer.  The SFR of this host is relatively modest ($\sim 13$\,M$_\odot$\,yr$^{-1}$) in comparison to its total mass ($\sim 10^{11}$\,M$_\odot$).  These values are broadly consistent with those reported by \cite{Hashimoto+2010}. 

\subsection{GRB 080607}

This exceptional event has been the subject of several previous papers by our collaboration detailing various aspects of the burst and its host environment.  Notable attributes of this burst include the extreme luminosity of its prompt emission and afterglow \citep{Perley+2011}, well-determined line-of-sight extinction properties ($A_V \approx 3$\,mag of dust with a clear 2175\,\AA\ absorption bump; \citealt{Perley+2011}), an afterglow spectrum rich in ionic and molecular absorption features from a molecular cloud within the host along the line of sight \citep{Prochaska+2009,Sheffer+2009}, and the relatively high redshift of the system ($z=3.038$ from the same spectra).  The host galaxy was previously studied by \cite{Chen+2010}; see also the Erratum by \citet{Chen+2010erratum}.

We have reanalyzed all of our ground-based photometry of this event for the study here.  By fixing the aperture at the host location we achieve marginal (2--3$\sigma$) detections of the host in both the Magellan $r$ and Keck $I$ bands.  The basic result of \cite{Chen+2010erratum} remains the same, which is that this is a massive, star-forming host.  Compared to other galaxies in our sample, however, none of its properties are particularly extreme (SFR $\approx 20$\,M$_\odot$\,yr$^{-1}$, $M_* \approx 4 \times 10^{10}$\,M$_\odot$).

\subsection{GRB 081109}

GRB 081109 is an event analogous to GRB 080607, a luminous burst occurring within a dusty environment which heavily reddened (but did not prevent detection of) its optical afterglow.  No spectra of the afterglow were available, but the redshift of 0.979 was established from later spectroscopy of its bright host galaxy (the location of which is coincident with the afterglow based on comparison of early-time and late-time GROND imaging).  While none of the optical measurements are below the $\beta_{\rm OX} < 0.5$ criterion, the GROND afterglow SED provides a strong constraint on the extinction of $A_V=3.4^{+0.4}_{-0.3}$ mag \citep{Kruehler+2011}, easily satisfying the criterion for inclusion in this sample.

Our observations are taken mostly from \cite{Kruehler+2011} and supplemented by our {\it Spitzer} photometry, which provides an improved (slightly higher) estimate of the stellar mass.  Our values are otherwise generally consistent with those of \citealt{Kruehler+2011} (SFR = 50\,M$_\odot$\,yr$^{-1}$, $M_* = 9 \times 10^{9}$\,M$_\odot$, $A_V = 1.3$\,mag) and indicate a moderate-mass, rapidly star-forming host.

\subsection{GRB 081221}

Another very bright \Swift\ burst, GRB 081221 was not observed rapidly, but deep observations with a variety of telescopes at later times reported no optical afterglow \citep{GCN8689,GCN8719} with the exception of NIR imaging with Gemini-North, which detected a faint source in only the $K$ band \citep{GCN8698}.  This source was observed to fade in our later Gemini epochs, leaving a slightly extended source underlying this position.  The source is also visible in optical images (where it does not fade), as also reported by \cite{GCN8711} and \cite{GCN8719}.   The region around the transient is fairly complicated, with several other galaxies visible within a few arcseconds.  None of these appear to be visibly interacting with the GRB host.

Photometry is reported in a variety of optical and NIR filters from Keck and Gemini observations and with {\it HST}. 
This galaxy was also observed with X-shooter; a secure redshift based on detection of the \OIII\ and H$\alpha$ emission lines at $z=2.260$ was reported by \cite{Salvaterra+2012}.

The host of this system is massive ($\sim 4 \times 10^{10}$\,M$_\odot$) but young and rapidly star forming ($\sim 170$\,M$_\odot$\,yr$^{-1}$), with only a very weak Balmer break evident in the SED.  As with all other rapidly star-forming hosts in our sample, the dust extinction is also large ($A_V = 1.7$\,mag).

\subsection{GRB 090404}

GRB 090404 was imaged rapidly by the Xinglong TNT and after about 8\,hr with the NOT and (in the NIR) the Calar Alto telescope; no transient was detected with any of these observations \citep{GCN9090,GCN9093,GCN9099}. However, an afterglow was detected at millimeter wavelengths \citep{GCN9100}.

Our imaging includes Keck $B$, $g$, and $I$ bands, as well as {\it HST} and Gemini/NIRI ($K$ band) taken 22 days after the burst.  The position of the afterglow is established by the millimeter observations of \cite{GCN9100}, which place it near the center of an extended, highly elongated source that is clearly detected with {\it HST} (in F160W only) and NIRI. It is marginally detected in most optical filters.  The field around this presumptive host is complex: a pair of bright interacting galaxies is evident slightly to the east (the near edge of one galaxy's disk is only 1.5\arcsec\ away) and a compact, blue source is visible 1\arcsec\ to the northeast.  These nearby sources potentially can introduce significant blending; fortunately, all images reported here were taken in very good seeing conditions, and the most nearby source is very blue and contributes negligibly to the {\it Spitzer} bands.  (Nevertheless, as with all other {\it Spitzer} observations in crowded fields, we subtracted all nearby sources before performing photometry.)

No spectroscopic redshift is currently available for this object.  The galaxy is consistent with a photometric redshift of $\sim 3$ based on a (questionable) Lyman break in our photometry; we do not significantly detect the host in a deep $B$-band observation despite a relatively strong detection in $g$.  This redshift is generally consistent with the other colors available for the object, in particular the suggestion of a Balmer break in the NIR photometry and the relatively red IRAC 3.6--4.5\,$\mu$m color.  However, quantitatively this is not highly significant, with a wide range of redshifts consistent with the data (EaZy gives $1.18 < z < 3.83$ at 95\% confidence, or $z=2.87^{+0.47}_{-1.08}$ at 1$\sigma$).

Assuming the Lyman-$\alpha$ break is real and the redshift is high, the bright IRAC detections indicate a quite massive galaxy:  $\sim 5 \times 10^{10}$\,M$_\odot$ at $z=3$, similar to the values measured for GRBs 080207, 080325, and 060923A.  The extinction is significant ($A_V \approx 1.8$\,mag); only a (relatively poorly constrained) upper limit can be placed on the UV star-formation rate of $<200$\,M$_\odot$\,yr$^{-1}$.

\subsection{GRB 090407}

Early-time observations of this GRB were limited to the small ROTSE telescopes (which did not detect a counterpart but are not deep enough to be constraining; \citealt{GCN9102}), and the X-ray afterglow was relatively faint at this time, but deep VLT $R$-band and $z$-band limits at 13.7\,hr \citep{GCN9108} are far below the extrapolated X-ray flux and identify this event as a dark burst.  GROND limits (including NIR observations; \citealt{GCN9109}) were contemporaneous with the VLT observation but, given the time since the burst, these are not particularly constraining.

The host galaxy of this event is extremely red, and to date we have clearly detected it only in the NIR, with {\it HST}, {\it Spitzer}, and Gemini-NIRI (despite partly cloudy conditions), although forced aperture photometry at the afterglow location does give weak (3$\sigma$) detections in our Keck observations.  An X-shooter redshift of 1.448 is available \citep{Kruehler+2012b} based on detection of H$\alpha$ and \OII.  Given the weak (and possibly uncertain) optical detections the properties of the system are difficult to constrain with any precision---although as with most other galaxies in our sample, moderate-to-large values for the SFR, extinction, and mass are preferred.

\subsection{GRB 090417B}

GRB 090417B was a moderately bright but very long burst; its prominent X-ray afterglow was immediately recognized to be coincident with an SDSS galaxy \citep{GCN9140} whose redshift was measured early to be 0.345 \citep{GCN9156}.  Given this bright host, afterglow upper limits reported in the GCN Circulars are not straightforward to interpret (image subtraction against a later reference epoch would be necessary); in the very conservative assumption that the afterglow is no brighter than the quoted host-galaxy magnitudes, the $A_V > 1$\,mag criterion is marginally satisfied, though in all likelihood the real limits are much deeper than anything currently quoted and so the actual extinction is probably significantly larger.   No thorough study of the optical afterglow has yet been performed; the claimed UVOT limit of $u > 24.9$\,mag at 11\,hr in \cite{Holland+2010} does not seem reasonable given typical limits attained by the UVOT in a few hours of exposure time.  (The specific estimate of $A_V > 11$\,mag in \citealt{Holland+2010} is based on the $N_{\rm H}$ measurement from the X-ray afterglow, not on optical limits.)

As the lowest-redshift event in our sample, the data on the host galaxy of this extremely obscured event are particularly extensive, including several UVOT detections (not shown in Figure \ref{fig:hostsedtile}) as well as a wide suite of detections through the optical--NIR bands and {\it Spitzer} photometry.  The majority of our measurements come from \cite{Holland+2010}, and are supplemented by our {\it HST} and {\it Spitzer} photometry.

Like some other bursts in the sample, the strong Balmer break points toward a declining SFR despite a relatively young stellar age.  For a constant SFR we derive a best-fit value of 3\,M$_{\odot}$\,yr$^{-1}$, but a significantly better fit is obtained if the current SFR is decoupled from the assumption of a constant star-formation history (in which case it converges to $0.35 \pm 0.2$\,M$_{\odot}$\,yr$^{-1}$, suggesting a recent decline in the SFR).  The stellar mass is moderate at $\sim 3.5 \times 10^9$\,M$_\odot$, intermediate between the SMC and the Large Magellanic Cloud (LMC).  Interestingly, while the host of GRB 090417B is a relatively low-mass galaxy (like other low-$z$ GRB hosts), it does not appear to be forming stars at a particularly high rate for its mass (quite unlike most other low-$z$ GRB hosts).  The host is dusty ($A_V = 0.9$\,mag) according to our fits, although this value is lower than for other events in our sample and is far less than the value of $A_V = 3.5^{+1.0}_{-0.5}$ mag quoted by \citet{Holland+2010}, which assumed an intrinsic power-law host SED (an assumption which is unreasonable for this galaxy given the strong Balmer break).

\subsection{GRB 090709A}

GRB 090709A was another very bright \Swift\ burst with a prominent X-ray afterglow.  The event triggered both the P60 and PAIRITEL, which automatically slewed to the location and detected only a very faint and red afterglow.  Our analysis of the afterglow was previously published by \cite{Cenko+2010}, in which we interpret the burst as a classical dark burst with a highly dust-reddened afterglow.  However, the possibly quasiperiodic prompt-emission behavior of this event led some to suggest that the event may actually be a Galactic source such as a magnetar \citep{GCN9645,GCN9649}.  This claim has been disputed by others on account of its high Galactic latitude, evidence for X-ray absorption, and the classical behavior of its afterglow \citep{deLuca+2010,Cenko+2010}.

Host-galaxy observations reported in our earlier work included deep integrations of the field with GTC and Keck (LGS-AO) which did not detect a host galaxy at the afterglow position.  Since that time we have integrated much deeper in both the optical (with LRIS) and NIR bands (with Gemini), as well as with {\it HST} (WFC3/IR) and {\it Spitzer}.  A faint object is detected in all of these observations (except in the $RG_{850}$ filter) directly at the position of the NIR transient.  In the {\it HST} observations this object is clearly elongated in the E-W direction, identifying it as a distant galaxy.

The {\it HST} photometry, combined with the deep $RG_{850}$-band limit from LRIS, clearly indicates the presence of a spectral break in the SED at around 11,000\,\AA\ which we associate with Balmer absorption at $z \approx 1.8$.   A fit to the broad-band photometry with EaZy gives a 95\% confidence redshift range of $1.14<z<2.34$, which also rules out the claim by \cite{deLuca+2010} of a moderately high redshift of $4.2 \pm 0.2$ based on the X-ray absorption properties.  (Indeed, the $B$-band detection of the host alone rules out this redshift.)

The properties of the host are relatively moderate among events in this sample: at $z=1.7$ we have $M_* \approx 10^{10}$\,M$_\odot$, SFR $\approx 8$\,M$_\odot$\,yr$^{-1}$, and $A_V \approx 1.4$\,mag.  The SFR is quite sensitive to the assumed redshift; if the redshift is at the bottom end of the allowed range then only the $g$ and $B$ filters measure the rest-frame UV (which are too similar to usefully constrain the spectral index and therefore the extinction correction to the SFR) and the SFR cannot be usefully constrained.  However, as long as $z \gtrsim 1.3$ (which is secure at $\sim 90$\% confidence) this is not a concern, and the properties are similar to those outlined above.

% S5
\section{Sample Properties}
\label{sec:properties}

\begin{figure}
\centerline{
\includegraphics[scale=0.6,angle=0]{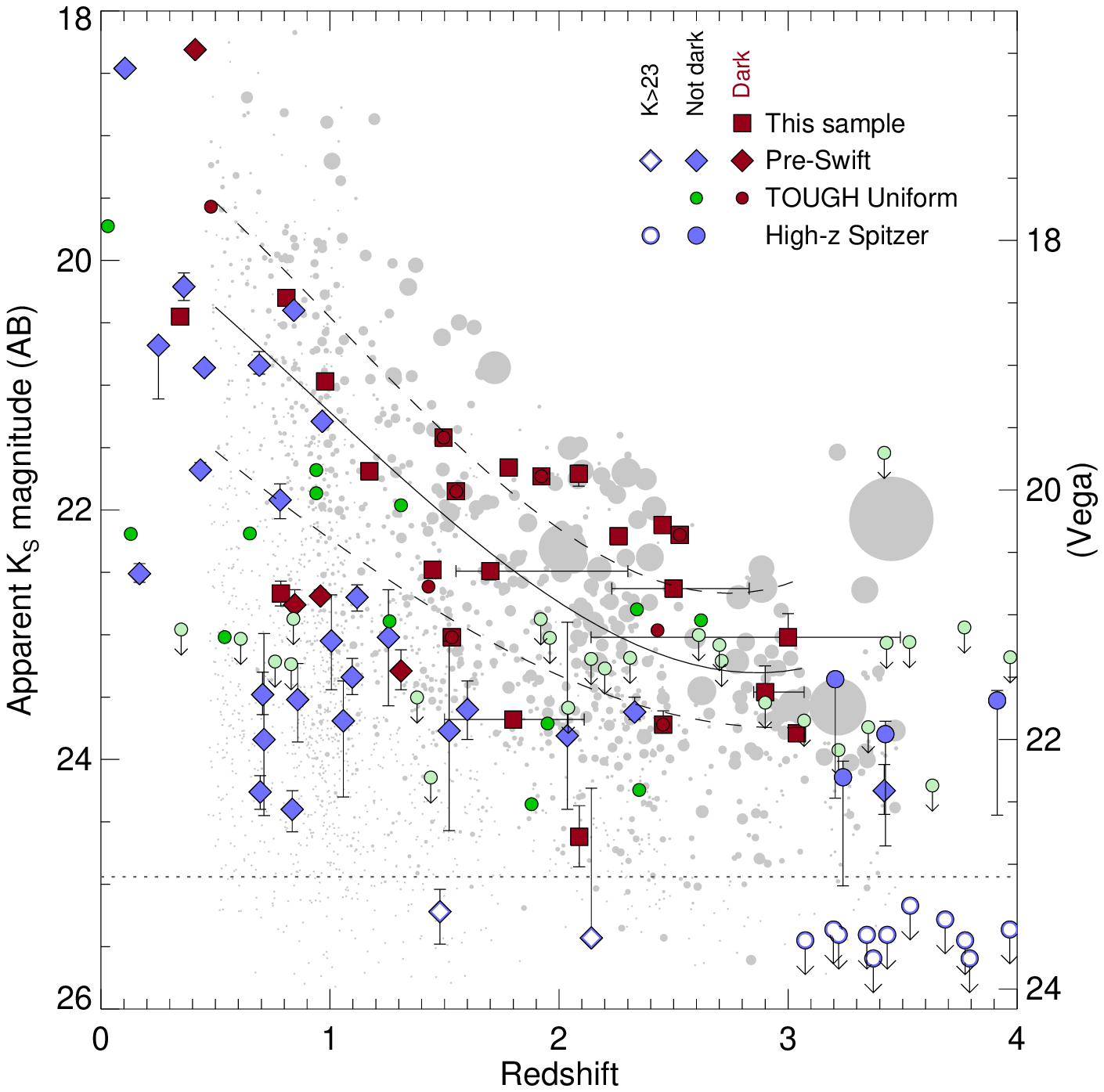}} 
\caption{$K$-band apparent magnitudes of GRB host galaxies from our sample ($A_V > 1$\,mag \Swift\ bursts, squares) and others: pre-\Swift\ GRBs from \citet{Savaglio+2009}, diamonds; a uniformly selected sample of \Swift\ GRBs from \citet{Hjorth+2012}, small circles; and optically bright $z>3$ GRBs from \citet{Laskar+2011}, large circles (we assume $K-3.6 = 0^{+0}_{-1}$ AB mag for these sources).   Colored symbols indicate GRB hosts, with dark red indicating dust-obscured events.  Hollow symbols indicate $K$-band faint hosts ($K > 23$ Vega mag).  Additionally, in light gray we plot field galaxies from the MOIRCS Deep Survey \citep{Kajisawa+2010} of the GOODS-N field, which is complete to $K=23.1$ mag at 5$\sigma$.  The area of each field-galaxy symbol is scaled by the galaxy's SFR, and so above the completion limit of the survey (dashed horizontal line) the GRB host distribution should follow the distribution of gray ``ink.''  We have quantified this by plotting dotted and solid curves representing the first/third quartile boundaries and the median (respectively) of the \emph{SFR-weighted} distribution of $K < 23$\,mag (Vega) field galaxies and, therefore, the \emph{expected} distribution for $K < 23$\,mag GRB hosts if they follow star formation uniformly.  This expectation is certainly not met at $z \approx 1$: a large majority of hosts occur in galaxies fainter than the expected median (but brighter than the MODS completeness limit).  }
\label{fig:zk}
\end{figure}

\begin{figure}
\centerline{
\includegraphics[scale=0.6,angle=0]{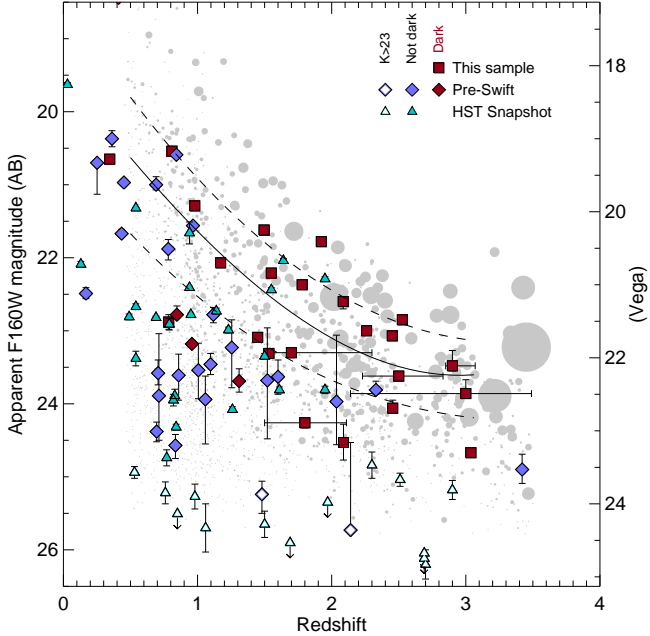}} 
\caption{F160W ($H$-band) apparent magnitudes of GRB host galaxies from our sample and pre-\Swift\ comparison sample, plus a sample of randomly selected \Swift\ GRB hosts with measured optical redshifts (representative of optically bright \Swift\ bursts) at $z<3$ from the {\it HST} Snapshot survey (Tibbets-Harlow et al., in prep.)  Symbol conventions are the same as in Figure \ref{fig:zk} except that we plot the Snapshot hosts (cyan triangles) in place of the TOUGH galaxies.  These hosts confirm the trend seen in Figure \ref{fig:zk} at $z \approx 1$ for GRBs to prefer subluminous galaxies, but at $z \approx 2$ both optically selected and dark GRBs commonly occur in luminous galaxies and show no strong deviations from the expected magnitude distribution. }
\label{fig:zh}
\end{figure}

\begin{figure}
\centerline{
\includegraphics[scale=0.6,angle=0]{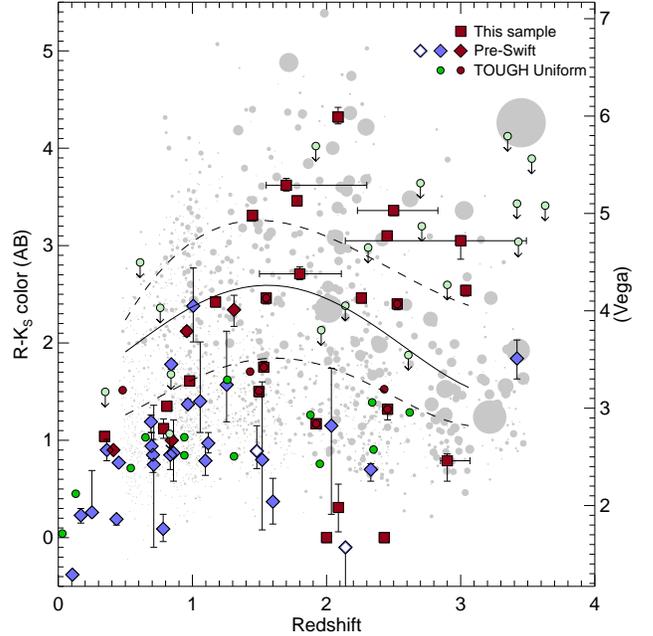}} 
\caption{$R-K$ color of GRB hosts and field galaxies.  Symbol conventions are the same as in Figure \ref{fig:zk}, and the curves indicate quartile boundaries of the expected host distribution \emph{for $K < 23$\,mag galaxies}.   The hosts of $z>1$ dust-obscured GRBs are much redder than those of unobscured hosts.  GRBs clearly appear to avoid red hosts at $z \approx 1$ compared to expectation from the relative SFRs of this population, although this effect is significantly less pronounced (or may even be absent entirely) at higher redshifts.}
\label{fig:zrmk}
\end{figure}

The properties of the host sample discussed above differ starkly from those of the optically selected host samples of the past.  While most such previous efforts have struggled to even detect $\sim50\%$ of the hosts in the rest-frame NIR \citepeg{LeFloch+2003,LeFloch+2006}, \emph{every} host in our sample was detected in the NIR and with {\it Spitzer} (Figures \ref{fig:irmosaic}--\ref{fig:iracmosaic} and \ref{fig:zk}--\ref{fig:zh}).  The host galaxies in our sample show a diverse range of colors but include two extremely red objects (EROs, with $R-K > 5$ mag: GRBs 080207 and 090407) and many additional objects that are just short of this threshold ($R-K > 4$\,mag: GRBs 090404, 090709A, 080607, 070521, 071021, 060306, 060923A, and 080325).  Only two GRB host galaxies with $R-K > 4$\,mag were known pre-\Swift\ \citep{Levan+2006,Berger+2007}, both of which were also dark bursts (Figure \ref{fig:zrmk}).

To facilitate direct comparison between our results and optically selected GRB host studies, we downloaded the photometric data for all pre-\Swift\ GRBs with known redshift and deep, previously published observations in both the optical and NIR bandpasses from the online compilation in the GHosts database\footnote{grbhosts.org .} \citep{Savaglio+2009}, and ran these SEDs through the same procedure as for the galaxies in our obscured-burst sample, fitting for the SFR, mass, and $A_V$ using our SED analysis code.  The sample (and the results of our SED fitting) is summarized in Table \ref{tab:hostmodelpar}.   Most of these events are optically bright bursts, although a few are ``dark'' and probably similar in nature to the objects in our sample.   Our derived values are generally consistent with those previously published by \cite{Savaglio+2009}, although we were also able to solve for (or at least constrain) $A_V$ for all hosts, which Savaglio et al.\ attempted only for a few objects.   We did not use any long-wavelength ($>10\,\mu$m) observations: an analysis of submillimeter/radio SFRs will be reserved for future work.

We did \emph{not} extend this comparison sample to \Swift-era bursts.  Only a few \Swift\ hosts have been studied at the level of detail necessary to perform SED fitting (at least 3 broad-band filters, including NIR observations), and these events may not be representative since the pursuit of multi-filter observations and publication of the resulting work was probably influenced by the properties of the hosts in the first place.  These same concerns probably apply to some extent to pre-\Swift\ hosts as well, but since the sample of known-redshift events was relatively small before 2005, observations were necessarily concentrated on a limited number of objects (specifically, our sample of 31 ``well-observed'' hosts constitutes more than 65\% of all pre-\Swift\ events with known redshifts)---and so the literature sample of published host photometry is much less susceptible to selective-observing and selective-publication bias than in the current era.\footnote{Still, even the pre-\Swift\ sample is probably not perfectly representative of the hosts of optically bright bursts; a bias against very faint hosts may still be present (since acquiring multiple filters may not be attempted if the first observation produces a deep nondetection, and emission-line redshifts are much easier to provide for brighter hosts).  As we continue to find an overabundance of \emph{faint} galaxies after incorporating the obscured population, we do not expect that this significantly impacts the primary conclusions of this work.}  

% Figure 7

% Figure 8

% Figure 9

% Figure 10
\begin{figure}
\centerline{
\includegraphics[scale=0.6,angle=0]{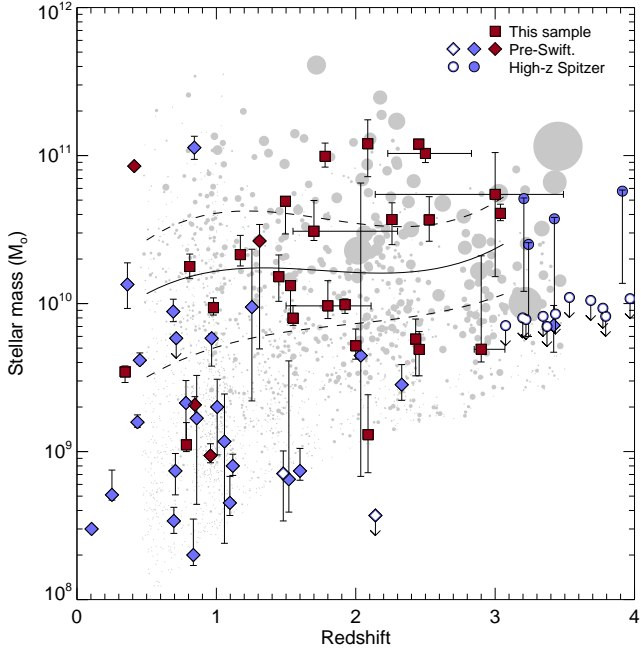}} 
\caption{Stellar mass versus redshift of GRB hosts and field galaxies; symbol conventions are the same as in Figure \ref{fig:zk} with curves indicating quartile boundaries in the expected distribution for $K < 23$\,mag galaxies.  Dust-obscured GRBs probe a much more massive host population than unobscured GRBs, an effect which produces the magnitude trend seen in Figure \ref{fig:zk}.   The connection between the GRB rate and the SFR appears significantly influenced by mass, especially at $z<1.5$ where the relative numbers of GRBs in ``high-mass'' ($>10^{10}$\,M$_\odot$) versus ``low-mass'' ($<10^{10}$\,M$_\odot$) galaxies differ by a factor of $\sim 5$ despite an expected ratio of unity.  While even deep field surveys are incomplete at the low-mass end, very few GRB hosts would have been missed in surveys to this depth, and this does not explain our result. Note: some additional scatter has been added to the MODS points to remove some minor nonphysical discreteness in the mass distribution introduced as a result of the mass-fitting procedure in MODS \citet{Kajisawa+2010,Kajisawa+2011}.}
\label{fig:zmass}
\end{figure}

% Figure 11
\begin{figure}
\centerline{
\includegraphics[scale=0.6,angle=0]{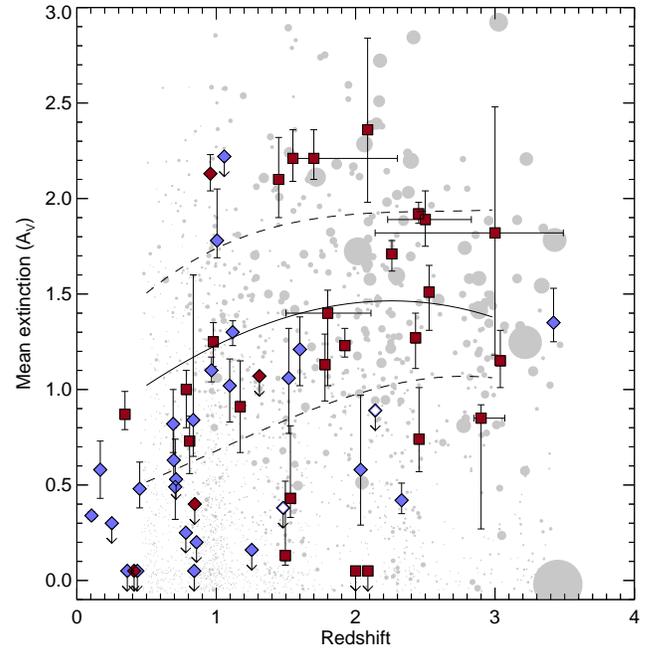}} 
\caption{Average dust attenuation for GRB hosts, as given by the results of our SED fitting.  Symbol conventions are the same as in Figure \ref{fig:zmass} with curves indicating quartile boundaries of the expected host distribution for $K < 23$\,mag galaxies.  Dust-obscured GRBs (unlike optically selected GRBs) tend to inhabit dust-obscured galaxies, although the trend is not universal---see also Figure \ref{fig:extcompare}.}
\label{fig:zav}
\end{figure}

% Figure 12
\begin{figure}
\centerline{
\includegraphics[scale=0.6,angle=0]{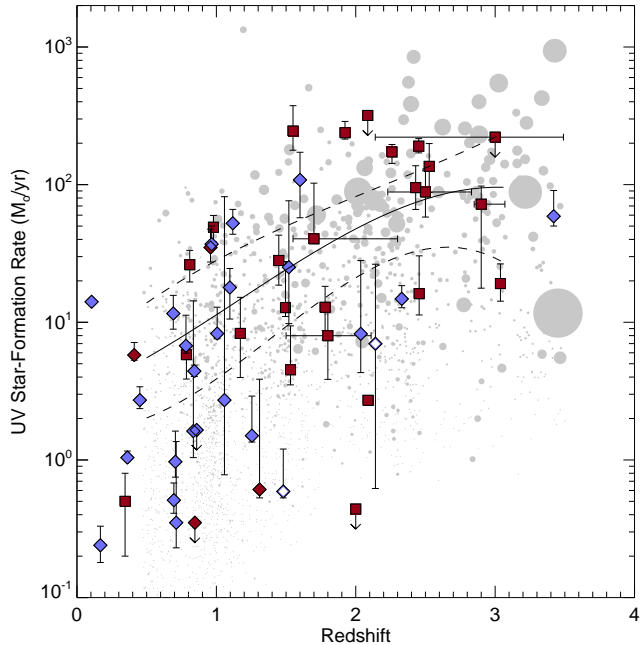}} 
\caption{Dust-corrected UV SFR (as inferred from SED fitting to the UV/optical/IR observations) for GRB hosts and field galaxies.  Symbol conventions are the same as in Figure \ref{fig:zmass} with curves indicating quartile boundaries of the expected host distribution for $K < 23$\,mag galaxies.  The symbol size for the MODS survey galaxies is proportional to the total star formation \emph{including contribution from 24\,$\mu$m-derived SFRs}, although for most galaxies this is similar to the UV dust-corrected SFR plotted on the ordinate.  Dust-obscured and optically selected GRBs occupy galaxies with a range of SFRs, although rapidly star-forming galaxies are much more likely to host obscured GRBs.  Several GRBs are hosted in probable ULIRGs given their extremely large SFRs.  Note that there are very few nondark comparison objects at $z>2$ since there is no way to determine SFRs from the one-filter and two-filter photometry of \cite{Hjorth+2012} and \cite{Laskar+2011}.}
\label{fig:zsfr}
\end{figure}

% Figure 13
\begin{figure}
\centerline{
\includegraphics[scale=0.6,angle=0]{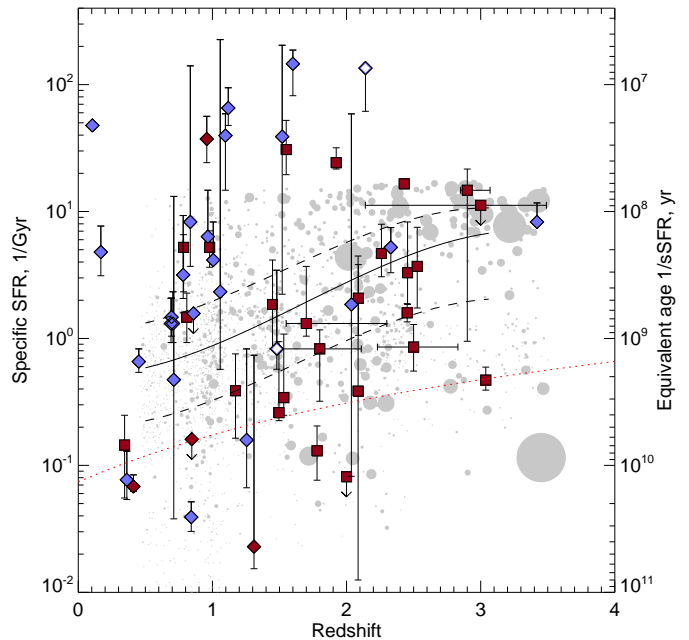}} 
\caption{Specific SFR for GRB hosts.  Symbol conventions are the same as in Figure \ref{fig:zmass} with curves indicating quartile boundaries of the expected host distribution for $K < 23$\,mag galaxies.  There are no obvious trends visible, and the uncertainty of many of our points is large.  A significant fraction of both obscured-GRB and unobscured-GRB hosts occur in galaxies with very high sSFRs, with ages clearly younger than the minimum permitted within the MODS modeling.  However, this population is not dominant (perhaps 20--30\% of GRB hosts).   A few hosts actually show significantly \emph{lower} sSFRs, even below 1/$t_{\rm Hubble}$ (dotted curve), indicating a mature galaxy whose apparent SFR has fallen with time.}
\label{fig:zssfr}
\end{figure}

Because of the lower sensitivity of most other GRB satellites relative to \Swift, the median redshift of pre-\Swift\ events is much lower ($z \approx 1$--1.5) than for \Swift\ events ($z \approx 2$--3; \citealt{Jakobsson+2006,Jakobsson+2012}).  Consequently, while the pre-\Swift\ sample is appropriate for comparisons at $z \approx 1$, it is much more limited at higher redshifts.   While there does not yet exist a completed survey of \Swift\ GRBs with both understandable selection effects and deep observations in enough filters to model the SED in any detail, some individual studies using only one or two filters have recently been published that are useful for extending our comparisons to higher redshifts in a more limited sense.  The Optically Unbiased GRB Host Project (TOUGH; \citealt{Hjorth+2012}) is a large ground-based VLT project centered on deep $R$ and $K$ imaging of 69 uniformly selected bursts that is 80\% redshift-complete.  (Since this sample is not optically selected it includes dust-obscured events; indeed, six dust-obscured events in this paper overlap with the TOUGH sample.)  The $K$-band limits in TOUGH only reach $K \approx 21$\,mag, but a much deeper survey of the hosts of a sample of randomly selected, optically bright \Swift\ bursts at $z < 3$ was conducted with {\it HST} in the F160W filter (Tibbets-Harlow et al., in prep.).  Looking to even higher redshifts, a recent IRAC study of optically selected GRBs at $z > 3$ \citep{Laskar+2011} provides deep observations of a large sample of high-$z$ hosts at 3.6\,$\mu$m.

In addition to comparing the GRB populations to each other, a major motivation of this work is to compare both populations to field-selected galaxies in order to reevaluate to what extent the inclusion of dust-obscured GRB hosts in the overall sample is likely to alter our perceptions of how well the GRB rate tracks the SFR at high redshift.  While a rich literature of deep comparison surveys of many different fields now exists, we have chosen as our primary comparison sample the catalogs provided from the MOIRCS Deep Survey (MODS) of \cite{Kajisawa+2010,Kajisawa+2011}, a small-area, ultradeep survey of the GOODS-N field, selected in the $K$-band and complete to $K = 23$ Vega mag with redshifts measured by a combination of spectroscopic and photometric techniques.  The primary driver of this choice of comparison is the depth of the MODS survey (reaching well below $L^*$) and its selection in a single, mostly dust-unbiased NIR band, avoiding the need to extrapolate or make large completeness corrections in order to quantify the importance of the faint galaxies which many GRBs are observed to inhabit.  Magnitudes and derived properties are taken directly from the online catalogs, with an adjustment to the SFRs and stellar masses of a factor of 1.8 (decrease) to convert from a Salpeter to a Chabrier IMF in order to fairly compare these results to ours \citep{Erb+2006,Michalowski+2012a}.

\subsection{Redshift Distribution}
\label{sec:zdist}

While the redshift distribution of our sample is not expected to necessarily reflect the parent \Swift\ distribution given our selection methods (\S \ref{sec:sample}), we actually observe the two to be reasonably similar; a K-S test against a catalog of all publicly available \Swift\ redshifts gives a probability value of 0.36 that the distributions are drawn from the same population.  This suggests that any redshift dependence in the intrinsic dust obscuration of GRBs is relatively minor.  There is a possible hint toward a dearth of relatively nearby GRBs:  half of the \Swift-triggered events (11 out of 22) are in a narrow redshift range of 1.5--2.5, an epoch centered on the peak of cosmic star-formation intensity and of the abundance of submillimeter galaxies \citepeg{Chapman+2005,Hopkins+2006,Wardlow+2011,Michalowski+2012c}.  Only about 25\% of all \Swift\ GRBs are in this same range \citepeg{Jakobsson+2012}.   More observations of a cleaner sample would be needed to confirm whether this represents a real effect (rather than a more subtle result of selection bias or a result of small-number statistics).

\subsection{Stellar Mass}
\label{sec:massdist}

The low observed stellar masses of the population of known GRB host galaxies has long been one of the strongest arguments for a GRB population that does not trace star formation uniformly in all environments.  Many $z \approx 0$ hosts have extremely low mass, falling well below the stellar masses of typical star-forming galaxies selected from SDSS or from the hosts of Type II or Type Ib/c supernovae found in untargeted surveys \citep{Modjaz+2008}.  While it is not completely clear how the low-redshift ($z<0.1$), underenergetic ($E_{\rm iso} < 10^{49}$\,erg) GRB population that dominates these samples relates to more typical\footnote{``Typical'' in terms of detected rates---while rarely observed, low-luminosity events are more common in a volumetric sense \citepeg{Soderberg+2007}.} cosmological GRBs, generally (if not ubiquitously) low masses are also observed in the case of the host of the nearby, cosmological-luminosity GRB~030329 ($M_* \approx 6 \times 10^7$\,M$_\odot$, similar to that of the SMC) and in samples of $0.3<z<1$ and $z \approx 1$ GRB hosts \citep{Levesque+2010g,CastroCeron+2010}.

The stellar masses of obscured GRB hosts (Figure \ref{fig:zmass}) are much larger than those of the unobscured population, a trend manifest in the remarkable fact that all dark hosts in our sample are detected in the NIR and with IRAC, a reversal of previous conclusions focused on optically bright bursts \citep{LeFloch+2006}. The dark-host population in our sample has a median mass of about $5 \times 10^9$\,M$_\odot$ at $z \approx 1$ and $3 \times 10^{10}$\,M$_\odot$ at $z \approx 2$, roughly an order of magnitude higher than the equivalent values for unobscured GRBs\footnote{For reference, the stellar mass of the LMC is $\sim 3 \times 10^9$\,M$_\odot$ and that of the Milky Way is $\sim 5 \times 10^{10}$\,M$_\odot$.}.  We know of only a few obscured GRB hosts at $z \approx 3$, but comparisons to even the maximum mass values derived for the host population of unobscured GRBs at $z=3$--4 by \cite{Laskar+2011} suggests that this difference persists up to at least this redshift range.

While unobscured GRB host galaxies show an obvious aversion for the most massive host galaxies identified in field surveys (only a single unobscured-GRB host has a mass $>2 \times 10^{10}$\,M$_\odot$), dark GRBs show no such restrictions and populate both high-mass and low-mass galaxies, especially at $z>1.5$ where massive hosts are quite common.  At the same time, it is notable that at $z \approx 1$ we still do not identify any bursts within our sample above $\sim 2 \times 10^{10}$\,M$_\odot$, a mass scale which still contributes substantially to the SFR density at this epoch.  Such galaxies are very easy to detect at this redshift (two are present in the pre-\Swift\ sample), and there is no reason we should be missing these objects if they are indeed common among GRB hosts.  We will return to this question in more detail in \S \ref{sec:discussion}.

\subsection{Average Attenuation}
\label{sec:extdist}

Given that our sample was selected on the basis of an obscured sightline to the host progenitor, it would be surprising if similar signatures of extinction were not manifest in the host galaxy itself.  In Figure \ref{fig:zav} the inferred extinction of each event in our sample is plotted against the comparison sample.

Indeed, the typical dark GRB host is quite dust obscured.  A rest-frame $A_V \approx 1$\,mag is typical, with extinctions often extending to $\sim 2$\,mag (although not much beyond).  At the same time, a few hosts show little or no evidence of extinction, so this trend is not universal.  These properties seem to differ from those of the unobscured GRBs in our sample, which typically have $A_V \approx 0.5$\,mag and rarely exceed $A_V = 1$\,mag (although this could, in part, reflect the redshift differences between the populations).  

Most high-$z$ star formation occurs in fairly dust-obscured regions. A majority of the luminous galaxies that dominate cosmic star formation at $z = 1$--4 are moderately dust-obscured and visibly dust-\emph{reddened} in the UV \citep{Madau+1998,Meurer+1999}, with UV attenuations of a few magnitudes.   The $A_V$ values observed in our dust-obscured hosts appear to be fairly typical of the sorts of galaxies that dominate the SFR density, whereas the hosts of optically selected GRBs prefer much more optically thin galaxies.

\subsection{Star-Formation Rate}
\label{sec:sfrdist}

Inferring the current SFR in a galaxy from photometry alone is intrinsically more complicated than determining most other observable properties because of its sensitivity to the star-formation history:  a recent shut-off (or surge) in the SFR is difficult to recognize, since stellar populations continue to produce significant UV flux during the first 50--100\,Myr.  All UV-continuum calibrations rely implicitly or explicitly on the assumption of constant SFR (or some other prescribed slowly changing star-formation history), as we also do here (at least during the last 50\,Myr; \S \ref{sec:modeling}.)   In addition, the UV bands are strongly affected by interstellar dust; for a typical luminous $z \approx 2$ galaxy, 75\% or more of the UV light is likely to be absorbed by dust before it is able to escape the galaxy.

In spite of these uncertainties, UV-based estimates of SFR are a standard part of the extragalactic toolkit and generally have been quite successful at matching SFR calibrators at other wavelengths \citepeg{Elbaz+2007,Pannella+2009,Salim+2009,Reddy+2012}.  The submillimeter galaxies (and low-redshift ULIRGs) represent an exception to this---in these galaxies most of the star formation occurs in optically thick regions; its presence can be inferred only from the reemitted light at long wavelengths \citep{Chapman+2005}.

Most of the host galaxies in our dust-obscured sample show very high SFRs (Figure \ref{fig:zsfr}).  The median SFR for dark GRB hosts at $z = 1$--2 is $\sim 10$\,M$_\odot$\,yr$^{-1}$; for those at $z>2$ it is $\sim 60$\,M$_\odot$\,yr$^{-1}$.  This appears significantly higher (by a factor of a few) than for the hosts of optically selected GRBs, although given the rapid evolution of SFR with redshift and the small number of published \emph{non}-dark GRBs at $z > 1.5$ it is difficult to be as certain of this trend as for the other parameters discussed previously, since the TOUGH and {\it Spitzer} high-$z$ samples cannot be used to constrain the SFR even approximately.  At minimum, however, it is clear that after inclusion of dark-GRB hosts, the supposed dearth of very luminous galaxies (LIRGs and ULIRGs; see \citealt{LeFloch+2005}) compared to field samples is at least reduced if not necessarily alleviated completely.

\subsection{Age and Specific Star-Formation Rate}
\label{sec:ssfr}

We are not actually able to constrain the age directly for most of our sample, beyond distinguishing crudely between very young galaxies with a weak or absent Balmer break (and therefore a very young age) and galaxies with a strong Balmer break (and therefore an age of a few hundred Myr or more) in the case of hosts that are not too obscured.
However, since we can measure the mass and SFR, this constrains the age indirectly in terms of the specific SFR (sSFR = SFR/$M$) and its inverse, which is effectively a characteristic timescale for forming the galaxy if star formation proceeded at its current rate since formation.

The sSFRs of our sample are plotted in Figure \ref{fig:zssfr}. Values of sSFR $\approx 1$\,Gyr$^{-1}$ (equivalently, formation times of about 1 Gyr) are typical for both dark and nondark hosts, seemingly regardless of redshift.  However, there are significant excursions in both directions:  a few dark GRBs have sSFRs about one order of magnitude above this, indicating very young (dominant) stellar populations.  Given the short timescales involved ($10^7$--$10^8$\,yr), these galaxies clearly are in the throes of a short-lived event such as a galaxy merger.  On the other side of the plot, older galaxies seem to produce dark GRBs almost exclusively: with perhaps one or two marginal exceptions, the only hosts with inverse sSFRs equivalent to the Hubble time (indicating long-term quiescent evolution over many Gyr or a declining sSFR) are those of dark GRBs.

Other than this, no obvious trends are visible.  Several hosts reach above the limit of maximum sSFR of the MOIRCS Deep Survey catalog of \cite{Kajisawa+2010,Kajisawa+2011}, although this is not necessarily a physical statement as the assumed star-formation history in the MODS modeling did not allow for younger stellar populations than 50\,Myr.  Much more insight can be gleaned from examination of the sSFR in relation to other parameters, which we will return to in the discussion.

% Figure 14
\begin{figure*}
\centerline{
\includegraphics[scale=0.5,angle=0]{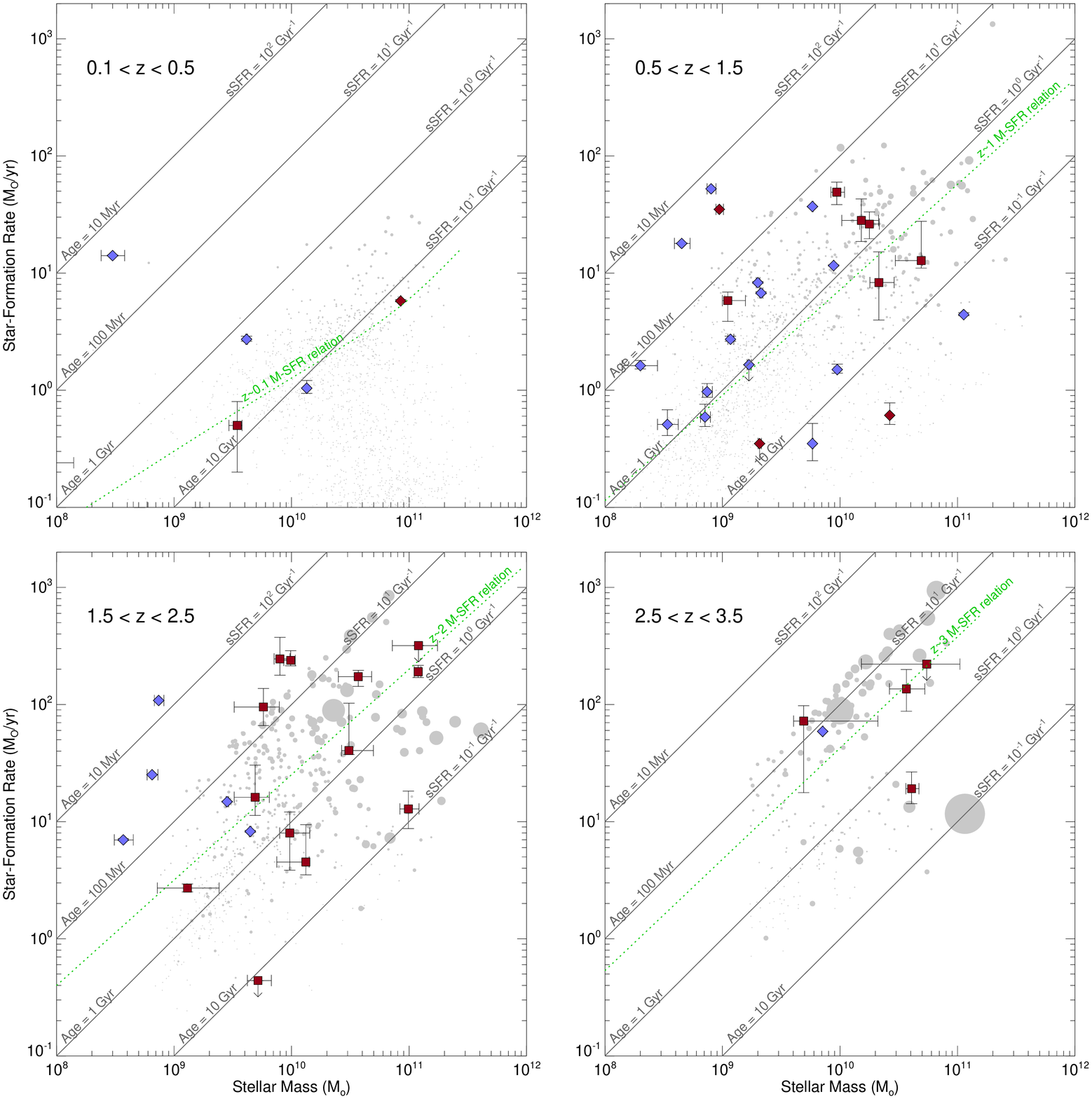}} 
\caption{SFR versus mass for obscured and unobscured GRB host galaxies compared to galaxies from SDSS and the MODS field survey at a variety of redshift ranges:  $z<0.5$, $z \approx 1$, $z \approx 2$, and $z \approx 3$.  (An SDSS slice of $0.1 < z < 0.15$ is used as the comparison sample at $z<0.5$; data are taken from the MPA-JHU catalog.)  We also show mass-SFR correlations roughly corresponding to each redshift range \citep{Salim+2007,Elbaz+2007,Daddi+2007,Magdis+2010}.  Despite the large sample of dark GRB hosts added to the population, we continue to see a notable deficiency of high-$M$, high-SFR galaxies at $z \approx 1$ (if not necessarily at higher redshifts).  Symbol conventions are the same as in previous plots.}
\label{fig:masssfr}
\end{figure*}

% S6
\section{Discussion}
\label{sec:discussion}

Previous, smaller-scale studies of the hosts of dark GRBs have led to some disagreement regarding the nature of this group of galaxies relative to the broader GRB host population.  In many cases, the detection of a relatively blue, unobscured host has led to suggestions that even the inclusion of dark GRBs is unlikely to change the overall GRB host population substantially \citepeg{Gorosabel+2003b,Tanvir+2008,Perley+2009b}.  But other studies have found very different results, with the detection of luminous, red host galaxies possibly necessitating a wholesale revision of our notion of what types of galaxies host GRBs \citepeg{Hashimoto+2010,Hunt+2011,Svensson+2012,Kruehler+2012b,Rossi+2012}.

With our new study of a much larger sample, the basic reason for this disagreement is now apparent:  the population itself is intrinsically very diverse, and small numbers of objects necessarily show only a subset of this diversity.  In fact, the host galaxies of dark GRBs cover many orders of magnitude in almost every property, ranging across most of the parameter space encountered among field-selected high-$z$ star-forming galaxies detected in current surveys.  Compounding this, the properties of both obscured and unobscured GRB hosts appear to evolve substantially with redshift, confusing any comparisons unless redshift differences in the populations are taken into account.

% S6.1
\subsection{Do GRBs Follow Star Formation?}

Given this diversity---and the fact that the dark GRB host-galaxy population seems to represent an almost perfect complement to the host population of low-$A_V$, optically bright GRBs (Figure \ref{fig:masssfr})---it is tempting to rush to the conclusion that GRBs follow star formation in a uniform way after all.  Broadly speaking, every type of high-$z$ star-forming galaxy appears capable of producing a GRB: the full range of masses, extinctions, SFRs, and other parameters now seems to be represented among the GRB host population.

However, the question is not just \emph{whether} all types of galaxies can produce GRBs, but whether they do so \emph{at the rate that we expect}.  Only if GRB formation is strictly limited to a small region of physical parameter space by a critical parameter (e.g., by a hard metallicity cutoff at a specific value) will GRBs avoid certain regions of galaxy parameter space entirely. The question is one of number statistics, matching the GRB rate as a function of various parameters (redshift, mass, luminosity, etc.) to the SFR.

To illustrate the point more clearly, in Figure \ref{fig:masssfr} we plot the mass of objects in our sample against the UV-inferred optical SFR (which is currently all we are able to measure without longer-wavelength data).  The plot is divided into four redshift bins:  $z < 0.5$,  $0.5 < z < 1.5$, $1.5 < z < 2.5$, and $2.5 < z < 3.5$ (SDSS is used as the comparison sample at $z<0.5$ instead of MODS, using the DR7 MPA-JHU value-added catalog of \citealt{Brinchmann+2004}.)  The distinction between the dark and nondark hosts is easily apparent, and the inclusion of dark GRBs greatly improves the consistency of the host population with the overall star-forming galaxy population probed by the MOIRCS Deep Survey. 

Since GRBs are star-formation selected, they should not occupy the same space filled by the field galaxies in the diagram uniformly but should prefer galaxies with higher SFR in direct proportion to their SFR.  We have represented this fact visually on our plots by scaling the area of the MOIRCS galaxies in proportion to their total SFR---loosely speaking, on our diagrams we expect the GRB distribution to follow the density of gray ``ink.''   In fact, even with dust-obscured bursts included, this is not how the GRB host population actually behaves: at least at lower redshifts (in particular the $0.5 < z < 1.5$ bin), the top end of the diagram remains underpopulated, with most events crowding toward the bottom and the left, indicating low masses and SFRs.  While the dust-obscured bursts are in more luminous and massive galaxies, these galaxies are not {\it sufficiently} different from the rest of the population to reverse this trend. Nor are they sufficiently common: events of the type analyzed here contribute only about 15\% of the population (Figure \ref{fig:avhist}), much less than the ratio shown in the figures above $z > 1$.

% S6.1.1
\subsubsection{As a Function of NIR Magnitude}

The above argument is not quantitative and, importantly, it ignores the incompleteness of the comparison survey. By definition, flux-limited galaxy field surveys only probe star formation above a certain limiting flux level, while GRB-targeted surveys probe all star formation, including that in galaxies below the sensitivity limit (which will show up as nondetections if observed to the same depth as the field survey).  This will produce a corresponding difference in the overall galaxy populations that will complicate detailed comparisons of the two samples if not corrected.

This problem can be avoided by simply mimicking the cut set by the survey depth threshold on the GRB host sample.  For the MOIRCS Deep Survey, the selection criterion was a magnitude cut at $K < 23.1$ Vega mag (the 5$\sigma$ threshold for their ``wide'' field).  This limit is significantly fainter than the flux of any of our dust-obscured hosts, suggesting that all of those galaxies would be recovered in a MODS-like field survey.  Many nondark GRBs, however, are in galaxies that are undetected in $K$, and these upper limits do not actually reach $K \approx 23$\,mag in any case (which is a very deep limit, difficult to achieve in practice).  However, we can calculate their expected $K$-band fluxes approximately via synthetic photometry of our best-fit model (in essence, extrapolating the flux measurements in the more-sensitive bluer filters).  Based on this procedure, we find that \emph{all except for two} of the GRB host galaxies in the (mostly lower-redshift) pre-\Swift\ comparison sample would also have been recovered in MODS, indicating that---as advertised---this very deep survey reaches flux levels that include the large majority of the star formation probed by GRBs at $z<1.5$.  Consequently, field-survey incompleteness has almost no effect on the comparison between our results and this survey\footnote{This is, in part, caused by the fact that the pre-\Swift\ comparison sample contains very few events at $z>2$.  A higher-redshift comparison sample would almost certainly contain a much larger number of galaxies below the MODS threshold---as evidenced by the work of \cite{Laskar+2011}, which we will consider in more detail in the next section.}.  Its effects can be eliminated completely by removing these two galaxies from consideration.

We illustrate this point explicitly in Figure \ref{fig:zk}.  The survey threshold is shown as a dashed horizontal line; nearly all GRB hosts we consider lie above this level.  This figure also includes a curve showing the median SFR-weighted galaxy magnitude of the comparison sample as a function of redshift (i.e., for a given redshift, we have calculated the magnitude at which 50\% of star formation probed by the survey\footnote{We emphasize that these median and quartile boundaries \emph{only} probe star formation in galaxies above the $K>23.1$ mag survey threshold.  Since we have discarded any GRBs occurring within fainter galaxies from consideration in this analysis, star formation in galaxies fainter than this limit does not concern us.} occurs in brighter galaxies and 50\% occurs in fainter galaxies).  Dotted lines show the first and third quartile boundaries (i.e., the 25th and 75th percentiles).  (To smooth out statistical variations and effects of cosmic variance in this small-angle survey, we have fit a polynomial over the full redshift range for each line.)  If GRB hosts are to represent uniform star-formation tracers, then the numbers of events should divide approximately evenly between the four lines, except below the bottom horizontal line at $K > 23$ mag---numbers in galaxies fainter than this limit cannot be predicted since the field survey does not probe beyond this depth.  (To emphasize this, we plot host galaxies fainter than this level with hollow points.)

The comparison is most easily made at $z \approx 1$, where there is a large pre-\Swift\ GRB comparison sample and where the incompleteness limit of MODS (and typical observed upper limits on faint GRB host galaxies) are far below the magnitudes of the galaxies at which most star formation is actually occurring.  We focus specifically on the redshift range $0.6 < z < 1.4$, which contains 31 events in total between our sample, the pre-\Swift\ sample, and TOUGH; 7 of these are ``dark,'' a fraction that is probably comparable to the actual contribution of obscured bursts to the overall population \citep{Cenko+2009,Perley+2009b,Greiner+2011}.  Of these events, only three host galaxies are brighter than the expected median line and \emph{none} are brighter than the 75th-percentile line!  All other GRB hosts (90\% of the sample) land below the expected median and a large majority (65\%) land in the bottom quartile.  With the exception of only three events (TOUGH nondetections not observed to deep limits), all of these hosts do lie above the magnitude limit of the MODS survey, however, meaning that their overabundance is \emph{not} simply due to GRBs probing star formation below the survey limit---GRBs genuinely seem to be overabundant in faint galaxies and underabundant in bright ones at this redshift.   A K-S test between the observed $K$-band magnitude distribution for detected galaxies at $0.6 < z < 1.4$ relative to the expected distribution for a SFR-weighted sample with the same redshifts (for $K<23$ mag galaxies) gives a probability of 0.003\% that the observed distribution is consistent with random chance.

This trend weakens significantly toward higher redshifts: at $z>1.5$ significant numbers of GRB hosts do populate the high-luminosity region of the plot and agreement with the assumption of an unbiased SFR-tracing population improves.  It is even possible that the ``bias'' disappears entirely---host galaxies detected in $K$-band in the uniform TOUGH survey at $1.5 < z < 2.5$ (the small circles in Figure \ref{fig:zk}, excluding upper limits) are distributed almost uniformly between the four SFR-weighted quartiles, seemingly quite consistent with an unbiased SFR-tracing population.  On the other hand, most of the nondetections in the TOUGH survey are 1--2 mag short of the MODS limit and a large population of faint (but MODS-detectable) hosts may have been missed by that effort.  In Figure \ref{fig:zh} we compare instead against a much deeper survey, the {\it HST} Snapshot project (Tibbetts-Harlow et al., in prep.), a randomly selected sample of the hosts of optically bright GRBs observed to very deep limits with WFC3-IR on {\it HST}.  While the number of events in the $z \approx 2$ redshift range is still relatively small, five out of seven hosts at $F160W > 23.5$ AB mag (roughly corresponding to the typical upper limit of $K > 21.5$ Vega mag in TOUGH) are in fact far below that threshold and would not have been recovered in MODS, suggesting that most of the TOUGH nondetections might also fall below the MODS survey limit and therefore not affect the comparison between the detected hosts and the MODS survey.  However, both comparison surveys are small, and it is quite possible that a larger sample would reveal a clearer trend toward fainter galaxies at $z \approx 2$ similar to (if weaker than) what is seen at $z \approx 1$.  (It is also worth noting that in terms of \emph{color}, even the most luminous hosts from the uniform TOUGH survey appear unusually blue relative to expectations for a $z \approx 2$ SFR-selected population; see Figure \ref{fig:zrmk}.)

We emphasize that this result is not dependent in any significant way on the details of our SED modeling or fitting procedures---it is a simple matter of number counts as a function of $K$ magnitude.  Our modeling is used only to extrapolate the SED for the faintest galaxies to confirm that they are not \emph{so} faint that they would not be recovered in deep field surveys.  Similarly, it is not particularly sensitive to the use of the pre-\Swift\ host comparison sample---it is noteworthy that even the population of dust-obscured GRB hosts alone, despite being \emph{relatively} massive compared to unobscured GRBs, does on its own still trend toward subluminous galaxies at $z \approx 1$.  

Our conclusion \emph{is} of course dependent to some degree on the modeling and cosmic volume of the MODS survey itself.  However, the strength of the discrepancy that we observe would require very large errors to explain.   In particular, the actual rate of star formation in faint (sub-25th percentile) $z \approx 1$ galaxies would need to be a factor of 2--4 higher than claimed relative to the SFR density in bright galaxies.  Dust extinction in faint galaxies is \emph{not} a viable explanation for why this hypothetical extra SFR was missed from MODS, since if this were the case the number of dust-obscured GRBs would be much larger than the $\sim 20$\% value that is currently reported.  Cosmic variance, in turn, is unlikely to significantly affect even a very narrow-angle slice covering such a wide redshift range ($\sim 2$ comoving Gpc between $z \approx 0.5$ and $z \approx 1.5$).  

Finally, we consider the possibility that our GRB sample is somehow still nonrepresentative of GRB host galaxies in a significant way.  Clearly our ``comparison'' sample is quite inadequate at $z \approx 2$ and beyond, but questions could also be raised about the biases involved in the reporting of the pre-\Swift\ $z \approx 1$ comparison sample.  Nevertheless, it is extremely difficult to imagine a situation in which our $z \approx 1$ sample would be biased to such an extent to reproduce the trends observed.  For example, we have illustrated as an example that out of 31 known GRBs we have considered in the range $0.6 < z < 1.4$, we identify \emph{no} events in hosts brighter than the 75th SFR percentile curve.  Hosts of that brightness ($K \approx 19$\,mag at $z=1$) are easy to find, even with relatively small (1--2\,m) telescopes.  Yet not a single host with $K<19$ mag and $z>0.9$ has been reported in the entire pre-\Swift\ era or within the unbiased TOUGH survey (which includes 10 events in the $0.6 < z < 1.4$ redshift range, none of which are brighter even than the expected 50th percentile in $K$ magnitude)---indeed, to our knowledge, no such hosts have been reported in the entire \Swift\ era.  Even if there are somewhat more ``dark'' GRB hosts than we realize at $z \approx 1$, even these events are just as likely to occupy faint galaxies at this redshift than bright ones, and so would only slightly improve the consistency between expectations and observation.

The conclusion that GRBs do not trace cosmic star formation (at least at $z<1.5$) therefore appears very difficult to avoid.

% S6.1.2
\subsubsection{As a Function of Stellar Mass}
\label{sec:massdiff}

While the analysis above in terms of observed magnitude provides strong evidence of a difference between the GRB rate and the SFR, it offers no direct insight into its physical origin.  However, the same argument presented above in terms of the observed $K$ magnitude can be recast in terms of physical properties quite easily.  While addressing the relative completeness of the populations is less straightforward, the fact that over 95\% of our host population would in fact be recovered to the MODS limit implies that this technique is still powerful when employed on a range of variables.  To avoid the risk of incompleteness, we simply discard the two events which would have fallen below the MODS threshold (they are plotted as hollow points).

The solid and dashed curves in Figure \ref{fig:zmass} correspond to the same concepts used in the magnitude-based analysis, indicating 25th, 50th, and 75th percentile of the SFR density as a function of mass at each redshift.  Unsurprisingly (since the NIR luminosity is a close tracer of stellar mass), this diagram shows a result very similar to that of the $K$-band analysis, with most GRB hosts (including many dust-obscured GRBs) falling below the 25th percentile line at $z \approx 1$.

It is worth noting that while the conclusion of \cite{Kajisawa+2011} that the cosmic SFR is dominated by moderate- to high-mass galaxies ($\gtrsim 10^{10}$\,M$_\odot$) at $z \approx 1$ has been supported by other recent work \citep{Zheng+2007,Santini+2009}, some other studies (in particular, optically selected samples) do suggest a larger role for star formation at lower masses: for example, \cite{Juneau+2005} and \cite{Mobasher+2009} suggest that the median mass is somewhere near or below $10^{10}$\,M$_\odot$ at $z \approx 1$--2.  However, the much shallower depths of these surveys, $K < 20.6$\,mag (Vega) in the case of \citet{Juneau+2005} and $i < 25$\,mag (SDSS/AB) for \citet{Mobasher+2009}, causes them to become incomplete at around $10^{10}$\,M$_\odot$, making it difficult to actually evaluate the amount of star formation occurring below these levels where most of our sample lurks.  Purely optically derived SFRs also miss contributions from very dusty galaxies which, while not necessarily dominant, are at minimum an important contributor to the SFR density at higher redshifts.

Mass, therefore, appears to be a primary driver of the deviation between the expected and observed characteristics of the GRB host population.  Assuming the balance of ``unobscured'' and ``dark'' GRBs in our combined sample at $z \approx 1$ is close to the intrinsic one, the difference in GRB rate per unit star formation varies by a factor of at least 5 and possibly by 10 or more between $10^{11}$\,M$_\odot$ galaxies and $10^{9}$\,M$_\odot$ galaxies.  This observation is easily interpreted under the current (if still-controversial) paradigm that the GRB progenitor is strongly metal-averse, since the mass and average metallicity of a galaxy are strongly correlated.  Nevertheless, mass correlates with a number of other observables as well, and so it is useful to check to determine what other bulk properties may contribute to the observations.

\subsubsection{As a Function of Luminosity and SFR}

It is natural to ask how the \emph{cosmic} SFR density is divided among the \emph{individual} SFRs of the galaxies that contribute to it, including the relative balance of the large numbers of individually insignificant low-SFR galaxies versus spectacular but rare systems such as submillimeter galaxies, ULIRGs, and the most luminous LBGs.  The actual answer to this question has remained controversial over the years and (often) highly dependent on the wavelength at which the study was conducted, although the most recent studies have been converging on a contribution of $\sim 20$\% from highly luminous and obscured super-star-forming galaxies (ULIRGs and SMGs), with the remaining fraction about evenly distributed between moderately luminous galaxies (LIRGs and luminous LBGs) and lower-luminosity systems \citep{Reddy+2008,Reddy+2009,Rodighiero+2010,Murphy+2011,Magnelli+2011}.  The importance of luminous galaxies peaks at $z=2$--3 and falls rapidly toward lower and (probably) higher redshifts \citep{Bouwens+2009}.

We are currently restricted to the dust-corrected UV as a probe of the SFR, which is insensitive to star formation in high-optical-depth regions.  Long-wavelength observations are being pursued and will be the basis of future work, but even without those constraints the problem can be framed in terms amenable to the data at hand using the same basic technique as in previous sections.  In Figure \ref{fig:zsfr} we calculate the median and 25th/75th percentile boundaries for all star formation in $K < 23$\,mag galaxies as a function of the UV-inferred, dust-corrected SFR, which corresponds to the same measure probed by our sample.  Despite reports of a GRB sample deficient in luminous galaxies and LIRGs \citep{LeFloch+2006}, we actually observe reasonably good agreement in this property with a significant number of events occupying the most luminous quartile at all redshifts, and even without direct long-wavelength constraints we predict (from the UV luminosities and mean dust attenuations) that most of our dust-obscured events are LIRGs and a few are (borderline) ULIRGs.  At $z \approx 1$ where we previously observed a clear trend among GRB hosts toward low-mass galaxies, we infer only a slight preference toward lower-luminosity galaxies, and this deviation is consistent with statistical variations.  (Again, at $z \gtrsim 1.5$ the nondark comparison sample is inadequate at present to claim that our overall sample is representative of all GRBs.)

Given that samples of high-$z$ galaxies show a correlation between mass and SFR \citepeg{Daddi+2007,Elbaz+2007}, the fact that the trend observed in stellar mass does not carry over to SFR is somewhat surprising.  Even if SFR were physically unconnected to the GRB rate, we would somewhat expect the tendency of GRBs to favor low-mass galaxies to also cause them to prefer low-SFR galaxies.  It is possible that a larger sample would recover this trend.  Alternatively, it could be actively counteracted by a third parameter---in particular, a tendency for GRBs to prefer high-sSFR galaxies, a possibility we explore in the next section.

\subsubsection{As a Function of Specific Star-Formation Rate}

As we noted in \S \ref{sec:ssfr}, a remarkable number of our host galaxies show very high sSFRs, evidenced directly by young SEDs with little or no hint of a Balmer break, indicating average stellar ages younger than 100\,Myr or (equivalently) sSFRs of $>10$\,Gyr$^{-1}$.  About 20\% of our sample lies above this threshold.  This tendency for GRBs to occur in young, high-sSFR galaxies has been noticed previously among nondark GRBs \citep{Christensen+2004,Mannucci+2011,Kocevski+2011}; the inclusion of our sample of dark GRBs extends the trend up to higher redshifts (and stellar masses), as shown in Figure \ref{fig:masssfr}.

It is important to note that the association of GRBs with very young galaxies is far from exclusive---the remaining 80\% of the population we examined occurs in galaxies with quite ``normal'' population ages for a SFR-selected population.  Nevertheless, it is interesting that a very significant fraction of GRBs occurs in extremely young galaxies whose contribution to overall cosmic star formation is almost negligible: \cite{Kajisawa+2011} find only a few percent of all SFR density occurring in $>10$\,Gyr$^{-1}$ systems at $z<2$, most of which comes from  galaxies with $10^{10} < M < 10^{10.5}$\,M$_\odot$.  The fraction rises with redshift, but is still low even at $z \approx 2$ (10\%; \citealt{Rodighiero+2011}, although see \citealt{Yoshikawa+2010}). 

This is not just a reflection of the mass difference discussed earlier (\S \ref{sec:massdiff})---the mass-sSFR relation is nearly flat \citep{Elbaz+2007,Daddi+2007,Whitaker+2012}, and so (at a given redshift) the apparent skew toward lower mass should not produce any comparable differences in sSFR.  Rather, for a galaxy \emph{of a given mass} at a given redshift, the GRB rate in high-sSFR galaxies seems to be substantially elevated above what is expected given the SFR.  The origin of this trend is not clear; while such an effect is predicted by the three-parameter mass-sSFR-metallicity correlation of \cite{LaraLopez+2010} and \cite{Mannucci+2010,Mannucci+2011}, in practice the sSFR-metallicity portion of the correlation is fairly weak and unlikely to produce the degree of the correlation observed here \citep{Kocevski+2011} \emph{unless} the actual metallicity dependence of the GRB rate is quite strong (and possibly continuous across the entire mass range---note that the apparent preference for high-sSFR systems is present across the entire mass range of star-forming galaxies, as can be seen in Figure \ref{fig:masssfr}).  More work will be necessary to determine if this sort of metallicity effect would be sufficient, or if dependence on another parameter more directly correlated to specific star formation (such as the UV radiation field of the galaxy) would need to be introduced to explain the apparently elevated rate in these systems.

\subsection{Correlating Afterglow and Host Extinction---The Distribution of Host Dust}

% Figure 15
\begin{figure}
\centerline{
\includegraphics[scale=0.68,angle=0]{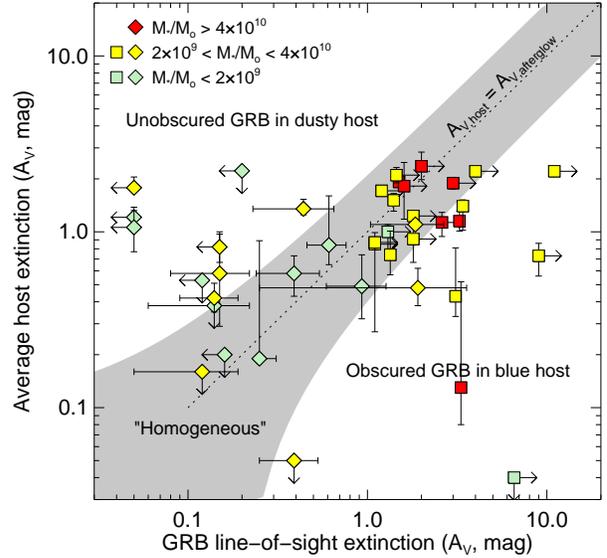}} 
\caption{Comparison of extinction measured along the afterglow sightline (from \citealt{Kann+2006} or this work) against the average extinction measured in the host from our SED-fitting procedure (which generally indicates the average attenuation toward young stars in the optically thin portions of the galaxy).  The dotted line indicates equality of the two values.  Most of the points are fairly consistent (within a factor of 2--3) with this line, indicating that the ISM in these systems is relatively homogeneous and that dustier galaxies do indeed tend to produce dustier bursts.  A minority of systems do show significant excursions, in all directions:  effectively unobscured GRBs in moderately dusty galaxies exist, as do highly obscured GRBs in blue, unobscured galaxies.  Such examples correspond to more heterogeneous galaxies where the sightline extinction to an individual star can deviate greatly from the average.  Nearly all of the most conspicuously heterogeneous galaxies are of intermediate mass.
}
\label{fig:extcompare}
\end{figure}

While GRBs may not be unbiased statistical tracers of star formation, they are still able to address questions regarding the detailed internal characteristics of high-$z$ host galaxies that cannot be answered by more traditional field-survey techniques at all.

In Figure \ref{fig:extcompare} we plot the extinction as measured from the \emph{afterglow} (determined either by our own preliminary analysis in \S \ref{sec:sample} for dark-burst hosts or by the study of \citealt{Kann+2006} for pre-\Swift\ hosts) against the inferred average $A_V$ of the galaxy's starlight (in optically thin regions) from our host SED fitting.    If every star in the galaxy were obscured by the same amount and type of extinction (a homogeneous screen), the points would all fall along the diagonal line plotted.   Excursions from this line (in either direction) indicate heterogeneity in the dust distribution with respect to the stars, while a preferential excursion would indicate a tendency for GRBs to inhabit either the dustier or less dusty parts of the host (relative to the typical optically thin region).

Most points in the diagram are in fact fairly close to the diagonal line (within a factor of $\sim 2$), indicating that in \emph{most} galaxies the extinction is consistent with a fairly homogeneous distribution that affects most stars in a similar way.  This result confirms that the simplified picture used to treat extinction in populations of high-redshift galaxies (namely, the treatment of extinction and reddening of galaxies as a simple screen with a roughly universal attenuation curve) is at least reasonable in many objects, and suggests that, when the amount of obscuration seen along a GRB sightline is low to moderate (0.2--2\,mag), this dust is probably dominated by the diffuse ISM of the galaxy (since most other stars are attenuated by a similar magnitude).   Given previous claims (by our group and others; e.g., \citealt{Perley+2009b}) that extinction may merely be a local effect, this correlation is remarkable.

On the other hand, this result is far from universal---a large number of events deviate far from the basic expectation of a linear relation in Figure \ref{fig:extcompare}.    In particular, the GRB sightline extinction can be much, much larger than the extinction obscuring a ``typical'' star in the optically thin portion of the galaxy---indicating that the GRB lies inside or behind a region of optically thick dust that is \emph{not} representative of the diffuse ISM filling the rest of the galaxy.  Typically, galaxies hosting these extremely obscured GRBs are already moderately dusty overall (the GRB sightline is just dustier than average), but in two cases there is clear evidence of a very heavily obscured GRB occurring in a galaxy with minimal or no obscuration.  Such events are rare and appear to be outliers from the rest of the population, but there is little question of their existence. (One particular event, GRB 061222A, is practically a prototype of the heavily obscured \Swift\ GRB: NIR-only detections, very red $H-K$ color, large $N_{\rm H}$, unambiguous galaxy association, host spectroscopic redshift.)

There are several possible interpretations of this observation, as follows.

\emph{(1) Extremely heterogeneous ISM}.  One possibility is that these (relatively uncommon) associations of obscured bursts in blue galaxies indicate random alignments between the GRB progenitor and an unassociated, dense molecular cloud in an otherwise mostly transparent galaxy.  Something like this probably happened with GRB 080607---spectra of the optical afterglow of this heavily obscured GRB indicate that the burst sightline penetrated a discrete molecular cloud along its sightline at a distance of a few hundred pc from the burst itself (as evidenced from the excitation of fine-structure lines and nondestruction of molecular lines in the spectrum; \citealt{Prochaska+2009,Sheffer+2009}).  The discrepancy between host and afterglow $A_V$ for this GRB (1.1\,mag vs. 3.3\,mag, respectively) is not as large as for some other examples, but could indicate that this scenario is widespread.

\emph{(2) Local dust}. The most recently-formed stars in a galaxy tend to be obscured significantly more heavily than slightly older stars \citep{CharlotFall2000}, since newly formed stars often remain embedded in the (dusty, optically thick) molecular cloud that formed them for a few Myr after formation, whereas the older stars that dominate the continuum UV flux have time to disperse these clouds entirely.  A sufficiently massive, short-lived progenitor star may explode on a short enough timescale that the surrounding gas and dust are not yet cleared.  Since only a few GRBs are heavily obscured, this effect cannot be very significant in most cases (either the optically thick covering fraction is quite small or the GRB event is able to disperse the cloud on its own via its X-ray and/or UV emission; \citealt{Waxman+2000,Fruchter+2001,Perna+2003}), but under the right circumstances such local dust could produce a large obscuring column. 

\emph{(3) Nuclear super-starbursts}.  The extinction inferred from the UV continuum only probes optically thin regions of the galaxy.  In fact, if a small, extremely heavily obscured portion of the galaxy were the site of most of its star formation (as in local ULIRGs such as Arp 220 and in submillimeter galaxies), then the $A_V$ and SFR determined from the optically thin regions would be entirely inappropriate for the galaxy overall.  Deep submillimeter or radio observations would be necessary to check this hypothesis; observations so far \citep{Perley+2013b} suggest that this applies only to a few cases.

Excursion in the other direction is also present---while the typical optically bright GRB has a blue, low-$A_V$ host, the nondark host population does contain a few moderately obscured galaxies as well.  This could simply be a reflection of geometry (since stars are embedded inside the diffuse interstellar dust and not actually behind a single screen, some should always be much less obscured and others somewhat more obscured), though if GRBs preferred the outer, less-obscured regions of their hosts (e.g., due to an aversion of the more metal-rich center), such a trend would be greatly amplified.  The submillimeter hosts of optically bright GRBs mentioned previously---in particular GRBs 000418 and 010222 \citep{Berger+2003,Tanvir+2004,Michalowski+2008}---probably represent much more extreme and unambiguous examples of this phenomenon. 

The situation is complex, and all of these effects may operate, affecting some bursts and not others.  More observations---in particular high-resolution observations of both dark GRBs and their host galaxies, and afterglow spectroscopy of fortuitous events like GRB 080607 luminous enough to be observable optically through their dust screens---will be necessary to say more about this issue.  Nevertheless, although a high-extinction afterglow sightline does not guarantee a dusty host and vice versa, the fact that the two properties are correlated does imply that the origin of most of the dust that is observed for at least moderately obscured GRBs (if not necessarily extremely obscured GRBs) is ordinary dust in the ISM; the darkness of a GRB is not purely geometric.

\subsection{Implications of the Mass-Obscuration Correlation}

Stellar mass is well known to correlate with the average attenuation of a star-forming galaxy within field-selected samples \citepeg{Brinchmann+2004,Pannella+2009,Garn+2010}, although there is significant scatter in the relation and these variables also correlate with many other properties, such as the SFR and metallicity.

Similar trends are, unsurprisingly, observed in our sample.  More notable, however, is that the correlation between mass and obscuration remains extremely strong when comparing host mass to \emph{afterglow} obscuration, in spite of the large number of outliers in the host vs. afterglow comparison of Figure \ref{fig:extcompare}.  To emphasize this, we have color-coded the points based on the mass of the host galaxy, with the most massive hosts ($> 4 \times 10^{10}$\,M$_\odot$) in red, the least massive ($< 2 \times 10^9$\,M$_\odot$) in green, and intermediate-mass objects in yellow.  The mass of the host galaxy is clearly a strong predictor of both measures of obscuration, but the strong separation between the sightline-extinction observed in the lowest-mass galaxies and the highest-mass galaxies is particularly noteworthy.  While some caution should be used in this comparison given that the effects of redshift evolution are not considered, we find that the least-massive galaxies host almost universally unobscured GRBs while the most massive galaxies host almost universally obscured GRBs.  Intermediate-mass galaxies, in contrast, show much greater diversity and represent all of the strong outliers on this plot.  No other property we have examined (SFR, sSFR, and the $A_V$ of the galaxy itself) shows such an obvious correlation at the extreme ends.

Why would mass be the primary driver of afterglow obscuration, more important than even the mean extinction in the galaxy itself?  Presumably, mass is correlated not just with the amount of dust but also with the degree of homogeneity or heterogeneity of that dust.  Low-mass galaxies must be low in dust throughout---there are very few sightlines through such objects that produce appreciable extinction, and so only the occasional rare alignment piercing an unassociated molecular cloud produces substantial absorption.  On the other end of the scale, high-mass ($M>2 \times 10^{10}$\,M$_\odot$) star-forming galaxies offer very few escape routes for a GRB beam to exit the host without encountering at least some significant amount of dust, suggesting that a substantial amount of obscuring dust is distributed relatively homogeneously within the galaxy (as also evidenced by its heavily reddened starlight).  Of course, such galaxies may also have their own localized clouds (or large nuclear starbursts) that may increase the attenuation of certain sightlines still further.  

This class of intermediate galaxies represents the bulk of cosmic SFR (though not, as we have discussed, the bulk of GRBs except at $z>1.5$) and includes a wide diversity of both host colors and afterglow colors, indicating a group of galaxies that is heterogeneous as a population and (often) also internally heterogeneous in terms of the distribution of dust and stars within its ISM.  On the other hand, all of the most massive star-forming galaxies in this sample obscure their entire stellar population (including the individual star that produced the GRB), while the least massive galaxies rarely obscure any of their stars.

% S7
\section{Conclusions}
\label{sec:conclusions}

We have compiled by far the largest sample of dust-obscured GRB host-galaxy results to date, targeting a total of 23 GRBs with afterglow-inferred host extinction columns of $A_V > 1$\,mag (corresponding to the dustiest $\sim 25$\% of all GRBs) with intensive follow-up observations at optical--NIR wavelengths both on the ground and in space.  This sample is compared against a compilation of optically selected, well-studied pre-\Swift\ GRB hosts as well as a recent deep, narrow-field galaxy survey.

We find large, unambiguous differences between the populations.  Dust-obscured GRB hosts are more massive than the hosts of optically bright GRBs within the same redshift range by about an order of magnitude, and show similar tendencies toward higher luminosities and SFRs, redder colors and higher dust extinctions, and older stellar populations on average.  

Despite the revision upward in the average mass and luminosity of the GRB host population that these results imply, dark GRB hosts are neither sufficiently frequent nor massive enough to bring the overall population into line with the expectations of a purely SFR-tracing population at $z \approx 1$.  The average mass of a GRB host is about one order of magnitude lower at this redshift than would be predicted from the population of galaxies inferred from deep multiwavelength field surveys.  Consistency improves significantly at $z > 1.5$; at these earlier epochs massive GRB hosts become fairly common, approaching the frequency that would be expected for an SFR-tracing population at these redshifts.

Among the parameters we analyzed, the primary factor affecting a galaxy's GRB rate per unit star formation is mass---GRBs avoid high-mass galaxies and prefer very low-mass ones.  Without additional data we can only speculate about the cause of these biases at present, although because mass and metallicity are well correlated \citep{Tremonti+2004,Erb+2006} it seems consistent with the general idea that metallicity is a fundamental driver of the GRB rate.   If so, metallicity is likely to act as a gradual modifier of the GRB rate, rather than impose a strict cutoff at a specific value---for example, we can definitely rule out the model of \cite{Kocevski+2009}, who predict that this metallicity effect should impose an equivalent mass cutoff of $10^{9}$\,M$_{\odot}$ at $z=1$.  In reality, of course, a metallicity cut would be somewhat softened by dispersion in the mass-metallicity relation and by chemical inhomogeneity within galaxies themselves \citep{Niino+2011,Campisi+2009}.  Still, the apparent amplification in the GRB-to-SFR rate in very low-mass galaxies relative to moderate-mass galaxies is more suggestive of a gradual trend, and the direct detection of a handful of GRBs in high-metallicity environments \citep{Prochaska+2009,Levesque+2010f,Levesque+2010g,Kruehler+2012a,Savaglio+2012} is difficult to explain with a simple cutoff.

The apparently improved consistency of the GRB and SFR toward $z \approx 2$ could reflect an overall decrease of metallicity in luminous galaxies compared to similar-mass objects at lower redshifts \citep{Erb+2006,Reddy+2010}.  The speed of this transition is notable---GRBs in the quartile of star formation representing the most massive galaxies are common at $z=1.5$--2.0 but almost entirely absent at $z<1.5$.  This rapid trend hints at a sharp change in the properties within the Universe's most massive galaxies at this redshift.  This could be interpreted as a consequence of the gradual buildup of metals in the most massive galaxies pushing the metallicity of these systems above a critical value beyond which GRB production is greatly suppressed.  Alternatively, perhaps this epoch in cosmic history corresponds to the end of the accretion of streams of metal-poor gas onto massive galaxies (permitting chemical homogenization and a sharp drop in the amount of metal-poor star formation).  

Independent of mass, GRBs may exhibit a secondary preference for systems with high sSFRs \emph{beyond that} predicted from their higher overall SFRs.  This could also be an imprint of metallicity:  high-sSFR galaxies have lower metallicity at a given mass \citep{Mannucci+2011}, although the surprising strength of this effect in the highest-sSFR system requires (at minimum) a very strong metallicity dependence to explain, and may also require consideration of additional factors affecting the GRB rate other than metallicity.

As an alternative to metallicity, large variations at the top end of the IMF in different galaxies could also produce significant variation in the GRB rate by affecting the relative numbers of extremely massive stars that (presumably) produce GRBs relative to the smaller (but still massive) stars which dominate the long-term UV luminosity.  However, if indeed the ratio of GRB rate to SFR varies across the entire mass range (and given the need to produce an order-of-magnitude change in the rate of production of the most massive stars), this effect would have to be so pervasive---or limited only to the most massive stars---that it is difficult to imagine such an effect would escape notice by other means.  A more subtle effect to recognize would be a change in the \emph{binarity} properties of the initial stellar population, an effect which would be observationally difficult to recognize directly but which could affect the production of GRBs dramatically if the progenitor is a binary system.  Nevertheless, the simplest interpretation of our results at present is a direct dependence of the formation rate of the progenitor on metallicity.

Our conclusion---a GRB rate (relative to that of star formation) that is a strong function of host-galaxy environment---differs from that of some other recent work reexamining the GRB rate in the context of the mass-sSFR-metallicity relation \citep{Mannucci+2011}.  We note that these authors considered \emph{only} the explanation of the apparent observed metallicity bias itself and did not address in detail the observed mass distribution of GRB host galaxies beyond comparison of a single mass bin to a single study.  While it is natural to expect an SFR-selected population to skew toward higher (s)SFRs (and therefore lower metallicities) than a galaxy-count-selected population, the magnitude of the effect should be relatively small \citep{Kocevski+2011} and has difficulty explaining the frequency of GRBs in extremely low-mass environments (and the low rate in $L^*$ galaxies.)  In addition, \cite{Graham+2013} have recently considered predictions for the mass and metallicity distributions of GRBs for SFR-weighted galaxy populations and SN hosts, and they continue to find large inconsistencies between the expected and observed GRB rates in the same $z<1$ population unless metallicity dependence (or analogous effects) are considered.

It is also important to note that our result does not directly contradict previous work showing general consistency between the GRB rate and SFR in high-redshift host populations (e.g., \citealt{Jakobsson+2005b,Fynbo+2008,Chen+2009,Savaglio+2009}).  These studies considered only (or primarily) optically bright bursts and only UV-based estimates of the SFR density.  The hosts in which GRBs are most dramatically underabundant tend to be heavily dust obscured, so consideration of the obscured GRB population (and dust-unbiased estimates of galaxy SFR) is essential to see these effects clearly.  In addition, since the effect is both intrinsically smaller and observationally more difficult to recognize at $z > 1.5$ relative to at $z \approx 1$ (and may even disappear completely at higher redshifts), it is no surprise that work focused on higher-redshift populations similarly did not show the same trends.

The clear dependence of the GRB rate on host-galaxy properties out to at least $z \approx 1$ provides a strong cautionary note about the use of GRBs as direct tracers of the cosmic SFR density before the intrinsic reason for this variation is better understood.  While it is encouraging that the magnitude of the trend for GRBs to prefer low-mass galaxies decreases at $z \approx 2$ (suggesting that GRBs become better tracers of star formation as the average cosmic metallicity drops), we cannot yet confirm that it becomes insignificant at any redshift.  One of the most exciting broader applications of GRB studies is to probe the contribution to the star-formation rate from low- to moderate-luminosity star-forming galaxies at $z>5$ \citep{Tanvir+2012}, a population that no other observational techniques can currently reach.  However, such work needs to be supported by better multiwavelength studies of GRB hosts at $z=2$--3 to more conclusively evaluate whether the tools we wish to apply to these earliest epochs of cosmic history work in the way that we expect at more familiar redshifts.

Even if GRBs do not represent ideal \emph{population} probes of the star-formation properties of the Universe, they certainly remain useful as \emph{individual} probes of specific galaxies and their internal structure.  The clear correlation between afterglow-sightline and host-averaged dust attenuation indicates that the dust within high-$z$ galaxies is usually distributed fairly homogeneously, suggesting that the adopted empirical reddening corrections are generally reasonable.  But there are several exceptions (of both obscured GRBs in blue hosts and unreddened GRBs in obscured hosts), indicating that dust in dense heterogeneous clouds is also present.  This heterogeneity is most pronounced in intermediate-mass galaxies, whereas low-mass galaxies are almost always homogeneously dust free and high-mass galaxies are ubiquitously dusty (although some sightlines are surely much dustier than others).  

On a similar note, we have now determined unambiguously that GRBs do routinely form in ULIRGs and related classes of massive, heavily obscured star-forming galaxies, environments that previously had been found to host only a few isolated events.  While this study shows that GRBs in these environments are rarer than might be expected (given their contribution to cosmic SFR) and their afterglows are almost always heavily obscured, for the most luminous bursts in the Universe this is not always a limitation, especially if large-telescope spectrographs can observe the event rapidly.  This has in fact occurred for two events already (GRBs 080607 and 070802), and while such events are rare, future rapid follow-up observations of GRB afterglows should identify more such targets, giving us perhaps the only means of probing deep inside high-redshift LIRGs and ULIRGs at this level of detail.

In addition, the tendency of GRBs to prefer faint galaxies actually has some advantages, since this is exactly the population that is difficult to understand from field-survey techniques.  For example, perhaps one of the most important results from our study is the fact that we did not find any dust-obscured GRBs within very faint or low-mass galaxies.  A long-standing concern in current estimates of the cosmic SFR density is that the presence of additional, unaccounted, heavily obscured star formation within very faint galaxies cannot be directly observationally ruled out.  If low-mass galaxies at $z \approx 2$ were somehow able to conceal a majority of their star formation within optically thick regions, this would have profound implications for the sites of cosmic star formation.  However, if this sort of star formation were cosmologically relevant, we would observe significant numbers of highly obscured GRBs from low-mass galaxies.  We do not see such a population; GRBs in low-mass, NIR-faint galaxies are unobscured with only occasional exceptions.  

In addition, if the reason for the GRB rate dependence can be understood (for example, if it can be firmly pinned on metallicity), then the variations we see can themselves be used as tracers of cosmic evolution.  For example, the apparent commonality of GRBs in massive galaxies at $z \approx 2$ compared to their relative rarity at $z \approx 1$ probably says something about the evolution of the characteristics of systems with redshift: massive $z \approx 1$ starbursts seem to have a higher metallicity (or other systematically different feature relevant to the GRB production rate) than their counterparts at higher redshifts.

In short, while the apparent noncorrespondence of the GRB host population with the simplest expectations is disappointing from the perspective of using them as easy probes of evolution of the SFR density, further study of these intriguing objects nevertheless may continue to provide important clues about the nature of the high-redshift universe.

\vskip 1cm

\acknowledgments

We thank D.~Malesani for early access to the VLT photometry and redshifts, and for additional assistance with the TOUGH data.
We acknowledge useful conversations with R.~Chary, and thank M.~Michalowski, D.~A.~Kann, and the referee for helpful commentary on the manuscript.
We are grateful to E.~Petigura for assistance with the Keck-NIRC observations and
S.~Kulkarni for some additional LRIS observations.  We also thank M.~Kajisawa for assistance with providing and interpreting the Subaru MODS observations.

Support for this work was provided by NASA through Hubble Fellowship grant HST-HF-51296.01-A awarded by the Space Telescope Science Institute (STScI), which is operated by the Association of Universities  for Research in Astronomy (AURA), Inc., for NASA, under contract NAS 5-26555.
The Dark Cosmology Centre is supported by the Danish National Science Foundation. 
J.P.U.F. and B.M.J. acknowledge support from ERC-StG grant EGGS-278202.
T.K. acknowledges support by the European Commission under the Marie  Curie Intra-European Fellowship Programme in FP7.
A.V.F. and his group acknowledge generous financial assistance from Gary \& Cynthia Bengier, the Richard \& Rhoda Goldman Fund, the Christopher R. Redlich Fund, NASA/\Swift\ grants NNX10AI21G and NNX12AD73G, the TABASGO Foundation, and NSF grant AST-1211916.
J.X.P. acknowledges support from NASA/\Swift\ grants NNX07AE94G and NNX12AD74G.

This work is based in part on observations made with the NASA/ESA {\it Hubble Space Telescope}, obtained from the Space Telescope Science Institute.  STScI is operated by the Association of Universities for Research in Astronomy, Inc. under NASA contract NAS 5-26555.  These observations are associated with programs GO-10908, 11343, 11840, 12307, 12378, and 12764, and 12949.
Support for {\it HST} programs GO-11840, GO-12378, and GO-12674 was  provided by NASA through a grant from STScI, which is operated by AURA, Inc., under NASA contract NAS 5-26555.
The W. M. Keck Observatory is operated as a scientific partnership among the California Institute of Technology, the University of California, and NASA; the Observatory was made possible by the generous financial support of the W. M. Keck Foundation.  We wish to extend special thanks to those of Hawaiian ancestry on whose sacred mountain we are privileged to be guests.  
This work is based in part on observations made with the  {\it Spitzer Space Telescope}, which is operated by the Jet  Propulsion Laboratory, California Institute of Technology, under a contract with NASA.  Partial support for this work was provided by NASA through an award issued by JPL/Caltech.
It is also based in part on observations obtained at the Gemini Observatory, which is operated by AURA, Inc., under a cooperative agreement with the NSF on behalf of the Gemini partnership: the NSF (United States), the National Research Council (Canada), CONICYT (Chile), the Australian Research Council (Australia), Minist\'{e}rio da Ci\^{e}ncia, Tecnologia e Inova\c{c}\~{a}o (Brazil), and Ministerio de Ciencia, Tecnolog\'{i}a e Innovaci\'{o}n Productiva (Argentina). Observations were acquired under Program IDs GN-2006A-Q-14, GN-2006B-Q-21, GN-2007A-Q-19, GS-2008A-Q-20, GN-2007B-Q-99, GN-2008B-Q-6, GN-2009A-Q-26, GN-2009A-Q-84, GN-2010A-C-8, and GN-2010B-C-2.
PAIRITEL is operated by the Smithsonian Astrophysical Observatory (SAO) and was made possible by a grant from the Harvard University Milton Fund, a camera loan from the University of Virginia, and continued support of the SAO and UC Berkeley. The PAIRITEL project is further supported by NASA/\Swift\ Guest Investigator grant NNX08AN84G.  

This work made use of data supplied by the UK \Swift\ Science Data Centre at the University of Leicester.  
This research also made use of the NASA/IPAC Extragalactic Database (NED), which is operated by the Jet Propulsion Laboratory, California Institute of Technology, under contract with NASA.
It is a pleasure to thank all members of the \Swift\ team, who built and continue to operate this highly
successful mission. 

\bigskip

{\it Facilities:} \facility{Swift}, \facility{Keck:I (LRIS,NIRC)},
\facility{HST (WFC3)}, \facility{Gemini:North (GMOS,NIRI)}, 
\facility{Spitzer (IRAC)}

% Figure sizes for one-per-page:
% 1,2: 0.85
% 3,4,5: 0.3
% 6-12: 0.85
% 13: 0.5

\begin{deluxetable*}{llrrl ll ll ll}  %*
%\rotate
\tabletypesize{\footnotesize}
\tablecaption{GRB Afterglow Properties}
\tablecolumns{11}
\tablehead{
\colhead{GRB} &
\colhead{$z$\tablenotemark{a}} &
\colhead{\scriptsize $T_{\rm 90}$} &
\colhead{\scriptsize $S_\gamma$\tablenotemark{b}} &
\colhead{\scriptsize $N_{\rm H}$ excess\tablenotemark{c}} &
\multicolumn{2}{c}{{$\beta_X$}\tablenotemark{d}} &  % making this a colhead creates errors
%\colhead{a} & 
%\colhead{b} & 
\colhead{{$\beta_{\rm OX}$ limit\tablenotemark{e}}} &
\colhead{{$A_{V}$ limit\tablenotemark{f}}} &
\colhead{$A_{V,{\rm fit}}$\tablenotemark{g}} %&
%\colhead{References}
\\
\colhead{} &
\colhead{} &
\colhead{\scriptsize s} & 
\colhead{\tiny $10^{-7}$\,erg\,cm$^{-2}$} & %15-150
\colhead{\scriptsize $10^{20}$\,cm$^{-2}$} &
\colhead{\scriptsize measured} &
\colhead{\scriptsize used} &
\colhead{\scriptsize } &
\colhead{\scriptsize mag} &
\colhead{\scriptsize mag} %&
%\colhead{}
}
\startdata
 050915A & 2.527 &  25 &   9 &$ 8.8^{ +7.7}_{ -6.8}$&$1.02^{+0.28}_{-0.25}$& 1    & 0.23  &  1.03 &  1.4  \\%JHK  &   \\
 051008  & 2.9   &  16 &  51 &$30.0^{ +5.7}_{ -5.1}$&$0.86^{+0.14}_{-0.13}$& 1    &-0.17  &  1.17 &  --   \\%--   &   \\
 051022  & 0.809 & 200 &1400 &$83.9^{ +8.5}_{ -7.8}$&$1.07^{+0.12}_{-0.11}$& 1    &-0.12  &  9.34 &  --   \\%--   &   \\
 060202  & 0.785 & 204 &  21 &$45.2^{ +6.6}_{ -6.1}$&$1.75^{+0.20}_{-0.18}$& 1.57 & 0.20  &  2.87 &  --   \\%K    &   \\
 060306  & 1.551 &  61 &  21 &$29.8^{ +8.3}_{ -7.3}$&$0.96^{+0.19}_{-0.17}$& 1    & 0.21  &  4.60 &  --   \\%--   &   \\ 
 060319  & 1.172 &  12 &   3 &$34.0^{ +8.3}_{ -7.1}$&$1.06^{+0.21}_{-0.18}$& 1    & 0.19  &  1.05 & $>$1.8 \\%& K  &   \\ 
 060719  & 1.532 &  55 &  15 &$33.7^{+12.5}_{-10.1}$&$1.40^{+0.34}_{-0.28}$& 1.12 &-0.27  &  2.14 &  3.1  \\%JHK  &   \\
 060814  & 1.923 & 146 & 150 &$27.0^{ +3.5}_{ -3.2}$&$1.14^{+0.10}_{-0.10}$& 1.04 &-0.21  &  1.60 & --    \\%     &   \\
 060923A & 2.6   &  52 &   9 &$ 6.9^{ +9.2}_{ -7.5}$&$0.83^{+0.28}_{-0.24}$& 1    &-0.09  &  2.26 & --    \\%--   &   \\
 061222A & 2.088 &  72 &  80 &$30.4^{ +2.7}_{ -2.6}$&$1.03^{+0.07}_{-0.06}$& 1    &-0.26  &  7.19 &  --   \\% K   &   \\
 070306  & 1.496 & 210 &  54 &$31.1^{ +5.8}_{ -5.2}$&$1.04^{+0.14}_{-0.13}$& 1    &-0.08  &  3.69 &  3.7  \\% HK  &   \\
 070521  & 1.7   &  28 &  80 &$40.4^{+10.9}_{ -9.6}$&$1.12^{+0.23}_{-0.21}$& 1    &-0.52  &  12.3 &  --   \\% --  &   \\
 070802  & 2.454 &  16 &   3 &$ 1.4^{+16.6}_{-15.4}$&$0.97^{+0.60}_{-0.56}$& 1    & 0.14  &  0.64 &  1.3  \\%grizJHK & \\
 071021  & 2.452 & 225 &  13 &$ 5.0^{ +8.1}_{ -7.1}$&$0.87^{+0.24}_{-0.22}$& 1    & 0.37  &  0.52 &  1.5  \\% z   &   \\
 080207  & 2.086 & 340 &  79 &$54.2^{+14.7}_{-13.1}$&$1.41^{+0.31}_{-0.27}$& 1.14 & 0.06  &  2.32 &  --   \\%--   &   \\   
 080325  & 1.78  & 128 &  49 &$ 1.0^{+17.7}_{-15.7}$&$0.43^{+0.56}_{-0.53}$& 1    & 0.18  &  2.67 &  --   \\%K    &   \\
 080607  & 3.038 &  79 & 240 &$14.2^{ +3.9}_{ -3.6}$&$1.16^{+0.13}_{-0.12}$& 1.04 & 0.04  &  1.05 &  3.3  \\%VRIYJHK  &  \\
 081109  & 0.979 & 190 &  36 &$32.1^{ +7.9}_{ -7.0}$&$1.32^{+0.24}_{-0.21}$& 1.11 & 0.30  &  0.55 &  3.4  \\%grizJHK  &  \\
 081221  & 2.26  &  34 & 181 &$39.3^{ +4.7}_{ -4.4}$&$1.67^{+0.15}_{-0.14}$& 1.53 & 0.00  &  1.35 &  --   \\%K    &   \\
 090404  & 3.0   &  84 &  30 &$50.6^{+10.3}_{ -8.8}$&$2.44^{+0.36}_{-0.31}$& 2.13 & 0.20  &  1.25 &  --   \\%--   &   \\
 090407  & 1.448 & 310 &  11 &$22.6^{ +4.4}_{ -4.1}$&$1.15^{+0.13}_{-0.12}$& 1.02 & 0.14  &  1.57 &  --   \\%--   &   \\
 090417B & 0.345 & 260 &  23 &$82.3^{+13.5}_{-12.2}$&$1.54^{+0.21}_{-0.20}$& 1.34 & 0.28  &  2.66 &  --   \\%--   &   \\
 090709A & 1.8   &  89 & 257 &$16.8^{ +2.7}_{ -2.5}$&$0.96^{+0.08}_{-0.08}$& 1    &-0.56  &  3.04 &  3.4  \\%rizJHK  &  \\
\enddata 
\tablenotetext{(a)}{Redshift of the GRB or host galaxy.  Host-galaxy photometric redshifts are given to one significant decimal place (see Table \ref{tab:hostmodelpar} for uncertainties).}
\tablenotetext{(b)}{Gamma-ray fluence in the \Swift\ BAT 15--150\,keV band, with the exception of GRB 051022 for which the HETE WXM/FREGATE 30--400\,keV band is used.}
\tablenotetext{(c)}{Excess X-ray absorption above the Galactic value, expressed as an equivalent neutral-hydrogen column at $z=0$.}
\tablenotetext{(d)}{X-ray spectral slope assumed in the afterglow extinction-fitting analysis.  The left column is the measured value with uncertainties (from the automated analysis of Butler et al.); the right column shows the ``minimum" value assumed in deriving limits on extinction.  It is assumed that $\beta_X \geq 1.0$.}
\tablenotetext{(e)}{\ Limiting value (upper limit) of the observed optical-to-X-ray index (corrected for Galactic extinction).  The optical or NIR point after $t>1000$\,s with the strongest constraint is chosen.}
\tablenotetext{(f)}{\ Limiting extinction ($A_V$ in the host frame) based on analysis of individual points relative to an extrapolation of the the X-ray flux assuming the spectral index indicated under the $\beta_X$ column.  The optical or NIR point after $t>1000$\,s with the strongest constraint is shown.}
\tablenotetext{(g)}{Measured or limiting extinction ($A_V$ in the host frame) based on a fit of at least two optical/NIR points in different filters.}
\label{tab:grbprop}
\end{deluxetable*}

\begin{deluxetable*}{lll lll llll}  %*
%\rotate
\tabletypesize{\footnotesize}
\tablecaption{Afterglow and Host Positions}
\tablecolumns{9}
\tablehead{
\colhead{} &
\multicolumn{4}{c}{Best Afterglow Position} &
\colhead{} &
\multicolumn{2}{c}{Host Position} &
\colhead{} &
\colhead{}
\\
\colhead{GRB} &
\colhead{RA} &
\colhead{Dec} &
\colhead{Unc.\tablenotemark{a}} &
\colhead{Source\tablenotemark{b}} &
\colhead{$E_{B-V}$\tablenotemark{c}} &
\colhead{RA} &
\colhead{Dec} &
\colhead{$r_{\rm ap, min}$\tablenotemark{d}} &
\colhead{$P_{\rm chance}$\tablenotemark{e}}
\\
\colhead{} &
\multicolumn{2}{c}{(J2000)} &
\colhead{\arcsec} &
\colhead{} &
\colhead{mag} &
\multicolumn{2}{c}{(J2000)} &
\colhead{\arcsec} &
\colhead{}
}
\startdata
050915A &  05:26:44.804 &$-$28:00:59.27 &  0.18 & PAIRITEL & 0.025 & 05:26:44.84 &$-$28:00:59.94 & 0.9 & 0.010  \\
051008  &  13:31:29.550 & +42:05:53.30 &  1.2  & XRT      & 0.012 & 13:31:29.49 & +42:05:53.40 & 1.1 & 0.058  \\
051022  &  23:56:04.115 & +19:36:24.04 &  0.33 & CXO      & 0.059 & 23:56:04.10 & +19:36:24.16 & 2.0 & 0.001  \\
060202  &  02:23:23.010 & +38:23:03.20 &  1.1  & XRT      & 0.048 & 02:23:22.93 & +38:23:04.41 & 1.0 & 0.029  \\
060306  &  02:44:22.910 &$-$02:08:54.00 &  1.3  & XRT      & 0.035 & 02:44:22.89 &$-$02:08:54.74 & 1.2 & 0.037  \\
060319  &  11:45:32.890 & +60:00:39.10 &  0.9  & XRT      & 0.022 & 11:45:33.05 & +60:00:39.32 & 0.8 & 0.028  \\
060719  &  01:13:43.700 &$-$48:22:50.60 &  1.4  & XRT      & 0.008 & 01:13:43.70 &$-$48:22:51.31 &     & 0.086  \\ 
060814  &  14:45:21.320 & +20:35:10.63 &  0.4  & UKIRT    & 0.040 & 14:45:21.31 & +20:35:10.96 & 1.5 & 0.003  \\
060923A &  16:58:28.160 & +12:21:38.90 &  0.25 & VLT      & 0.060 & 16:58:28.14 & +12:21:38.74 & 1.0 & 0.004  \\
061222A &  23:53:03.419 & +46:31:58.60 &  0.2  & Gemini   & 0.102 & 23:53:03.42 & +46:31:58.97 & 0.6 & 0.005  \\
070306  &  09:52:23.320 & +10:28:55.40 &  1.1  & XRT      & 0.027 & 09:52:23.31 & +10:28:55.49 &     & 0.018  \\
070521  &  16:10:38.620 & +30:15:22.40 &  1.4  & XRT      & 0.027 & 16:10:38.68 & +30:15:22.86 & 1.0 & 0.072  \\
070802  &  02:27:35.680	&$-$55:31:38.9  &  0.3  & GROND    & 0.027 & 02:27:35.73 &$-$55:31:38.78 & 1.0 & 0.005  \\ 
071021  &  22:42:34.310 & +23:43:06.50 &  0.5  & TNG+NOT  & 0.064 & 22:42:34.31 & +23:43:06.02 & 0.9 & 0.006  \\
080207  &  13:50:02.980 & +07:30:07.40 &  0.4  & CXO      & 0.023 & 13:50:02.97 & +07:30:07.51 &     & 0.003  \\
080325  &  18:31:34.230 & +36:31:24.80 &  0.2  & Subaru   & 0.065 & 18:31:34.24 & +36:31:24.14 & 1.0 & 0.012  \\
080607  &  12:59:47.221 & +15:55:10.86 &  0.3  & KAIT     & 0.022 & 12:59:47.25 & +15:55:10.92 & 1.0 & 0.006  \\
081109  &  22:03:09.720 &$-$54:42:39.5  &  1.0  & REM      & 0.019 & 22:03:09.64 &$-$54:42:40.36 &     & 0.001  \\ 
081221  &  01:03:10.168 &$-$24:32:51.67 &  0.15 & Gemini   & 0.022 & 01:03:10.19 &$-$24:32:51.30 & 1.1 & 0.003  \\
090404  &  15:56:57.520 & +35:30:57.50 &  0.3  & PdBI     & 0.021 & 15:56:57.52 & +35:30:57.50 & 0.6 & 0.004  \\
090407  &  04:35:54.980 &$-$12:40:45.50 &  0.4  & CXO      & 0.067 & 04:35:55.03 &$-$12:40:45.38 & 1.0 & 0.006  \\
090417B &  13:58:46.590 & +47:01:05.00 &  1.0  & XRT      & 0.017 & 13:58:46.65 & +47:01:04.37 &     & 0.006  \\
090709A &  19:19:42.640 & +60:43:39.30 &  0.4  & P60      & 0.090 & 19:19:42.71 & +60:43:39.54 & 0.9 & 0.011  \\
\enddata 
\tablenotetext{(a)}{90\% confidence radius of the error circle.}
\tablenotetext{(b)}{Telescope providing the most precise afterglow position.  See text for references.}
\tablenotetext{(c)}{Foreground extinction, from \cite{Schlegel+1998}.}
\tablenotetext{(d)}{Minimum aperture radius used for photometry of the host galaxy.}
\tablenotetext{(e)}{Probability of inclusion of an equivalent or brighter galaxy in the error circle due to chance.}
\label{tab:positions}
\end{deluxetable*}

\begin{deluxetable*}{llrllrllll}  %*
\tabletypesize{\small}
\tablecaption{Host-Galaxy Photometry}
\tablecolumns{10}
\tablehead{
\colhead{GRB} &
\colhead{Filter} &
\colhead{} &
\colhead{Mag.} &
\colhead{Unc.} &
\colhead{} &
\colhead{Flux} &
\colhead{Unc.} &
\colhead{Telescope} &
\colhead{Reference}
\\
\colhead{} &
\colhead{} &
\colhead{} &
\colhead{} &
\colhead{} &
\colhead{} &
\colhead{($\mu$Jy)} &
\colhead{} &
\colhead{} &
\colhead{}
}
\startdata
050915A & $g$  & $=$ & 25.20 &  0.17 & $=$ &  0.33 &  0.06 & Keck-I/LRIS     &      \\
        & $V$  & $=$ & 25.06 &  0.11 & $=$ &  0.38 &  0.04 & Keck-I/LRIS     &      \\
        & $R$  & $=$ & 24.56 &  0.16 & $=$ &  0.49 &  0.08 & VLT/FORS2       & (a)  \\
        & $I$  & $=$ & 23.86 &  0.09 & $=$ &  0.73 &  0.06 & Keck-I/LRIS     &      \\
        & $K$  & $=$ & 20.69 &  0.24 & $=$ &  3.57 &  0.87 & VLT/ISAAC       & (a)  \\
        & 3.6  & $=$ & 18.88 &  0.10 & $=$ &  7.86 &  0.76 & {\it Spitzer}-IRAC    &      \\
        & 4.5  & $=$ & 18.38 &  0.10 & $=$ &  7.97 &  0.77 & {\it Spitzer}-IRAC    &      \\
\enddata
\tablenotetext{(*)}{Host photometry for all other galaxies can be found in the accompanying online supplement.  Magnitudes are given in the standard calibration system for each filter (Vega or SDSS; the AB system is used for HST magnitudes) with 1$\sigma$ uncertainty. Observations with no reference specified are from this work, although in some cases images are taken from previous studies and reanalyzed---see text for details.  Magnitudes are not corrected for foreground extinction, while flux values in $\mu$Jy are corrected for foreground extinction.}
\tablenotetext{(a)}{\cite{Hjorth+2012}; Malesani et al.\ 2013, in prep.}
\label{tab:hostphotometry}
\end{deluxetable*}

\begin{deluxetable*}{llrrrrc}  %*
%\rotate
\tabletypesize{\footnotesize}
\tablecaption{GRB Host-Galaxy Properties}
\tablecolumns{7}
\tablehead{
\colhead{GRB\tablenotemark{a}} &
\colhead{$z$\tablenotemark{b}} &
\colhead{SFR\tablenotemark{c}} &
\colhead{$M_*$\tablenotemark{d}} &
%\colhead{Maximum age\tablenotemark{d}} &
\colhead{$A_V$\tablenotemark{e}} &
\colhead{$\chi^2$/dof}
\\
\colhead{} &
\colhead{} &
\colhead{M$_\odot$\,yr$^{-1}$} &
\colhead{$10^9$\,M$_\odot$} &
%\colhead{Gyr} &
\colhead{mag} &
\colhead{}
}
\startdata
050915A & $2.53                $ & $135.8^{+ 63.1}_{- 48.2}$ & $ 36.7^{+ 16.0}_{-10.4}$ & $1.51^{+0.14}_{-0.20}$ & 4.8/4   \\
051008  & $2.90^{+0.28}_{-0.15}$ & $ 72.1^{+ 25.5}_{- 54.4}$ & $  4.9^{+ 16.1}_{- 0.9}$ & $0.85^{+0.07}_{-0.58}$ & 9.4/6   \\
051022  & $0.81                $ & $ 26.2^{+  7.1}_{-  6.6}$ & $ 17.8^{+  3.8}_{- 3.1}$ & $0.73^{+0.13}_{-0.17}$ & 18.2/12 \\
060202  & $0.79                $ & $  5.8^{+  1.1}_{-  2.0}$ & $  1.1^{+  0.5}_{- 0.1}$ & $1.00^{+0.10}_{-0.20}$ & 15.5/3  \\
060306  & $1.55                $ & $245.0^{+129.6}_{- 67.2}$ & $  8.0^{+  1.7}_{- 0.9}$ & $2.21^{+0.15}_{-0.12}$ & 11.5/6  \\
060319  & $1.17                $ & $  8.3^{+  6.9}_{-  4.3}$ & $ 21.4^{+  7.5}_{- 3.5}$ & $0.91^{+0.24}_{-0.24}$ & 11.5/3  \\
060719  & $1.53                $ & $  4.5^{+  4.9}_{-  1.0}$ & $ 13.2^{+  0.9}_{- 5.8}$ & $0.43^{+0.38}_{-0.10}$ & 0.7/2   \\
060814  & $1.92                $ & $238.2^{+ 49.6}_{- 24.0}$ & $  9.8^{+  0.9}_{- 1.2}$ & $1.23^{+0.09}_{-0.06}$ & 3.4/5   \\
060923A & $2.50^{+0.58}_{-0.52}$ & $ 88.4^{+ 37.6}_{- 30.3}$ & $103.3^{+  6.5}_{-13.7}$ & $1.89^{+0.15}_{-0.14}$ & 11.3/13 \\
061222A & $2.09                $ & $  2.7^{+  0.2}_{-  0.2}$ & $  1.3^{+  1.1}_{- 0.6}$ & $0.00^{+0.00}_{-0.00}$ & 13.3/6  \\
070306  & $1.50                $ & $ 12.8^{+ 14.9}_{-  1.8}$ & $ 49.1^{+  1.1}_{-19.6}$ & $0.13^{+0.39}_{-0.05}$ & 29.4/17 \\
070521  & $1.70^{+1.04}_{-0.36}$ & $ 40.4^{+ 62.1}_{-  3.0}$ & $ 30.8^{+ 18.9}_{- 4.1}$ & $2.21^{+0.15}_{-0.11}$ & 12.7/9  \\
070802  & $2.45                $ & $ 16.1^{+ 14.2}_{-  4.8}$ & $  4.9^{+  1.6}_{- 1.6}$ & $0.74^{+0.27}_{-0.17}$ & 10.4/5  \\
071021  & $2.45                $ & $190.3^{+ 25.6}_{- 20.3}$ & $119.6^{+  6.6}_{- 8.8}$ & $1.92^{+0.06}_{-0.05}$ & 33.2/7  \\
080207  & $2.09                $ & $ 46.2^{+271.9}_{- 44.7}$ & $120.2^{+ 54.3}_{-48.1}$ & $2.36^{+0.48}_{-0.38}$ & 48.3/15 \\
080325  & $1.78                $ & $ 12.9^{+  5.4}_{-  4.1}$ & $ 98.7^{+ 22.7}_{-15.2}$ & $1.13^{+0.16}_{-0.19}$ & 35.2/10 \\
080607  & $3.04                $ & $ 19.1^{+  7.4}_{-  4.9}$ & $ 40.7^{+  6.3}_{- 4.3}$ & $1.15^{+0.16}_{-0.14}$ & 9.5/5   \\
081109  & $0.98                $ & $ 49.0^{+ 10.7}_{- 10.6}$ & $  9.4^{+  1.5}_{- 1.0}$ & $1.25^{+0.10}_{-0.13}$ & 19.1/16 \\
081221  & $2.26                $ & $172.8^{+ 22.8}_{- 30.1}$ & $ 37.0^{+ 11.0}_{-12.0}$ & $1.71^{+0.07}_{-0.09}$ & 22.0/9  \\
090404  & $3.00^{+0.83}_{-1.82}$ & $ 98.8^{+122.4}_{- 98.8}$ & $ 54.7^{+ 50.2}_{-39.5}$ & $1.82^{+0.66}_{-0.64}$ & 17.1/8  \\
090407  & $1.45                $ & $ 28.1^{+ 14.8}_{-  9.5}$ & $ 15.2^{+  6.1}_{- 4.8}$ & $2.10^{+0.22}_{-0.20}$ & 12.8/6  \\
090417B & $0.34                $ & $  0.5^{+  0.3}_{-  0.3}$ & $  3.5^{+  0.3}_{- 0.5}$ & $0.87^{+0.12}_{-0.08}$ & 21.4/15 \\
090709A & $1.80^{+0.46}_{-0.71}$ & $  8.0^{+  4.1}_{-  4.1}$ & $  9.6^{+  4.6}_{- 1.7}$ & $1.40^{+0.12}_{-0.38}$ & 13.7/7  \\
\hline
970228  & $0.69                $ & $  0.5^{+  0.2}_{-  0.1}$ & $  0.3^{+  0.1}_{- 0.1}$ & $0.63^{+0.17}_{-0.15}$ & 11.3/4  \\
970508  & $0.83                $ & $  1.6^{+ 12.7}_{-  0.6}$ & $  0.2^{+  0.2}_{- 0.0}$ & $0.84^{+0.76}_{-0.19}$ & 8.3/3   \\
970828  & $0.96                $ & $ 35.0^{+ 12.6}_{-  7.6}$ & $  0.9^{+  0.2}_{- 0.1}$ & $2.13^{+0.10}_{-0.09}$ & 12.3/4  \\
971214  & $3.42                $ & $ 58.9^{+ 31.8}_{-  8.9}$ & $  7.1^{+  2.6}_{- 2.4}$ & $1.35^{+0.18}_{-0.10}$ & 4.3/3   \\
980613  & $1.10                $ & $ 17.9^{+  6.7}_{-  7.3}$ & $  0.4^{+  0.2}_{- 0.1}$ & $1.02^{+0.14}_{-0.19}$ & 20.1/2  \\
980703  & $0.97                $ & $ 37.0^{+ 13.1}_{-  3.3}$ & $  5.8^{+  0.4}_{- 2.0}$ & $1.10^{+0.07}_{-0.06}$ & 23.5/5  \\
990123  & $1.60                $ & $108.2^{+ 63.6}_{- 50.8}$ & $  0.7^{+  0.3}_{- 0.1}$ & $1.21^{+0.17}_{-0.19}$ & 4.8/4   \\
990506  & $1.31                $ & $  0.6^{+  3.2}_{-  0.1}$ & $ 26.5^{+  7.7}_{-21.5}$ & $0.00^{+1.07}_{-0.00}$ & 0.0/0   \\
990705  & $0.84                $ & $  4.4^{+  0.5}_{-  0.4}$ & $113.0^{+ 22.0}_{-18.8}$ & $0.00^{+0.00}_{-0.00}$ & 5.2/0   \\
990712  & $0.43                $ & $  0.0^{+  0.0}_{-  0.0}$ & $  1.6^{+  0.2}_{- 0.1}$ & $0.00^{+0.00}_{-0.00}$ & 13.7/5  \\
991208  & $0.71                $ & $  1.0^{+  0.6}_{-  0.2}$ & $  0.7^{+  0.2}_{- 0.2}$ & $0.49^{+0.25}_{-0.17}$ & 4.6/3   \\
000210  & $0.85                $ & $  0.0^{+  0.3}_{-  0.0}$ & $  2.1^{+  0.3}_{- 0.3}$ & $0.05^{+0.35}_{-0.05}$ & 17.0/6  \\
000418  & $1.12                $ & $ 52.4^{+ 13.7}_{-  8.8}$ & $  0.8^{+  0.2}_{- 0.1}$ & $1.30^{+0.06}_{-0.07}$ & 12.8/5  \\
000911  & $1.06                $ & $  2.7^{+ 79.0}_{-  1.9}$ & $  1.2^{+  1.3}_{- 0.9}$ & $0.80^{+1.42}_{-0.80}$ & 1.3/3   \\
000926  & $2.04                $ & $  8.2^{+ 19.9}_{-  3.9}$ & $  4.4^{+ 60.8}_{- 3.8}$ & $0.58^{+0.39}_{-0.29}$ & 0.5/2   \\
010222  & $1.48                $ & $  0.6^{+  0.6}_{-  0.1}$ & $  0.7^{+  0.3}_{- 0.4}$ & $0.05^{+0.33}_{-0.05}$ & 11.6/4  \\
010921  & $0.45                $ & $  2.7^{+  0.7}_{-  0.4}$ & $  4.1^{+  0.5}_{- 0.2}$ & $0.48^{+0.14}_{-0.10}$ & 21.6/11 \\
011121  & $0.36                $ & $  1.0^{+  0.1}_{-  0.1}$ & $ 13.5^{+  5.4}_{- 4.2}$ & $0.00^{+0.00}_{-0.00}$ & 2.2/2   \\
011211  & $2.14                $ & $  7.0^{+ 19.4}_{-  0.0}$ & $  0.1^{+  0.3}_{- 0.0}$ & $0.19^{+0.70}_{-0.00}$ & 8.1/0   \\
020405  & $0.69                $ & $ 11.6^{+  4.1}_{-  2.7}$ & $  8.8^{+  1.9}_{- 1.3}$ & $0.82^{+0.18}_{-0.15}$ & 18.5/3  \\
020813  & $1.25                $ & $  1.5^{+  1.4}_{-  0.2}$ & $  9.5^{+ 13.8}_{- 7.3}$ & $0.00^{+0.16}_{-0.00}$ & 6.7/1   \\
020819B & $0.41                $ & $  5.8^{+  1.4}_{-  0.5}$ & $ 84.9^{+  2.1}_{- 2.0}$ & $0.00^{+0.05}_{-0.00}$ & 43.5/6  \\
020903  & $0.25                $ & $  0.0^{+  0.0}_{-  0.0}$ & $  0.5^{+  0.2}_{- 0.0}$ & $0.34^{+0.00}_{-0.34}$ & 3.2/0   \\
021004  & $2.33                $ & $ 14.8^{+  3.7}_{-  2.0}$ & $  2.8^{+  1.0}_{- 0.6}$ & $0.42^{+0.09}_{-0.07}$ & 20.1/6  \\
021211  & $1.01                $ & $  8.3^{+  4.6}_{-  0.7}$ & $  2.0^{+  1.1}_{- 1.0}$ & $1.78^{+0.27}_{-0.09}$ & 2.4/0   \\
030328  & $1.52                $ & $ 25.1^{+ 51.2}_{- 15.4}$ & $  0.6^{+  3.5}_{- 0.3}$ & $1.06^{+0.26}_{-0.29}$ & 0.8/5   \\
030329  & $0.17                $ & $  0.2^{+  0.1}_{-  0.1}$ & $  0.1^{+  0.0}_{- 0.0}$ & $0.58^{+0.15}_{-0.15}$ & 11.2/11 \\
030528  & $0.78                $ & $  6.8^{+  4.5}_{-  0.8}$ & $  2.1^{+  0.9}_{- 1.1}$ & $0.00^{+0.25}_{-0.00}$ & 4.2/4   \\
031203  & $0.10                $ & $ 14.1^{+  0.5}_{-  0.3}$ & $  0.3^{+  0.0}_{- 0.0}$ & $0.34^{+0.02}_{-0.02}$ & 239.8/1 \\
040924  & $0.86                $ & $  0.9^{+  0.7}_{-  0.9}$ & $  1.7^{+  1.6}_{- 1.2}$ & $0.00^{+0.20}_{-0.00}$ & 1.0/6   \\
041006  & $0.71                $ & $  0.3^{+  1.0}_{-  0.1}$ & $  0.8^{+  5.1}_{- 0.7}$ & $0.00^{+0.53}_{-0.00}$ & 0.1/1   \\
\enddata 
\tablenotetext{(a)}{Burst name.  Events below the horizontal rule indicate the pre-\Swift\ comparison sample.}
\tablenotetext{(b)}{Redshift.  Uncertainties indicate a photometric redshift and correspond to the 95\% confidence range.}
\tablenotetext{(c)}{Dust-corrected star-formation rate, calculated from the UV/optical/IR SED fit.  Uncertainties on all parameters except redshift are 1$\sigma$.}
\tablenotetext{(d)}{Total stellar mass.}
\tablenotetext{(e)}{Average dust attenuation.}
\label{tab:hostmodelpar}
\end{deluxetable*}

%\bibliography{ref}{}
%\bibliographystyle{apj}

\end{document}